\documentclass{lmcs}

\usepackage{amsmath,amsfonts,amssymb,amsthm}
\usepackage{xcolor}
\usepackage{mathtools}
\usepackage{enumitem,booktabs}
\usepackage{tikz}
\usetikzlibrary{positioning,shapes.geometric,calc,arrows.meta,decorations.pathreplacing}

\newtheorem{cla}{Claim}[thm]

\newcommand{\UIp}[2]{{\mathcal{U}^{\smash{#1}}_{#2}}}
\newcommand{\equivp}[2]{\mathrel{{\equiv^{\smash{#1}}_{#2}}}}
\newcommand{\classp}[3]{{[#1]^{\smash{#2}}_{#3}}}

\newcommand{\gsig}{\tau}

\newcommand{\sig}{\sigma}
\newcommand{\priv}{\rho}

\newcommand{\Cons}{\textup{Cons}}
\newcommand{\signat}{\textup{sig}}

\newcommand{\set}[1]{\left\{#1\right\}}
\newcommand{\abs}[1]{\left|#1\right|}
\newcommand{\eqdef}{\mathrel{\vcentcolon=}}

\newcommand{\Nc}{\overline{N}}
\newcommand{\supp}{\operatorname{supp}}

\newcommand{\PC}{P}
\newcommand{\QC}{\mathcal Q_C}
\newcommand{\QCS}[1]{\mathcal Q_C[#1]}

\newcommand{\ML}{\textup{ML}}
\newcommand{\M}{\mathfrak{M}}
\newcommand{\N}{\mathfrak{N}}
\newcommand{\F}{\mathfrak{F}}
\renewcommand{\phi}{\varphi}
\newcommand{\Prop}{\textit{Prop}}
\newcommand{\PMod}{\textup{PMod}}
\newcommand{\Cov}{\mathfrak{K}}

\newcommand{\MForm}[3]{\Phi(#3)} 
\newcommand{\CForm}[3]{\Gamma(#3)}%
\newcommand{\Atom}[2]{{\textup{At}_{#2}^{\smash{#1}}}}
\newcommand{\CellAt}[3]{\textup{Cell}_{#2,#3}^{\smash{#1}}}

\newcommand{\dens}{\mathrm{UnifInt}}
\newcommand{\DCraig}{\mathrm{DCraig}}
\newcommand{\SetClash}{\mathrm{InCon}}
\newcommand{\SetCraig}{\mathrm{Craig}}
\newcommand{\NoCraig}{\mathrm{NoCraig}}

\newcommand{\Diff}{\mathsf{Diff}}
\newcommand{\Sfour}{\mathsf{S4}}

\newcommand{\Grz}{\mathsf{Grz}}
\newcommand{\Grzd}{\Grz_{[\Delta]}}

\newcommand{\sh}{\mathrm{sh}}

\newcommand{\nsets}{\Lambda}
\newcommand{\pin}[2]{#2^{#1}}
\newcommand{\fpin}[2]{#2^{-#1}}
\newcommand{\afpin}[2]{#2^{\downarrow#1}}
\newcommand{\pinf}{\pi}
\newcommand{\musig}{\mu}

\newcommand{\normalised}[1]{\textup{norm}(#1)}
\newcommand{\restrict}[2]{#1{\restriction_{#2}}}
\newcommand{\atomcolours}{F}
\newcommand{\scol}{s}
\newcommand{\rcol}{r}
\newcommand{\flexcompcol}[1]{\mathsf{Flex}(#1)}
\newcommand{\rigcompcol}[1]{\mathsf{Rigid}(#1)}

\newcommand{\somerigid}{\mathcal{S}}
\newcommand{\product}[1]{\boldsymbol#1}		
\newcommand{\totweight}{w}		
\newcommand{\rigweight}{w^{\textsf{R}}}
\newcommand{\flxweight}{w^{\textsf{F}}}	
\newcommand{\intervals}{\mathcal{J}}

\newcommand{\shpi}{\sh}

\newcommand{\nb}[1]{\hbox to 0pt{\textcolor{red}{!}}\marginpar{\footnotesize\color{red}#1}}

\usepackage{tikz}
\usetikzlibrary{positioning,calc,arrows.meta,decorations.pathreplacing}

\colorlet{greencolour}{green!16}
\colorlet{bluecolour}{blue!12}
\colorlet{orangecolour}{orange!25}
\colorlet{darkgreencolour}{green!60}
\colorlet{darkbluecolour}{blue!40}
\colorlet{darkorangecolour}{orange!70}

\tikzset{
	wnode/.style   = {circle, draw=black, fill=white, semithick,
		minimum size=6mm, inner sep=0.5pt, font=\small},
	rel/.style     = {thick, black!75, ->, >=latex},
	lvl/.style     = {densely dotted, black!30},
	lvlab/.style   = {font=\footnotesize, black!55, anchor=east},
	brk/.style     = {font=\footnotesize, black!60},
	imlab/.style   = {font=\scriptsize},
}

\title{The Density of Interpolants in Modal Logic: First Results}

\author{
	Roman Kontchakov \and
	Louwe Kuijer \and
	Frank Wolter
}

\date{}

\begin{document}
\maketitle

\begin{abstract}
	We initiate a quantitative study of interpolation in propositional modal logic. Instead of asking only whether a logic has Craig or uniform interpolation, we measure the density of formulas, or pairs of formulas, admitting interpolants. For quasi-normal modal logics, formulas of bounded modal depth over finite signatures are counted modulo logical equivalence. We show that interpolation failure is local to cells determined by the retained signature, yielding product decompositions of both uniform and Craig interpolation densities. These decompositions allow us to analyse their asymptotic behaviour as either the retained or the forgotten signature grows. When the retained signature grows, we obtain general results for all quasi-normal modal logics; when the forgotten signature grows, the behaviour is subtler, particularly when formulas are counted after existential quantification over forgotten variables. For locally tabular logics, we introduce cell incompatibility graphs and use them for a fine-grained analysis covering tabular logics, the difference logic $\Diff$, and bounded-depth extensions of $\Grz$.  
\end{abstract}

\section{Introduction}

The Craig interpolation property states that whenever an implication $\varphi\to\psi$ is valid, there exists a formula~$\theta$, using only the propositional variables common to $\varphi$ and $\psi$, such that both~$\varphi\to\theta$ and $\theta\to\psi$ are valid~\cite{craig_1957}. Uniform interpolation strengthens this requirement: given a formula $\varphi$ and a retained signature $\sigma$, one seeks a single formula over $\sigma$ that captures all consequences of $\varphi$ expressible in $\sigma$ and variables not used in $\varphi$~\cite{DBLP:journals/jsyml/Pitts92}. Both properties have been extensively studied in modal logic; see~\cite{tenCate2026Theory} for a recent introduction and surveys of interpolation and its applications.

The traditional study of interpolation is qualitative: a logic either enjoys an interpolation property or it does not. While fundamental, this distinction provides little information about the prevalence of interpolation failure. A logic may fail interpolation because of a small collection of exceptional configurations, or failures may be widespread among its formulas. This raises a natural quantitative question:
\begin{quote}
	\emph{How frequent are formulas, or pairs of formulas, for which interpolants exist?}
\end{quote}
In this paper, we initiate a systematic investigation of this question for quasi-normal propositional modal logics. We introduce densities of uniform and Craig interpolants by counting formulas modulo logical equivalence and study their asymptotic behaviour as the number of retained or forgotten propositional variables grows. We also investigate an alternative counting method that identifies formulas after existential quantification of the forgotten variables.
Our results show that the choice of counting method matters. After existential quantification the densities can exhibit substantially different behaviour, revealing distinctions between modal logics that are invisible under ordinary counting.

To obtain a finite counting space, we fix two disjoint finite signatures $\sigma$ and $\rho$ of retained and forgotten propositional variables, respectively, and a modal depth bound $d$. At bounded depth, formulas over a finite signature form only finitely many equivalence classes. For locally tabular logics, the depth bound can be dropped.

Formulas of depth at most $d$ over $\sigma\cup\rho$ can be identified modulo logical equivalence with subsets of the finite set of complete $(\sigma\cup\rho,d)$-types realised at distinguished worlds of Kripke models. We call these types $(\sigma\cup\rho,d)$-\emph{atoms}. We sample formulas uniformly modulo logical equivalence by choosing a subset of these atoms uniformly at random. Writing $n=|\sigma|$ and~$m=|\rho|$, we define the uniform interpolation density
\[
\dens^{d}_{n,m}
\]
as the proportion of these formulas that admit a uniform interpolant over $\sigma$.
 
For Craig interpolation, we work with pairs of formulas $\varphi,\psi$ over signatures $\sigma\cup\rho$ and $\sigma\cup\rho'$, respectively, such that $|\rho|=|\rho'|$. We consider pairs for which $\varphi\land\psi$ is inconsistent and ask whether their inconsistency can be witnessed by a formula over the common signature $\sigma$. This is equivalent to the usual formulation of Craig interpolation for valid implications. We define $\DCraig^{d}_{n,m}$ as the proportion of such inconsistent pairs admitting a Craig interpolant.

A first observation of the paper is that these densities can be decomposed into densities over $\sig^{\uparrow\priv}$-cells, where a $\sig^{\uparrow\priv}$-cell is a set of $(\sigma\cup\rho,d)$-atoms which cannot be distinguished, up to depth $d$, using the language over $\sigma$ alone. We prove that interpolation failure is local to cells and the global densities can be computed as products of local densities:
\[
\dens^{d}_{n,m}
=
\prod_C \dens_C,
\qquad
\DCraig^{d}_{n,m}
=
\prod_C \DCraig_C,
\]
where $C$ ranges over the $\sig^{\uparrow\priv}$-cells. This basic \emph{cell reduction principle} underlies all of our asymptotic results.

We begin with the regime in which $\rho$ is fixed and $\sigma$ grows. Suppose that interpolation fails for $\sigma_{0},\rho_{0}$, and modal depth $d$. For any $\sigma\supseteq \sigma_{0}$ of size $n$, the original `obstruction' can be copied into $2^{n}$ many distinct cells of the same size as the original one. Each copy contributes a fixed factor strictly below $1$ to the product decomposition, giving, for all $m\geq|\rho_{0}|$,
\begin{align*}
\lim_{n\to\infty}\dens_{n,m}^{d}& =0 \quad \text{ if uniform interpolation fails},
\intertext{and}
\lim_{n\to\infty}\DCraig_{n,m}^{d} & =0 \quad \text{ if Craig interpolation fails}.
\end{align*}
The opposite regime, in which $\sigma$ is fixed and $\rho$ grows, presents a different picture. As we show below, uniform interpolation admits a general convergence theorem in this regime, whereas the asymptotic behaviour of Craig interpolation remains open for arbitrary quasi-normal modal logics. For uniform interpolation, inside every cell there is a large family of atoms ($2^{m}$~many) which cannot occur in any formula where uniform interpolation fails. Consequently the failure probability inside a cell is at most
$2^{-2^m}$. Since the number of cells depends only on $\sigma$, it follows that, for all $n\geq 0$,
\[
\lim_{m\to\infty}\dens^{d}_{n,m}=1.
\]
For Craig interpolation, we focus on locally tabular quasi-normal modal logics, where no modal depth bound is needed. Within each cell $C$, inconsistent pairs correspond to ordered bicliques in an incompatibility graph $H_C$ on its atoms. Unlike in the uniform case, the large family of universally compatible atoms does not suffice to control the density, since large bicliques may still occur. To address this, we introduce the \emph{canonical profile of $H_{C}$} 
as its coarsest quotient preserving and reflecting
compatibility. 
Craig interpolation holds precisely when every cell
has a singleton canonical profile. Relaxing this
condition, we say that a logic~$L$ has the
\emph{uniformly bounded intra-cell incompatibility
	property \textup{(}uniform BICIP\textup{)}} if the sizes of canonical
profiles are bounded by a constant depending only
on $L$. We say that $L$ has \emph{BICIP} if the
bound may also depend on $n$, but not on $m$.
Both properties imply that, for every fixed $n$,
\[
\lim_{m\to\infty}\DCraig_{n,m}=1.
\]
We show that every tabular quasi-normal modal
logic has uniform BICIP.

\medskip

A natural alternative way of counting is to abstract from distinctions between formulas that disappear once the forgotten or non-shared variables are existentially quantified away. Thus, instead of counting the number of $\varphi$ itself, we count the number of $\exists\vec\rho.\varphi$ (up to logical equivalence).
For Craig interpolation, the two components are treated separately, using $\exists\vec\rho.\varphi$ and $\exists\vec\rho^{\,\prime}.\psi$. We denote the corresponding densities by
\[
\dens^{\Pi,d}_{n,m}
\qquad\text{and}\qquad
\DCraig^{\Pi,d}_{n,m}.
\]
We call these the uniform and Craig interpolation densities
\emph{after existential quantification}. Unlike ordinary modal
equivalence, equivalence after existential quantification
may depend on the chosen class of frames since two classes
determining the same modal logic may induce different
equivalence relations on existentially quantified formulas.
Thus, counting up to equivalence after existential
quantification depends not only on the logic but also on
the chosen frame class. The natural choice is the class
of all frames validating the logic, although we shall
also consider well-motivated alternatives.

For this new counting measure, the cell decomposition
remains valid, but the asymptotic behaviour can change
substantially. When $m$ is fixed and $n$ grows, the
previous asymptotic results continue to hold. The more
interesting case is that of fixed $n$ and growing $m$,
where existential quantification can identify formulas
that are distinct under ordinary modal equivalence.

Table~\ref{table:beh} summarises the results for this regime. 
With the exception of the difference logic $\Diff$ (see below),
our results on densities after existential quantification
are shown for the class of all frames validating the logic.
For tabular logics, there are only finitely many
equivalence classes of existentially quantified formulas
for any fixed $n$. Hence, no substantive asymptotic analysis
is required: the density sequences $\smash{\dens^\Pi_{n,m}}$
and $\smash{\DCraig^\Pi_{n,m}}$ eventually stabilise as
$m\to\infty$. This contrasts with their ordinary
counterparts, which converge to $1$. The difference
arises because large families of distinct atoms that
determine convergence under ordinary counting collapse to
a single equivalence class after existential quantification.

In case studies, we investigate the locally tabular logics $\Grzd$, for $\Delta\geq 1$, and the difference logic~$\Diff$. The former are determined by finite partial orders of depth at most $\Delta$. Craig interpolation fails for $\Grzd$ if and only if $\Delta\geq 3$~\cite[Chapter~8]{MGabbay2005-MGAIAD}. Counterexamples are, intuitively, rare as they require intricate incompatible arrangements of paths. By contrast, $\Diff$, the logic of difference frames $(W,\ne)$~\cite{vonWright79,Segerberg1980-SEGANO-2}, admits straightforward counterexamples to Craig interpolation enforcing models of incompatible sizes~\cite{DBLP:conf/aiml/KuruczWZ24}.

This distinction is not directly visible under ordinary counting,
where all the corresponding interpolation densities converge
to $1$. For $\Grzd$, convergence of the Craig interpolation
density follows from BICIP (and even uniform BICIP when
$\Delta\le3$). For $\Diff$, it follows from a more involved
analysis of quotients of incompatibility graphs.

The distinction becomes apparent after existential
quantification. For $\Grzd$, the quantified uniform
interpolation density still converges to $1$ and
the quantified Craig interpolation density remains
bounded away from $0$ (whether it tends to 1 is open). By contrast, for $\Diff$, both
quantified densities converge to $0$ over the class
of difference frames. We note that not every frame
validating $\Diff$ is a difference frame, but leave
the asymptotic analysis over the full class of
$\Diff$-validating frames for future work.

\begin{table}[t]
	\centering%
	\renewcommand{\tabcolsep}{4pt}%
	\renewcommand{\arraystretch}{1.35}%
	\begin{tabular}{lcccc}
		\toprule
		& \multicolumn{2}{c}{\textbf{ordinary counting}}
		& \multicolumn{2}{c}{\textbf{after existential quantification}}
		\\
		\textbf{logic}
		& $\dens_{n,m}$
		& $\DCraig_{n,m}$
		& $\dens^\Pi_{n,m}$
		& $\DCraig^\Pi_{n,m}$
		\\
		\midrule
		
		all quasi-normal $L$
		& $\to 1$
		& ?
		& ---
		& ---
		\\
		
		$L$ with BICIP
		& $\to 1$
		& $\to 1$
		& bounded away from $0$
		& bounded away from $0$
		\\
		
		tabular $L$
		& $\to 1$
		& $\to 1$
		& stabilises 
		& stabilises 
		\\
		
		$\Grzd, \Delta\geq 3$
		& $\to 1$
		& $\to 1$
		& $\to 1$
		& bounded away from $0$
		\\

		$\Diff$
		& $\to 1$
		& $\to 1$
		& $\to 0^*$
		& $\to 0^*$
		\\
		
		\bottomrule
	\end{tabular}%
\caption{Asymptotic behaviour of interpolant densities for fixed $n=|\sigma|$ as $m=|\rho|\to\infty$. The superscript $\Pi$ denotes counting modulo logical equivalence after existential quantification over the forgotten or non-shared variables. A question mark indicates that convergence to $1$ is open. A dash indicates that no universal result holds. The results for $\Diff$ marked with $^*$  hold for the class of difference frames.}
\label{table:beh}
\end{table}

\subsection{Related work}

Quantitative and asymptotic questions about logical properties have been
studied along two broadly independent lines; the density of interpolant
existence considered here sits at their intersection but does not seem to
have been addressed before. Also related is work on the complexity of interpolant existence for logics without Craig or uniform interpolation and work on repairing Craig interpolation minimally.

\paragraph{Density of truth.}
Closest in method is the \emph{density of truth} programme, which fixes a
signature and asks for the fraction of length-$n$ formulas that are
tautologies as \mbox{$n\to\infty$}~\cite{MoczuradTyszkiewiczZaionc2000,%
	Zaionc2006,KostrzyckaZaionc2004,KostrzyckaZaionc2008,%
	FournierGardyGenitriniZaionc2007,FournierGardyGenitriniZaionc2010}.
Of particular relevance, Kostrzycka and Zaionc compute the density of truth
for the $\{\to,\Box\}$ fragments of $\mathsf{S5}$ and
$\mathsf{S4}$~\cite{KostrzyckaZaioncModal2006}. Related fragment
computations include Dummett's logic and monadic intermediate
logics~\cite{Kostrzycka2003,Kostrzycka2007}. In contrast to our approach, counting is \emph{not} done modulo logical equivalence in this programme. In fact, since all tautologies are logically equivalent, this assumption would trivialise the density of truth as one would count only one tautology.

\paragraph{Zero-one laws.}
The older strand concerns randomness over \emph{structures} rather than
formulas. Glebskii et al.~\cite{Glebskii1969} and Fagin~\cite{Fagin1976} established that every first-order
sentence holds in almost all or almost no finite
structures. In the modal setting,
Halpern and Kapron proved zero-one laws for model validity in
$\mathsf{K}$, $\mathsf{T}$, $\mathsf{S4}$, $\mathsf{S5}$~\cite{HalpernKapron1994}; see also~\cite{LeBars2002,GorankoKapron2003,HalpernKapron2003}. More recently Verbrugge established
zero-one laws for provability logic and Grzegorczyk's logic~\cite{Verbrugge2021}. Our results are analogous in spirit to zero-one laws, although our probability spaces consist of formulas modulo logical equivalence rather than finite structures.

\paragraph{Deciding interpolant existence.}
The property whose density we measure has been studied qualitatively, as a decision problem, when Craig or uniform interpolation fails. For decidability of Craig interpolant existence, there are results for modal logics with nominals~\cite{DBLP:journals/tocl/ArtaleJMOW23,DBLP:conf/lics/JungKW26}, decidable fragments of FO~\cite{DBLP:conf/lics/JungW21}, difference logic and weak $\mathsf{K4}$~\cite{DBLP:conf/aiml/KuruczWZ24}, and logics extending $\mathsf{K4.3}$~\cite{kurucz2026non}. Decidability of uniform interpolant existence has mainly been investigated in description logic~\cite{DID_I_DAMAGE,DBLP:conf/kr/LutzSW12}. 

\paragraph{Repairing Craig interpolation.} Another way to deal with the failure of interpolation is to restore it by extending the expressive power of the language~\cite{DBLP:journals/apal/ArecesBM03}. This raises the natural question of whether there is a minimal extension, in a suitable sense, that has interpolation, and if so, what it is. This question has been studied for, among others, modal logic with nominals~\cite{DBLP:journals/jsyml/Cate05}, linear temporal logic~\cite{DBLP:conf/csl/GheerbrantC09}, and fragments of first-order logic~\cite{DBLP:journals/lmcs/CateC25}.

\section{Preliminaries}

We introduce basic notation for modal logic~\cite{DBLP:books/daglib/0030819,goranko20075}.
Fix a countably infinite set $\Prop$ of \emph{propositional variables}. A \emph{signature}
$\gsig$ is a finite subset of $\Prop$. The language $\ML$ of \emph{modal formulas} is built
from propositional variables using $\top$, $\neg$, $\wedge$ and the modal operator~$\Box$;
we use standard connectives $\bot$, $\vee$ and $\rightarrow$ as abbreviations and set $\Diamond\phi\eqdef\neg\Box\neg\phi$. We also use $\Box^{\leq d}$ defined as $\bigwedge_{i = 0}^d\Box^i\varphi$, where $\Box^i$ stands for $i$ operators $\Box$.  We write $\signat(\phi)$ for the \emph{signature of $\varphi$}, the finite set
of propositional variables occurring in~$\phi$, and $\textup{md}(\phi)$ for the \emph{modal depth} of $\phi$,
the maximal nesting of~$\Box$. For a signature $\gsig\subseteq \Prop$ and a depth $d\ge 0$, we put
\begin{equation*}
\ML^{d}_{\gsig}\eqdef\bigl\{\phi\in\ML\bigm| \signat(\phi)\subseteq\gsig,\ \textup{md}(\phi)\le d\bigr\}
\quad\text{ and }\quad
\ML_{\gsig}\eqdef\bigcup_{d\ge 0}\ML^{d}_{\gsig}.
\end{equation*}

Modal formulas are interpreted in Kripke frames, or simply frames. A \emph{frame} is a triple
$\F=(W,R,A)$, where $W$ is a non-empty set of \emph{worlds}, $R\subseteq W \times W$ is a binary \emph{accessibility relation} on $W$, and $A\subseteq 2^{W}$ is a set of \emph{admissible sets} containing $W$, closed under the Boolean operations
and unary $F$ defined by taking $F(U)=\bigl\{w\in W \mid   w'\in U,\linebreak\text{for all } w' \in W \text{ with } wRw'  \bigr\}$. If every subset of $W$ is admissible (that is, $A=2^W$),
we omit~$A$ and write~$(W,R)$. Unless stated otherwise,
we make this assumption whenever $W$ is finite. A \emph{model}
based on~$\F$ is a pair $\M=(\F,V)$ with a \emph{valuation}~$V\!\colon\Prop\to A$. A \emph{pointed model} is a pair of the form $(\M,w)$ with~$w\in W$. The
\emph{satisfaction relation} between pointed models~$(\M,w)$ and formulas is defined as follows:
\begin{itemize}
	\item $\M,w\models p$ iff $w\in V(p)$;
	\item $\M,w\models \top$;
	\item $\M,w\models\neg\phi$ iff $\M,w\not\models\phi$;
	\item $\M,w\models\phi\wedge\psi$ iff $\M,w\models\phi$ and $\M,w\models\psi$;
	\item $\M,w\models\Box\phi$ iff $w\in F(\{ w'\in W \mid \M, w'\models\phi\})$.
\end{itemize}

A formula $\phi$ is \emph{valid} in a pointed frame $(\F,w)$, written $\F,w\models\phi$, if
$\M,w\models\phi$ for every model $\M$ based on $\F$. For a class $\mathcal K$ of pointed
frames,
\begin{equation*}
\textup{Log}(\mathcal K)\eqdef\bigl\{\phi\in\ML\bigm| \F,w\models\phi \text{ for all }(\F,w)\in\mathcal K\bigr\}
\end{equation*}
is the \emph{logic determined by $\mathcal K$}; it is a quasi-normal modal logic. Unless stated otherwise, we assume that pointed frames $(\F,w)$ and models $(\M,w)$ are \emph{rooted} with \emph{root} $w$ in the sense that all $v\in W$ are reachable from $w$ along $R$-paths. Every quasi-normal modal logic is determined by rooted frames~\cite{DBLP:books/daglib/0030819}. 

In this article, many results are shown for classes of logics such as the class of all quasi-normal modal logics, the class of all tabular such logics, and the class of all such logics with (uniform) BICIP. However, we also consider a few concrete logics including
	\begin{itemize}
		\item Grzegorczyk's logic, $\Grz$, determined by the class of finite partial orders
		(reflexive, transitive, and antisymmetric frames);
	    \item for $\Delta\geq 1$, the logics $\Grzd$ determined by
	    finite partial orders of depth at most $\Delta$  (here, depth
	    counts worlds in a strict chain; in particular, a singleton
	    has depth $1$, see~\cite{Segerberg1971} and \cite[Section~8.6]{DBLP:books/daglib/0030819});
        \item the logic $\textsf{Diff}$ determined by all difference frames $(W,\ne)$, which was first considered as the `logic of elsewhere'~\cite{vonWright79,Segerberg1980-SEGANO-2}; see also~\cite{DBLP:journals/jsyml/Rijke92}. 
	 \end{itemize}   
Throughout, we fix a quasi-normal logic $L$ and, unless stated otherwise, let $\mathcal K$ denote the class of all pointed rooted frames in which $L$ is valid. Let
$\PMod_{L}$ denote the class of \emph{pointed models} over the full set of variables $\Prop$ based on frames in~$\mathcal{K}$. Semantic
consequence is taken over this class: for $\phi,\psi\in\ML$, we write $\phi\models\psi$
if every $(\M,w)\in\PMod_{L}$ with~$\M,w\models\phi$ also satisfies $\M,w\models\psi$, and
$\phi\equiv\psi$ if $\phi\models\psi$ and $\psi\models\phi$. So, all entailments below are over~$L$, but we omit the subscript to avoid clutter.

For pointed models $(\M,w),(\M',w')\in\PMod_{L}$, a signature $\gsig$ and a depth $d$, put
	\[
	(\M,w)\equiv^{d}_{\gsig}(\M',w')
	\qquad\text{ iff }\qquad
	\M,w\models\phi\Leftrightarrow \M',w'\models\phi, \text{ for all } \phi\in\ML^{d}_{\gsig}.
	\]
Each $\equiv^{d}_{\gsig}$ is an equivalence relation on $\PMod_{L}$.
For every signature $\gsig$ and every $d\geq 0$, the set $\ML^{d}_{\gsig}$ is finite up to $L$-equivalence, and
	so~$\equiv^{d}_{\gsig}$ has finitely many equivalence classes on $\PMod_{L}$.
This is the standard normal-form bound: over a finite set of variables and bounded depth, there are,
up to provable equivalence in the minimal quasi-normal logic, finitely many formulas, and
$L$-equivalence can only identify more. 

\begin{defi}[atom]\label{def:atoms}
	The $\equiv^{d}_{\gsig}$-class of $(\M,w)$ is denoted $[\M,w]^{d}_{\gsig}$ and called a
	\emph{$(\gsig,d)$-atom}. The \textup{(}finite\textup{)} set of all $(\gsig,d)$-atoms is denoted $\Atom{d}{\gsig}$; in other words, each $\equiv^d_{\gsig}$ partitions $\PMod_{L}$, and so $\Atom{d}{\gsig} = \PMod_{L}/_{\equiv^d_{\gsig}}$.
\end{defi}
Atoms are denoted by $\mathcal M,\mathcal N,\dots$; being equivalence
classes, they are literally sets of pointed models, and we use set-theoretic notation for them accordingly: for instance, we write
$\mathcal M\cap\mathcal{N}=\emptyset$ to say that no pointed model lies in both 
$\mathcal M$ and $\mathcal N$.
Also, 
for $\phi\in\ML^{d}_{\gsig}$ and a $(\gsig,d)$-atom~$\mathcal M$, we write
\begin{equation*}
\mathcal M\models\phi \qquad\text{ iff }\qquad \M,w\models\phi, \text{ for some (equivalently, all) }
	(\M,w)\in\mathcal M 
\end{equation*}	
because all members of $\mathcal M$ agree on
	$\ML^{d}_{\gsig}$-formulas.
\begin{defi}[characteristic formula]\label{def:char}
	The \emph{characteristic formula} of an atom $\mathcal M\in\Atom{d}{\gsig}$ is
	\[
	\MForm{d}{\gsig}{\mathcal M}\eqdef\bigwedge\bigl\{\phi\in\ML^{d}_{\gsig} \bigm| \mathcal  M\models\phi\bigr\},
	\]
	which is a finite (up to equivalence) conjunction. Note that we omit sub- and superscripts for $\Phi$ as they should be clear from $\mathcal{M}$.
\end{defi}
The characteristic formula of an atom is true exactly on its pointed models:
\begin{prop}\label{eq:char}
For any $\mathcal M\in\Atom{d}{\gsig}$,
we have $\M,w\models\MForm{d}{\gsig}{\mathcal M}$
	iff  $(\M, w)\in\mathcal{M}$, for every $(\M,w)\in\PMod_{L}$.
\end{prop}	
	
We extend the notion of characteristic formulas to sets $D\subseteq\Atom{d}{\gsig}$ of atoms:
\begin{equation*}
	\MForm{d}{\gsig}{D}\eqdef\bigvee_{\mathcal M\in D}\MForm{d}{\gsig}{\mathcal M}.
\end{equation*}
The map $D\mapsto\MForm{d}{\gsig}{D}$ is a bijection between subsets of $\Atom{d}{\gsig}$ and
$\ML^{d}_{\gsig}$ modulo $L$-equivalence: every $\phi\in\ML^{d}_{\gsig}$ is equivalent to
$\MForm{d}{\gsig}{D_{\phi}}$ for the unique
$D_{\phi}\eqdef\bigl\{\mathcal M\in\Atom{d}{\gsig}\mid \mathcal M\models\phi\bigr\}$. We may therefore
identify a ``formula of depth $\le d$ over $\gsig$'' with a ``subset of $\Atom{d}{\gsig}$''.

\subsection{Locally Tabular Logics}

Call a logic $L$ \emph{locally tabular} if, over every signature, there are only finitely
many formulas up to $L$-equivalence. For a locally tabular logic~$L$, for each signature $\gsig$, the descending chain of
$\equiv_\gsig^d$-equivalences on pointed models
\[
\equiv^{0}_{\gsig}\ \ \supseteq\ \ \equiv^{1}_{\gsig}\ \ \supseteq\ \ \cdots
\]
stabilises: there is a depth $d^{\ast}$ such that $\equiv^{d^{\ast}}_{\gsig}$ coincides with $\equiv^{d}_{\gsig}$, for all
$d\ge d^{\ast}$. It is an open question whether for every locally tabular $L$ there exists a $d^{\ast}$ such that the sequence  $\equiv^{d}_{\gsig}$ stabilises at $d^{\ast}$ for \emph{all} signatures~$\gsig$~\cite[Problem~2]{shapirovsky2026generalizationsfiniteheightcriterion}, see also~\cite{Shehtman2016,DBLP:journals/jsyml/Shapirovsky25}. This is the case for all modal logics $L$ of transitive frames and such an $L$ is locally tabular if and only if its frames have bounded depth~\cite[Theorem~12.21]{DBLP:books/daglib/0030819}. In $\textsf{Diff}$, every modal formula is logically equivalent to a formula of modal depth at most two.  

When dealing with locally tabular logics, we only consider frames $(W,R,A)$ with $A=2^{W}$ and we define
$\tau$-atoms over the set of \emph{all} formulas in $\ML_{\tau}$ without the depth restriction, hence $(\M,w)\equiv_{\gsig}(\M',w')$ iff $(\M,w)\equiv^{d}_{\gsig}(\M',w')$ for all~$d\geq 0$, and a $\tau$-atom $\mathcal{M}$ is defined as the $\equiv_{\tau}$-class of some $(\M,w)$ and then denoted~$[\M,w]_{\gsig}$. Accordingly, the characteristic formula  $\MForm{}{}{\mathcal{M}}$ is also defined over all formulas in $\ML_{\tau}$. Results obtained in the general case of bounded modal depth transfer to the locally tabular case by reading $\equiv_{\tau}$ as $\equiv_{\tau}^{d}$, where $d$ is such that $\equiv_{\tau}^{d'}\, = \,\equiv_{\tau}^{d}$ for all $d' \geq d$; such a $d$ exists by local tabularity.

A logic $L$ is \emph{tabular} if $L=\mathrm{Log}(\mathcal{K})$ for a finite class of finite pointed frames $\mathcal{K}$. Every tabular logic is locally tabular, but the converse does not hold as witnessed, for example, by $\Grzd$ for $\Delta\geq 2$ and \textsf{Diff}.

When describing the structure of $\gsig$-atoms, we refer to maximal consistent sets of $\gsig$-literals as \emph{$\gsig$-colours};  in other words, a $\gsig$-colour is a set containing either $p$ or $\neg p$, for each propositional variable $p\in\gsig$. Also, given a set of $\gsig$-colours $c_1,\dots, c_k$, we denote  
\begin{equation*}
\nabla (c_1,\dots,c_k) \ \ \ \eqdef \ \ \ \Bigl[\Box \bigvee_i \bigl(\bigwedge c_i\bigr)\Bigr] \ \ \ \land \ \ \ \Bigl[\bigwedge_i \Diamond \bigl(\bigwedge c_i\bigr)\Bigr].
\end{equation*} 

We illustrate these definitions on the following example of a tabular logic.

\begin{exa}[3-fork]\label{ex:3-fork} Let $\mathcal{K}$ consist of three pointed frames $((W_k, R_k), w_0)$, $k = 1, 2, 3$, where
\begin{equation*}
W_i = \{w_0, w_1, \dots, w_k\} \quad \text{ and }\quad w_0R_kw_i, \text{ for }1 \leq i \leq k;
\end{equation*} 
the frame with three successors is depicted in Fig.~\ref{fig:3-fork}a.
It should be clear that there are three groups of $\gsig$-atoms, with each successor having the same $\gsig$-colour, one of two $\gsig$-colours and one of three $\gsig$-colours; see Fig.~\ref{fig:3-fork}b, where we illustrate representative pointed models and the characteristic formulas of their atoms. One can easily verify that the number of $\gsig$-atoms is
\begin{equation*}
	2^{|\gsig|}
	\left[
	\binom{2^{|\gsig|}}{1}
	+\binom{2^{|\gsig|}}{2}
	+\binom{2^{|\gsig|}}{3}
	\right]
	\ \ =\ \ 
	\frac{1}{6}2^{4|\gsig|}
	+\frac{5}{6}2^{2|\gsig|}.
\end{equation*}
Obviously, atoms can contain pointed models with identical successors, but it will be useful for us to identify so-called standard pointed models for atoms without such duplicates (as depicted in Fig.~\ref{fig:3-fork}b).

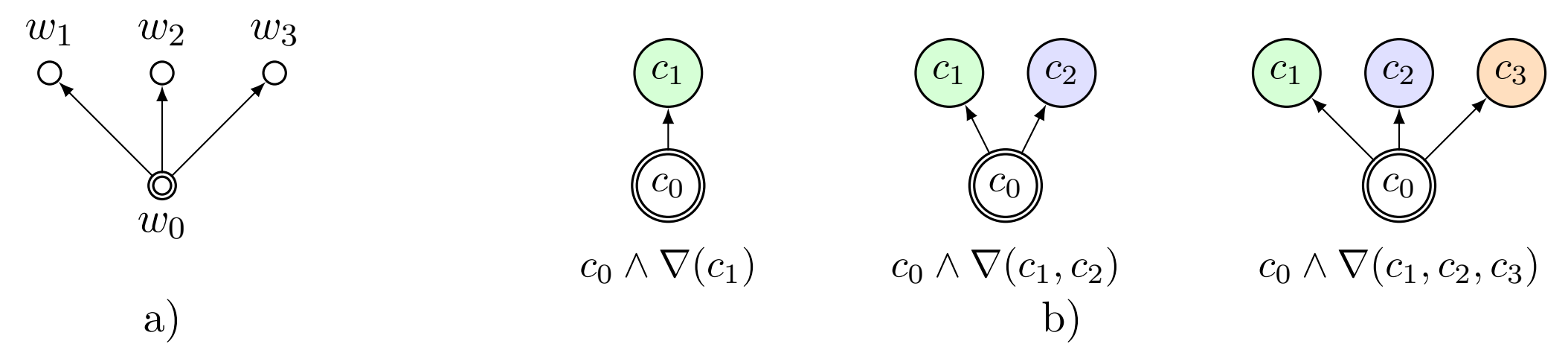
\begin{figure}[t]\centering%
\begin{tikzpicture}[>=latex,
nd/.style={inner sep=1pt,minimum size=2mm,draw,circle,semithick},
cnd/.style={inner sep=1pt,minimum size=6mm,draw,circle,semithick}
]
\begin{scope}
\node[nd,double,label=below:{$w_0$}] (w0) at (0,0) {};
\node[nd,label=above:{$w_1$}] (w1) at (-1,1) {};
\node[nd,label=above:{$w_2$}] (w2) at (0,1) {};
\node[nd,label=above:{$w_3$}] (w3) at (1,1) {};
\draw[->] (w0) -- (w1);
\draw[->] (w0) -- (w2);
\draw[->] (w0) -- (w3);
\node at (0, -1.2) {a)};
\end{scope}
\begin{scope}[xshift=105mm]
\node[cnd,double,label=right:{}] (w0s) at (-6,0) {\small$c_0$};
\node[cnd,label=above:{},fill=greencolour] (w1s) at (-6,1) {\small$c_1$};
\draw[->] (w0s) -- (w1s);
\node[below=1mm of w0s] {\small $c_0 \land \nabla (c_1)$};
\node[cnd,double,label=right:{}] (w0d) at (-3,0) {\small$c_0$};
\node[cnd,label=above:{},fill=greencolour] (w1d) at (-3.5,1) {\small$c_1$};
\node[cnd,label=above:{},fill=bluecolour] (w2d) at (-2.5,1) {\small$c_2$};
\draw[->] (w0d) -- (w1d);
\draw[->] (w0d) -- (w2d);
\node[below=1mm of w0d] {\small $c_0 \land \nabla(c_1, c_2)$};
\node[cnd,double,label=right:{}] (w0) at (0.5,0) {\small $c_0$};
\node[cnd,label=above:{},fill=greencolour] (w1) at (-0.5,1) {\small$c_1$};
\node[cnd,label=above:{},fill=bluecolour] (w2) at (0.5,1) {\small$c_2$};
\node[cnd,label=above:{},fill=orangecolour] (w3) at (1.5,1) {\small$c_3$};
\draw[->] (w0) -- (w1);
\draw[->] (w0) -- (w2);
\draw[->] (w0) -- (w3);
\node[below=1mm of w0] {\small $c_0 \land \nabla(c_1, c_2, c_3)$};
\node at (-2.5, -1.2) {b)};
\end{scope}
\end{tikzpicture}
\caption{a) The 3-fork pointed frame and b) its three groups of $\gsig$-atoms, for distinct $\gsig$-colours $c_1, c_2, c_3$, with their characteristic formulas.}\label{fig:3-fork}
\end{figure}
\end{exa}
We show that, for locally tabular logics $L$, atoms can always be represented by a single finite pointed model. We use the well-known notion of p-morphism (or, equivalently, bounded morphism)~\cite{goranko20075}.
	\begin{defi}[$\gsig$-p-morphism]\label{def:pmor}
		Let $\M=(W,R,V)$, $\M'=(W',R',V')$ be models. A map $f\colon W\to W'$ is a
		\emph{$\gsig$-p-morphism from $\M$ to $\M'$} if, for all $w,v\in W$, $u'\in W'$, the following holds:
		\begin{enumerate}[label=(p\arabic*)] 
			\item $\M, w\models p$ iff $\M', f(w)\models p$, for all $p \in \gsig$;
			\item $wRv$ implies $f(w)R'f(v)$;
			\item $f(w)R'u'$ implies $f(u)=u'$ for some $u\in W$ with $wRu$.
		\end{enumerate}
		A p-morphism of pointed models additionally maps the
		distinguished world to the distinguished world.
	\end{defi}
If $f$ is a $\gsig$-p-morphism 
from pointed model $(\M,w)$ to $(\M',w')$ and $(\M',w')$ is rooted,
then (p3) implies that $f$ is surjective. We use the following invariance property of p-morphisms~\cite[Corollary~16]{goranko20075}.
	\begin{lem}\label{lem:pmortype}
		If $f$ is a $\gsig$-p-morphism from $\M$ to $\M'$, then $w$ and $f(w)$ satisfy the same
		$\gsig$-formulas, for every $w\in W$. 
	\end{lem}
For any pointed $\gsig$-model $(\M,w)$, let
$(\M,w)/_{\equiv_{\gsig}}$ denote the pointed model obtained by
taking the quotient of $(\M,w)$ by $\equiv_{\gsig}$ and discarding all worlds
not reachable from the equivalence class of $w$.
Since $L$ is locally tabular, $\gsig$-equivalence coincides with
$\gsig$-bisimilarity. Thus, this construction is also known as the
\emph{bisimulation quotient} of a model~\cite[Definition~44]{goranko20075}.
Moreover, any two pointed models belonging to the same
$\gsig$-atom have isomorphic quotients.
This allows us to make the following definition.

\begin{defi}[$\gsig$-standard pointed model]
	Let $\mathcal{M}$ be a $\gsig$-atom for a locally tabular logic~$L$.
	The \emph{$\gsig$-standard pointed model}
	$(\mathfrak{S}_{\mathcal{M}},w_{\mathcal{M}})$
	is the quotient $(\M,w)/_{\equiv_{\gsig}}$ of any
	$(\M,w)\in\mathcal{M}$.
\end{defi}

We summarise the main properties of standard pointed models.
\begin{lem}\label{lem:standardmodel}
Let $\mathcal{M}$ be a $\gsig$-atom for a locally tabular logic $L$.
Then the $\gsig$-standard pointed model $(\mathfrak{S}_{\mathcal{M}},w_{\mathcal{M}})$ of $\mathcal{M}$ has the following properties:
	\begin{enumerate}
		\item $(\mathfrak{S}_{\mathcal{M}},w_{\mathcal{M}})\in \mathcal{M}$;
		\item $(\mathfrak{S}_{\mathcal{M}},w_{\mathcal{M}})$ is extensional, that is, distinct worlds are not $\equiv_{\gsig}$-equivalent;
		\item a rooted model $(\N,u)$ based on a frame for $L$ is in $\mathcal{M}$ iff 
		there is a $\gsig$-p-morphism from $(\N,u)$ onto $(\mathfrak{S}_{\mathcal{M}},w_{\mathcal{M}})$. 
	\end{enumerate}
\end{lem}

\section{Uniform Interpolants}

We now fix two disjoint signatures $\sig$ (retained) and $\priv$ (forgotten) and a modal depth $d\geq 0$. 
	Let $\phi\in\ML^{d}_{\sig\cup\priv}$.
	A formula $\theta\in\ML_\sig$ is a
	\emph{uniform $\sig$-interpolant} of $\phi$ if
	\begin{enumerate}
		\item $\phi\models\theta$;
		\item for every $\chi\in\ML$ with
		$\signat(\chi)\cap\priv=\emptyset$,
		if $\phi\models\chi$, then $\theta\models\chi$.
	\end{enumerate}
	A logic $L$ has the \emph{uniform interpolation property}
	if every formula has a uniform $\sig$-interpolant
	for every choice of retained signature $\sig$.
For $\phi\in\ML$, define the \emph{strongest $\sig$-consequence of~$\phi$ of depth $\le d$} as follows:
\begin{equation*}
\Cons^{d}_{\sig}(\phi)\eqdef\bigwedge\bigl\{\psi\in\ML^{d}_{\sig} \bigm| \phi\models\psi \bigr\};
\end{equation*}
this is a finite conjunction up
to equivalence. The sets grow with depth: $\Cons^{e}_{\sig}(\phi)\models\Cons^{d}_{\sig}(\phi)$
for $e\ge d$.
Then a uniform $\sig$-interpolant of $\phi$, if it exists, is equivalent to $\Cons^{e}_{\sig}(\phi)$ for some
$e$: taking $e$ at least its modal depth, the interpolant $\theta$ is a 
$\sig$-consequence of $\phi$ of depth $\leq e$, hence $\Cons^{e}_{\sig}(\phi)\models\theta$; and $\theta$ entails
every $\priv$-free consequence, in particular, every conjunct of $\Cons^{e}_{\sig}(\phi)$, so
$\theta\models\Cons^{e}_{\sig}(\phi)$. Therefore, non-existence means precisely that, for every
$e$, the candidate $\Cons^{e}_{\sig}(\phi)$ fails to entail some $\priv$-free consequence~$\chi_e$ of
$\phi$; recording $\chi_e$ over $\sig\cup\priv'$ with $\priv'\eqdef\signat(\chi_e)\setminus\sig$
disjoint from $\priv$ gives us the following working characterisation:
\begin{lem}[uniform interpolation criterion]\label{fact:unfold}
	A formula $\phi\in\smash{\ML^{d}_{\sig\cup\priv}}$ has \emph{no} uniform $\sig$-interpolant if and only if,
	for every $e\ge d$, there is a $\rho$-free formula $\chi_e$ such that
\begin{equation*}
	\phi\models\chi_e
	\qquad\text{and}\qquad
	\Cons^{e}_{\sig}(\phi)\not\models\chi_e.
\end{equation*}
\end{lem}

	\begin{exa}\label{exm:formulas}
	Consider the following two formulas for the 3-fork logic from Example~\ref{ex:3-fork}:
\begin{align*}
	\varphi & \ \ = \ \ 
	p\land q \ \ \land \ \
	\nabla(p\land q,\ p\land\neg q,\ \neg p\land q),\\
	\psi & \ \ = \ \ 
	p\land q' \ \ \land \ \
	\nabla(p\land q',\ \neg p\land q',\
	\neg p\land\neg q').
\end{align*}
The first requires two successors satisfying
$p$ and one satisfying $\neg p$, whereas the
second requires one successor satisfying $p$
and two satisfying $\neg p$. Each formula
describes a single atom over its respective
signature.

	Let $\sig=\{p\}$. Then $\varphi$ has no uniform $\sigma$-interpolant:
	its strongest $\sigma$-consequence is
	$\theta=p\land\nabla(p,\neg p)$. 
	However, $\theta$ is not a uniform
	$\sig$-interpolant of $\varphi$ because
	$\varphi\models\neg\psi$ but
	$\theta\not\models\neg\psi$.
	Thus, no $\sig$-formula captures all
	consequences of $\varphi$ over signatures
	disjoint from~$\{q\}$.
\end{exa}

\subsection{Cell Factorisation}

We note that standard criteria for uniform interpolants in the literature are bisimulation-based~\cite{Visser96,DBLP:conf/aiml/DAgostinoH96,French06,DBLP:conf/ijcai/LutzW11}. For our purposes, it is more convenient to work with logical indistinguishability directly and use atoms. We first lift the relation $\equiv^{d}_{\sig}$ from pointed models to atoms ($\equiv^{d}_{\sig}$ is used for both pointed models and atoms).
\begin{defi}[lifted equivalence]\label{def:lift}
	Let $\mathcal M_1\in\Atom{d}{\sig\cup\priv_1}$ and $\mathcal M_2\in\Atom{d}{\sig\cup\priv_2}$. We write $\mathcal M_1\equiv^{d}_{\sig}\mathcal M_2$ if
	$(\M_1,w_1)\equiv^{d}_{\sig}(\M_2,w_2)$, for some $(\M_i,w_i)\in\mathcal M_i$, for $i = 1, 2$.
\end{defi}
Although the lifted relation $\equiv_\sig^d$ is defined
by requiring some representatives to agree, it is
in fact independent of the choice of representatives.
\begin{prop}\label{rem:slice}
Let $\mathcal M_1\in\Atom{d}{\sig\cup\priv_1}$ and $\mathcal M_2\in\Atom{d}{\sig\cup\priv_2}$. If $\mathcal M_1\equiv^{d}_{\sig}\mathcal M_2$, then $(\M_1,w_1) \equiv^{d}_{\sig} (\M_2,w_2)$, for \emph{every} $(\M_i,w_i)\in \mathcal{M}_i$, $i = 1,2$. 
\end{prop}
Indeed, since each $(\sig\cup\priv_i, d)$-atom refines the $\equiv^{d}_{\sig}$-equivalence on pointed models, and so all members of $\mathcal M_i$ are pairwise
	$\equiv^{d}_{\sig}$-equivalent, $\mathcal{M}_1\equiv^{d}_{\sig}\mathcal{M}_2$ implies that all pointed models in~$\mathcal{M}_1$ and~$\mathcal{M}_2$ are $\equiv^{d}_{\sig}$-equivalent. 
	It also follows that
	$\equiv^{d}_{\sig}$ is an equivalence relation on~$\Atom{d}{\sig\cup\priv}$. We now introduce the corresponding partitioning of the set of $(\sig\cup\priv,d)$-atoms.
\begin{defi}[cell]\label{def:sliceset}
	The $\equiv^{d}_{\sig}$-class of a $(\sig\cup\priv,d)$-atom $\mathcal{M}$ is denoted $[\mathcal{M}]^{d}_{\sig}$ and called a
	\emph{$\sig^{\uparrow\rho}$-cell}. The set of all $\sig^{\uparrow\priv}$-cells is denoted $\CellAt{d}{\sig}{\priv}$; in other words, $\CellAt{d}{\sig}{\priv} = \Atom{d}{\sig\cup \priv}/_{\equiv^{d}_{\sig}}$.
\end{defi}
%
%
Since the pointed models of the $(\sig\cup \priv, d)$-atoms in a given $\sig^{\uparrow\rho}$-cell $C\in\smash{\CellAt{d}{\sig}{\priv}}$ are all $\equiv_\sig^{d}$-equivalent, 
they form a $(\sigma,d)$-atom themselves:
\begin{prop}\label{prop:second-part}
For any $C\in\smash{\CellAt{d}{\sig}{\priv}}$, we have $\bigcup C = [\M, w]_{\sig}^{d}$, for any $(\M, w) \in \mathcal{M}$ with $\mathcal{M}\in C$. 
\end{prop}
It follows, in particular, that the \emph{cell characteristic formula} $\CForm{d}{\sig\cup \priv}{C} = \MForm{d}{\sig\cup \priv}{\bigcup C}$  is an $\ML^{d}_\sig$-formula. 
Observe that the $\sig$-formula $\CForm{d}{\sig\cup \priv}{C}$ is $L$-equivalent to the $(\sig\cup\priv)$-formula
$\bigvee_{\mathcal{M}\in C} \MForm{d}{\sig\cup\priv}{\mathcal{M}}$.
Since the formulas $\CForm{d}{\sig\cup \priv}{C}$, $C\in \smash{\CellAt{d}{\sig}{\priv}}$, are pairwise
inconsistent, and their disjunction is valid, we have 
\begin{equation*}
\varphi\ \ \equiv\bigvee_{C\in\smash{\CellAt{d}{\sig}{\priv}}}\hspace*{-0.5em}(\varphi\land \CForm{d}{\sig\cup \priv}{C}), \qquad\text{ for any } \varphi\in\ML_{\sig\cup\priv}^{d}.
\end{equation*}
Note also that
$\MForm{d}{\sig\cup\priv}{D}\land \CForm{d}{\sig\cup \priv}{C} \equiv
\MForm{d}{\sig\cup\priv}{D\cap C}$, for any $D\subseteq\Atom{d}{\sig\cup \priv}$.
We are now in a position to state our main cell factorisation result:
\begin{lem}[cell factorisation]\label{lem:slicedecomp}
	A formula $\varphi\in\ML_{\sig\cup\priv}^{d}$
          has a uniform
	$\sig$-interpolant if and only if
	$\varphi\land\CForm{d}{\sig\cup \priv}{C}$ has a uniform
	$\sig$-interpolant,  for each $\sig^{\uparrow\priv}$-cell $C\in\CellAt{d}{\sig}{\priv}$.
\end{lem}

\begin{proof}
	Suppose first that $\theta$ is a uniform
	$\sig$-interpolant of $\varphi$. We show that
	$\theta\land\CForm{d}{\sig\cup \priv}{C}$ is a uniform
	$\sig$-interpolant of $\varphi\land\CForm{d}{\sig\cup \priv}{C}$, for each $C$.
	Trivially,
	$\varphi\land \CForm{d}{\sig\cup \priv}{C}\models\theta\land\CForm{d}{\sig\cup \priv}{C}$.
	Take any $\priv$-free formula $\psi$ with $\varphi\land \CForm{d}{\sig\cup \priv}{C}\models\psi$.
        Then $\varphi\models \CForm{d}{\sig\cup \priv}{C}\to\psi$.
	As
	$\CForm{d}{\sig\cup \priv}{C}\to\psi$ is $\priv$-free,
	uniform interpolation gives
	$\theta\models\CForm{d}{\sig\cup \priv}{C}\to\psi$, whence
	$\theta\wedge\CForm{d}{\sig\cup \priv}{C}\models\psi$.
	
	Conversely, suppose that 
	$\varphi\land\CForm{d}{\sig\cup \priv}{C}$ has a uniform
	$\sig$-interpolant $\theta_C$, for each $C$. We show that $\bigvee_C\theta_C$
	is a uniform $\sig$-interpolant of
	$\varphi$.
	Since $\varphi\equiv\bigvee_C(\varphi\land\CForm{d}{\sig\cup \priv}{C})$,
	we have
	$\varphi\models\bigvee_C\theta_C$.
	If $\varphi\models\psi$ for a $\priv$-free formula~$\psi$, then
	$\varphi\land\CForm{d}{\sig\cup \priv}{C}\models\psi$ for every $C$.
	Thus, $\theta_C\models\psi$ for every
	$C$, and consequently
	$\bigvee_C\theta_C\models\psi$.
\end{proof}

\begin{exa}\label{ex:3-forks-cells}
Let us revisit the logic of 3-forks from Example~\ref{ex:3-fork}. 
For a signature $\tau$, let~$\nsets^k_\tau$
denote the set of all $k$-element subsets of
the set of $\tau$-colours (we slightly abuse notation and assume that $\nsets^1$ contains $\tau$-colours themselves rather than their singleton sets). Thus
\begin{equation*}
|\nsets^k_\tau|
=\binom{2^{|\tau|}}{k}.
\end{equation*}
Given a $\sig$-colour $s$ and a $\priv$-colour~$r$, we denote by $sr$ the $(\sig\cup\priv)$-colour that is obtained by taking the union of $s$ and $r$ (which corresponds to the conjunction of their literals if used in a formula); in the examples, we often write, say, $pq$ for a colour comprising $p$ and $q$.  
With this notation at hand, we can describe the three groups of $\sig^{\uparrow\priv}$-cells for the logic of 3-forks; see Fig.~\ref{fig:3-fork-slices}.
\begin{description}
\item[$C^{\priv}_{s_0, s_1}$] for $s_0, s_1\in \nsets_\sigma^1$, containing the following atoms with characteristic formulas:
\begin{align*}
& \mathcal{M}_{s_0r_0, s_1r_1} &&  s_0 r_0 \land \nabla(s_1 r_1), && \text{for } r_0, r_1\in\nsets_\rho^1,\\
& \mathcal{M}_{s_0r_0,s_1 \{r_1, r_2\}} && s_0 r_0 \land \nabla (s_1  r_1, s_1  r_2),&& \text{for }r_0\in\nsets_\rho^1 \text{ and } \{ r_1, r_2\} \in \nsets_\rho^2,\\
& \mathcal{M}_{s_0r_0, s_1\{r_1, r_2,r_3\}} && s_0 r_0 \land \nabla (s_1  r_1, s_1  r_2, s_1  r_3),&& \text{for }r_0\in\nsets_\rho^1 \text{ and } \{ r_1, r_2, r_3\}\in\nsets_\rho^3;
\end{align*}
\item[$C^{\priv}_{s_0, \{ s_1, s_2\}}$] for $s_0\in\nsets_\sigma^1$ and $\{s_1, s_2\}\in\nsets_\sigma^2$, containing the following  atoms with characteristic formulas:
\begin{align*}
& \mathcal{M}_{s_0r_0, \{s_1r_1, s_2r_2\}} && s_0r_0\land \nabla(s_1r_1, s_2r_2), && \text{for } r_0, r_1, r_2\in\nsets_\rho^1,\\
& \mathcal{M}_{s_0r_0, s_1\{r_1, r_2\}, s_2r'} && s_0r_0 \land \nabla (s_1r_1, s_1r_2, s_2r'), && \text{for } r_0,r'\in\nsets_\rho^1 \text{ and }\{r_1, r_2\}\in\nsets_\rho^2,\\
& \mathcal{M}_{s_0r_0, s_1r',s_2\{r_1, r_2\}} && s_0r_0 \land \nabla (s_1r', s_2r_1, s_2r_2), && \text{for } r_0,r'\in\nsets_\rho^1 \text{ and }\{r_1, r_2\}\in\nsets_\rho^2;
\end{align*}
\item[$C^{\priv}_{s_0,\{s_1,s_2,s_3\}}$] for  $s_0\in\nsets_\sigma^1$ and $\{s_1,s_2,s_3\}\in\nsets_\sigma^3$, containing  the following atoms with characteristic formulas:
\begin{align*}
& \mathcal{M}_{s_0r_0, \{s_1r_1, s_2r_2, s_3r_3\}} && s_0r_0 \land \nabla(s_1r_1, s_2r_2, s_3r_3),&& \text{for } r_0, r_1, r_2, r_3\in\nsets_\rho^1.
\end{align*}
\end{description}
\begin{figure}[t]\centering%
\begin{tikzpicture}[>=latex,
nd/.style={inner sep=1pt,minimum size=3mm,draw,circle,thick,fill=white},
pnd/.style={inner sep=1pt,minimum size=4mm,draw,regular polygon, regular polygon sides=6,thick}
]
\begin{scope}[rounded corners=2mm]\small
\draw[fill=gray!10,draw=violet] (-0.75, -1.3) rectangle +(5.55, 2.6); \node at (2, -1.7) {$C^{\priv}_{s_0, \{s_1\}}$};
\draw[fill=gray!10,draw=violet] (5, -1.3) rectangle +(6.3, 2.6); \node at (8, -1.7) {$C^{\priv}_{s_0, \{s_1, s_2\}}$};
\draw[fill=gray!10,draw=violet] (11.5, -1.3) rectangle +(3, 2.6); \node at (13, -1.7) {$C^{\priv}_{s_0, \{s_1, s_2, s_3\}}$};
\end{scope}
\node[nd,double,label=below:{\small$\mathcal{M}_{s_0r_0, s_1r_1}$}] (w0) at (0.25,0) {};
\node[nd,rectangle,fill=greencolour] (w01) at (0.25,1) {};
\draw[->] (w0) -- (w01);
\node[nd,double,label=below:{\small\rule{0pt}{22pt}$\mathcal{M}_{s_0r_0,s_1 \{r_1, r_2\}}$}] (w1) at (1.75,0) {};
\node[nd,rectangle,fill=greencolour] (w11) at (1.25,1) {};
\node[nd,rectangle,fill=bluecolour] (w12) at (2.25,1) {};
\draw[->] (w1) -- (w11);
\draw[->] (w1) -- (w12);
\node[nd,double,label=below:{\small$\mathcal{M}_{s_0r_0, s_1\{r_1, r_2,r_3\}}$}] (w2) at (3.5,0) {};
\node[nd,rectangle,fill=greencolour] (w21) at (3,1) {};
\node[nd,rectangle,fill=bluecolour] (w22) at (3.5,1) {};
\node[nd,rectangle,fill=orangecolour] (w23) at (4,1) {};
\draw[->] (w2) -- (w21);
\draw[->] (w2) -- (w22);
\draw[->] (w2) -- (w23);
\node[nd,double,label=below:{\small$\mathcal{M}_{s_0r_0, \{s_1r_1, s_2r_2\}}$}] (w3) at (6.25,0) {};
\node[nd,rectangle,fill=greencolour] (w31) at (5.75,1) {};
\node[pnd,fill=orangecolour] (w32) at (6.75,1) {};
\draw[->] (w3) -- (w31);
\draw[->] (w3) -- (w32);
\node[nd,double,label=below:{\small\rule{0pt}{22pt}$\mathcal{M}_{s_0r_0, s_1\{r_1, r_2\}, s_2r'}$}] (w4) at (8,0) {};
\node[nd,rectangle,fill=greencolour] (w41) at (7.5,1) {};
\node[nd,rectangle,fill=bluecolour] (w42) at (8,1) {};
\node[pnd,fill=orangecolour] (w43) at (8.5,1) {};
\draw[->] (w4) -- (w41);
\draw[->] (w4) -- (w42);
\draw[->] (w4) -- (w43);
\node[nd,double,label=below:{\small$\mathcal{M}_{s_0r_0, s_1r',s_2\{r_1, r_2\}}$}] (w5) at (10,0) {};
\node[nd,rectangle,fill=orangecolour] (w51) at (9.5,1) {};
\node[pnd,fill=greencolour] (w52) at (10,1) {};
\node[pnd,fill=bluecolour] (w53) at (10.5,1) {};
\draw[->] (w5) -- (w51);
\draw[->] (w5) -- (w52);
\draw[->] (w5) -- (w53);
\node[nd,double,label=below:{\small$\mathcal{M}_{s_0r_0, \{s_1r_1, s_2r_2, s_3r_3\}}$}] (w6) at (13,0) {};
\node[nd,rectangle,fill=greencolour] (w61) at (12.5,1) {};
\node[pnd,fill=bluecolour] (w62) at (13,1) {};
\node[pnd,fill=orangecolour,regular polygon sides=5] (w63) at (13.5,1) {};
\draw[->] (w6) -- (w61);
\draw[->] (w6) -- (w62);
\draw[->] (w6) -- (w63);
\end{tikzpicture}
\caption{$\sig^{\uparrow\priv}$-cells for 3-forks: the $\sig$- and $\priv$-colours are indicated by the shape and the shading of the node, respectively.}\label{fig:3-fork-slices}
\end{figure}
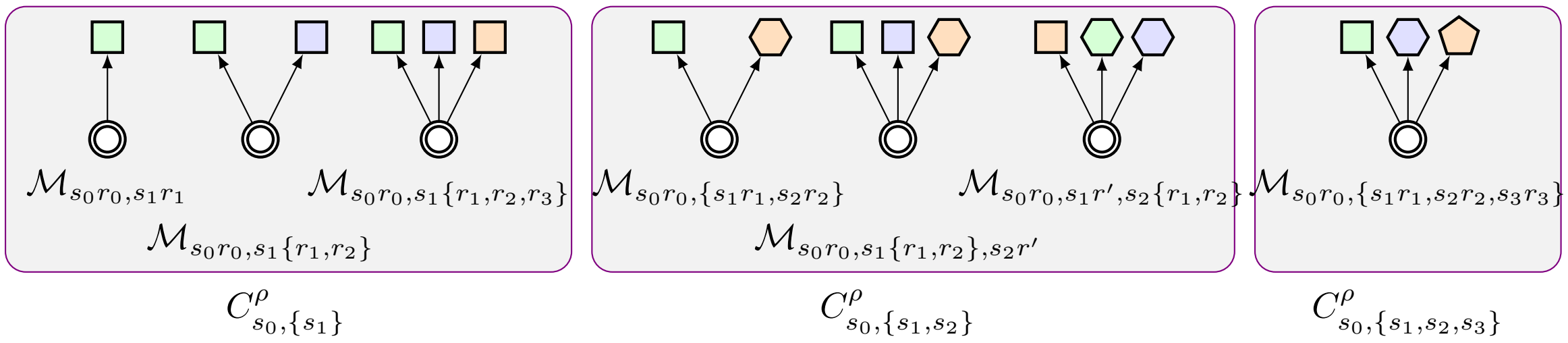
For $\sig=\{p\}$ and $\priv=\{q\}$, consider the atom
$\mathcal M_\varphi = \mathcal{M}_{pq, p\{q,\neg q\}, \neg p q}$
corresponding to $\varphi$ in Example~\ref{exm:formulas}.
This atom belongs to the cell
$\smash{C^{\priv}_{p,\{p,\neg p\}}}$, since its root
satisfies $p$ and its successors realise precisely
the two $\sig$-colours $\{p\}$ and $\{\neg p\}$.
Hence, cells of the form~$\smash{C^{\priv}_{s_0,\{s_1,s_2\}}}$ contain
counterexamples to uniform $\sig$-interpolation.
Moreover, one can show that counterexamples occur
only in cells of this form.
\end{exa}

\subsection{Pinning Signatures and Flipping Variables}\label{sec:pinning:flipping}

In the constructions below, we require two operations: one globally assigns either $\top$ or $\bot$ to each variable in a given signature, while the other flips the truth value of each variable in a given signature. Let $\gsig$ be a fixed signature. 

\begin{defi}[pinning a signature]\label{def:forgettruefalse}
A \emph{pinning of~$\gsig$} is a function $\pinf\colon \gsig \to \{\top, \bot\}$.
For a model $\M=(\F,V)$, let $\pin{\pinf}{\M} = (\F, \pin{\pinf}{V})$ be the model that agrees with $\M$ on symbols outside~$\gsig$ and sets each $s\in \gsig$ to a constant value according to~$\pinf$: 
\[
\pin{\pinf}{V}(s)\eqdef \begin{cases}
W, & \text{if } \pinf(s) = \top,\\
\emptyset, & \text{if } \pinf(s)= \bot,
\end{cases}
\qquad \text{ for each } s\in \gsig,
\]
which is admissible since $\emptyset,W\in A$. Similarly, given a formula $\varphi$, we write $\fpin{\pinf}{\varphi}$ for the simultaneous substitution of $\pinf(s)$ for each $s\in \gsig$. 
\end{defi}

Observe that 
\begin{equation*}
\signat(\fpin{\pinf}{\phi})\subseteq\signat(\phi)\setminus \tau\qquad \text{ and }\qquad
\textup{md}(\fpin{\pinf}{\phi})\le\textup{md}(\phi), \qquad\text{ for any formula } \varphi. 
\end{equation*}
The relevant preservation properties of pinning are as follows:
\begin{lem}\label{lem:constants} 
Let $\pinf$ be a pinning of $\gsig$. 
\begin{enumerate}[label=\textup{(\arabic*)}]
\item $(\M,w)\models\fpin{\pinf}{\varphi}$ if and only if $(\pin{\pinf}{\M},w)\models\varphi$, for every $\varphi$ and every $(\M,w)\in\PMod_{L}$\textup{;}	
		\item $(\M,w)\equiv^{d}_{\musig}(\N,u)$
		if and only if
		$(\pin{\pinf}{\M},w)\equiv^{d}_{\musig\cup\gsig}(\pin{\pinf}{\N},u)$,  for every $(\M,w),(\N,u)\in\PMod_{L}$ and every $\musig$ disjoint from $\gsig$.
\end{enumerate}
\end{lem}
\begin{proof}
(1) By induction on the structure of $\varphi$.

	(2) Suppose $(\M,w)\equiv^{d}_{\musig}(\N,u)$. Take any $\varphi\in\ML^{d}_{\musig\cup\gsig}$. Since
	$\fpin{\pinf}{\varphi}\in\ML^{d}_{\musig}$, we have	$\M,w\models\fpin{\pinf}{\varphi}$ iff $\N,u\models\fpin{\pinf}{\varphi}$; two
	applications of Item~(1) turn this into
	$\pin{\pinf}{\M},w\models\varphi$ iff $\pin{\pinf}{\N},u\models\varphi$. 
	
	Conversely, let $\psi\in\ML^{d}_{\musig}\subseteq\ML^{d}_{\musig\cup\gsig}$.
	As $\psi$ has no variable of $\gsig$, $\fpin{\pinf}{\psi}=\psi$, so Item~(1) gives
	$\pin{\pinf}{\M},w\models\psi$ iff $\M,w\models\psi$ and likewise for
	$\N$. Hence
	$\M,w\models\psi$ iff $\pin{\pinf}{\M},w\models\psi$
	iff $\pin{\pinf}{\N},u\models\psi$ iff $\N,u\models\psi$.
\end{proof}

We now extend the pinning operation to atoms: for $\mathcal{M}\in\Atom{d}{\musig}$ and a pinning $\pinf$ of $\gsig$ (disjoint from $\musig$), set
\begin{equation*}
\pin{\pinf}{\mathcal{M}} \ \ = \ \ \classp{(\pin{\pinf}{\M},w)}{d}{\musig\cup\gsig} \ \ \in\ \ \Atom{d}{\musig\cup\gsig},\qquad \text{ for some } (\M,w)\in \mathcal{M}.
\end{equation*}
By Lemma~\ref{lem:constants}(2), $\pin{\pinf}{\mathcal{M}}$ is well-defined and does not depend on the choice of $(\M,w)\in\mathcal{M}$.
Pinning can then be naturally lifted to sets of atoms and cells: for $D\subseteq\Atom{d}{\musig}$ and a pinning~$\pinf$ of $\gsig$  (disjoint from $\musig$), set
\begin{equation*}
\pin{\pinf}{D} \ \ = \ \ \bigl\{ \pin{\pinf}{\mathcal{M}} \bigm| \mathcal{M}\in D \bigr\} \ \ \subseteq \ \ \Atom{d}{\musig\cup\gsig}.
\end{equation*}
\begin{lem}\label{lem:constants:atomsets}
Let $\musig$ be a signature disjoint from $\gsig$ and $\pinf$ a pinning of $\tau$. 
\begin{enumerate}[label=\textup{(\arabic*)}]
\item $\MForm{}{}{\pin{\pinf}{D}} \equiv \MForm{}{}{D} \land \bigwedge_{p\in\gsig} \Box^{\leq d} (p \leftrightarrow \pinf(p))$,
for every $D\subseteq\Atom{d}{\musig}$.
\item The map $\pin{\pinf}{\cdot}\colon \mathcal M\mapsto\pin{\pinf}{\mathcal M}$ is a bijection from $D$ onto $\pin{\pinf}{D}$, for every $D\subseteq\Atom{d}{\musig}$.
\item $\pin{\pinf}{C}\in\CellAt{d}{\sig^*\cup\gsig}{\priv}$, for each cell $C\in\CellAt{d}{\sig^*}{\priv}$, for pairwise disjoint signatures $\sig^*$, $\gsig$ and $\priv$.
\end{enumerate}
\end{lem}
\begin{proof}
(1) Follows from Lemma~\ref{lem:constants}(1).

\smallskip
	
(2)	Surjectivity is by definition, while injectivity is by observing that the restriction of $\pin{\pinf}{\mathcal{M}}$ to $\musig$ is $\mathcal{M}$.

\smallskip

(3) By definition, $\pin{\pinf}{C}$ is a set of $(\sig^*\cup\gsig\cup\priv, d)$-atoms with,  by~(1), $\MForm{}{}{\pin{\pinf}{C}} \equiv \MForm{}{}{C} \land  \bigwedge_{p\in\gsig} \Box^{\leq d} (p \leftrightarrow \pinf(p))$.
On the other hand, consider $(\sig^*,d)$-atom $\mathcal{M} = \bigcup C$.  By definition, $\pin{\pinf}{\mathcal{M}}$ is a $(\sig^*\cup\gsig, d)$-atom with, by~(1), $\MForm{}{}{\pin{\pinf}{\mathcal{M}}} \equiv \MForm{}{}{\mathcal{M}} \land  \bigwedge_{p\in\gsig} \Box^{\leq d} (p \leftrightarrow \pinf(p))$. It follows that $\MForm{}{}{\pin{\pinf}{C}} \equiv\MForm{}{}{\pin{\pinf}{\mathcal{M}}}$, and so $\bigcup \pin{\pinf}{C} = \pin{\pinf}{\mathcal{M}}$. Thus, $\pin{\pinf}{C}$ is a $\smash{(\sig^*\cup\gsig)^{\uparrow \priv}}$-cell.
\end{proof}

We next introduce the second operation. 

\begin{defi}[flipping]
Let $v\subseteq \gsig$. For any model $\M = (\mathfrak{F}, V)$, let $\M^{v} = (\mathfrak{F}, V^v)$ flip the variables selected by $v$:
\[
V^{v}(p)\eqdef \begin{cases}
W\setminus V(p), & \text{if } p \in v,\\
V(p), & \text{otherwise},
\end{cases}
\]
which is admissible since $A$ is closed under complement. Similarly, for a formula $\varphi$,  denote by $\varphi^v$ the result of replacing each
$p\in v$ by $\neg p$.
\end{defi}
Observe that $\cdot^v$ is involutive and preserves both the signature and the modal depth of formulas. Moreover, we have the following:
\begin{lem}
$(\M^v,w)\models\varphi$ if and only if $(\M,w)\models\varphi^v$, for every formula $\varphi$ and every $(\M,w)\in\PMod_{L}$.
\end{lem}

We now extend the flipping operation to atoms: for $\mathcal{M}\in \Atom{d}{\musig\cup\gsig}$ with $\musig$ disjoint from $\gsig$ and $v\subseteq\gsig$, let 
\begin{equation*}
\mathcal M^{v} \ \ = \ \ \bigl\{(\M^{v},w)\bigm|(\M,w)\in\mathcal M\bigr\} \ \ \in \ \ \Atom{d}{\musig\cup\gsig}.
\end{equation*}
Next, for any $D\subseteq \Atom{d}{\musig\cup\gsig}$, set $D^{v}=\bigl\{\mathcal{M}^{v}\bigm| \mathcal{M}\in D\bigr\}\subseteq \Atom{d}{\musig\cup\gsig}$. 

\begin{lem}\label{lem:flipping:atomsets}
Let $\musig$ be a signature disjoint from $\gsig$ and $v \subseteq \gsig$. 
\begin{enumerate}[label=\textup{(\arabic*)}]
\item $\MForm{}{}{D^v}\equiv (\MForm{}{}{D})^v$, for every  $D \subseteq \Atom{d}{\musig\cup\gsig}$.

\item  The map $\cdot^v \colon \mathcal M\mapsto \mathcal{M}^v$ is a bijection from $D$ onto $D^v$, for every $D\subseteq\Atom{d}{\musig\cup\gsig}$. 

\item $C^v\in\CellAt{d}{\gsig}{\musig}$, for all cells $C\in\CellAt{d}{\gsig}{\musig}$. Moreover, $C^{v_1} \ne C^{v_2}$, for distinct $v_1,v_2\subseteq \gsig$.
\end{enumerate}
\end{lem}
\begin{proof}
(1) and (2) are immediate from the definitions.

(3) Two different flips already differ at depth $0$: $(\M^{v_{1}},w)\not\equivp{0}{\gsig}(\M^{v_{2}},w)$, for distinct $v_1,v_2\subseteq \gsig$,
since some $p\in\gsig$ receives opposite truth values.
Consequently, the $2^{|\gsig|}$ flips of a fixed atom lie in $2^{|\gsig|}$ \emph{distinct} $\gsig^{\uparrow\musig}$-cells, all of the same size as the original by the same relabelling bijection. 
\end{proof}

\section{Density of Uniform Interpolants}
\label{sec:density}

In this section, we introduce the density of uniform
interpolants and show that it factorises into a product
of intra-cell densities. We then investigate its asymptotic
behaviour as either the number of retained variables
or the number of forgotten variables tends to infinity.

Fix a logic $L$, a modal depth $d \geq 0$ and disjoint retained and forgotten signatures, $\sig$ and $\priv$.
Identifying, as in Definition~\ref{def:char}, a formula of
depth $\le d$ over $\sig\cup\priv$ with the subset of~$\Atom{d}{\sig\cup\priv}$ it defines, set
\[
\UIp{d}{\sig,\priv}\eqdef\bigl\{D\subseteq \Atom{d}{\sig\cup\priv}\bigm| \MForm{d}{\sig,\priv}{D}\text{ has a uniform }\sig\text{-interpolant}\bigr\}.
\]
The \emph{density of uniform interpolation} is
the proportion of $L$-equivalence classes of
formulas of depth at most~$d$ over
$\sig\cup\priv$ that admit a uniform
$\sig$-interpolant.
Equivalently, we equip the powerset of
$\Atom{d}{\sig\cup\priv}$
with the uniform distribution and put
\[
\dens^{d}_{\sig,\priv}\eqdef\frac{|\UIp{d}{\sig,\priv}|}{2^{|\Atom{d}{\sig\cup\priv}|}} .
\]
For a cell $C\in\CellAt{d}{\sig}{\priv}$, we define the corresponding \emph{intra-cell density} by restricting to subsets of
$C$:
\[
\mathcal U_{C}\eqdef\bigl\{D\subseteq C\bigm| \MForm{d}{\sig,\priv}{D}\text{ has a uniform }\sig\text{-interpolant}\bigr\}
\qquad\text{ and }\qquad
\dens_{C}\eqdef\frac{|\mathcal U_{C}|}{2^{|C|}} .
\]

\begin{thm}[cell reduction principle]\label{thm:slicereduction}
	\[
	\dens^{d}_{\sig,\priv}\ =\ \prod_{C\in\CellAt{d}{\sig}{\priv}}\dens_{C} .
	\]
\end{thm}
	\begin{proof}
		Since the cells partition
		$\Atom{d}{\sig\cup\priv}$, the map
		$D\mapsto (D\cap C)_{C\in
			\CellAt{d}{\sig}{\priv}}$
		is a bijection from
		$2^{\Atom{d}{\sig\cup\priv}}$ onto
		$
		\prod_{C\in
			\CellAt{d}{\sig}{\priv}}
		2^C$.
		By Lemma~\ref{lem:slicedecomp}, this bijection
		restricts to a bijection
		$\UIp{d}{\sig,\priv}
		\rightarrow
		\prod_{C\in
			\CellAt{d}{\sig}{\priv}}
		\mathcal U_C$.
		Consequently,
		$|\UIp{d}{\sig,\priv}|
		=
		\prod_C|\mathcal U_C|$.
		Since
		$2^{|\Atom{d}{\sig\cup\priv}|}
		=\prod_C 2^{|C|}$,
		division gives the claimed identity.
\end{proof}
As $0 \leq \dens_{C}\leq 1$, it follows that a single cell $C$ with $\dens_{C}<1$ ensures $\smash{\dens}^{d}_{\sig,\priv}<1$.
We also write $\smash{\overline{\dens}}^{\smash{d}}_{\sig,\priv}\eqdef 1-\smash{\dens}^{\smash{d}}_{\sig,\priv}$ and $\overline{\dens}_{C}\eqdef 1-\dens_{C}$ for the
	densities of interpolation \emph{failure}. 

\begin{exa}\label{ex:33}
We return to Example~\ref{ex:3-forks-cells}, where we presented the $\sig^{\uparrow\priv}$-cells for the logic of 3-forks. As we have observed, cells  $C$ of the form $\smash{C^{\priv}_{s_0, s_1}}$ and $\smash{C^{\priv}_{s_0, \{s_1, s_2, s_3\}}}$ contain no counterexamples to uniform $\sig$-interpolation, and so $\smash{\dens}_C = 1$. On the other hand, each of the cells $C$ of the form $C^{\priv}_{s_0, \{ s_1, s_2\}}$ contains counterexamples. It can be seen that every $\MForm{}{}{D}$, for a non-empty $D\subseteq C$, has no uniform $\sig$-interpolation iff $D$ consists entirely of atoms of the form $\mathcal{M}_{s_0r_0, s_1\{r_1, r_2\}, s_2r'}$ or entirely of atoms of the form $\mathcal{M}_{s_0r_0, s_1r', s_2\{r_1, r_2\}}$. Let~$M = 2^{|\priv|}$. Then $C$ contains  $M^3$ atoms  of the form $\mathcal{M}_{s_0r_0, \{s_1r_1, s_2r_2\}}$,  $\frac{1}{2}(M-1)M^3$ atoms of the form $\mathcal{M}_{s_0r_0, s_1\{r_1, r_2\}, s_2r'}$ and the same number of the form $\mathcal M_{s_0r_0,s_1r',s_2\{r_1,r_2\}}$; in total, $C$ contains $M^4$ atoms. Since the $D$ without uniform $\sig$-interpolation coincide with the non-empty subsets of the two latter types of atoms, we have
\begin{equation*}
\smash{\overline{\dens}}_{C} \ \ = \ \ 2 \cdot \frac{2^{\frac{1}{2}(M-1)M^3} - 1}{2^{M^4}}.
\end{equation*}
Let now $N = 2^{|\sig|}$. Since there are $\frac{1}{2}(N-1)N^2$ cells of the form $C^{\priv}_{s_0, \{ s_1, s_2\}}$, we obtain
\begin{equation*}
\dens_{\sig,\priv} \ \ = \ \ \Biggl(1 - \frac{2^{\frac{1}{2}(M-1)M^3} - 1}{2^{M^4 - 1}} \Biggr)^{\frac{1}{2}(N-1)N^2}.
\end{equation*}
(We drop the modal depth $d$ in $\dens_{\sig,\priv}$ since every formula in the logic of 3-forks is equivalent to a formula of modal depth at most $1$.)
\end{exa}

We next consider the asymptotic behaviour of $\dens^d_{\sig_n,\priv_m}$ on increasing signatures, which are the initial segments of infinite lists:
\[
\sig_{n}\eqdef\set{p_{1},\dots,p_{n}}
\qquad\text{ and }\qquad
\priv_{m}\eqdef\set{q_{1},\dots,q_{m}}.
\]
The modal depth
$d \geq 0$, on the other hand, is fixed. We abbreviate $\smash{\dens^{\smash{d}}_{\sig_n,\priv_m}}$ and  $\smash{\smash{\overline{\dens}}^{\smash{d}}_{\sig_n,\priv_m}}$
by $\dens^d_{n,m}$ and $\smash{\overline{\dens}}^{\smash{d}}_{n,m}$, respectively.

\subsection{Asymptotics for $n$ fixed and $m\to\infty$}\label{sec:uniform-mtoinfty}

We begin by proving that, within any single~$\sig^{\uparrow\priv}$-cell, the density of failure of uniform interpolation is at most $2^{\smash{-2^{|\priv|}}}$, and deduce that $\dens^d_{n,m}\to 1$ as $m\to\infty$, for each fixed $n \geq 0$.

Consider two disjoint signatures $\sig$ and $\priv$.
Recall that a \emph{$\priv$-colour} $r$ is a set containing either $q$ or
$\neg q$, for each $q\in\priv$. Each $\priv$-colour $r$ clearly defines a pinning of $\priv$, which we also denote by $r$.
The next lemma identifies atoms that cannot occur in a subset $D$ whose characteristic formula lacks a uniform interpolant. 
\begin{lem}\label{lem:typefan}
	For every cell $C\in \CellAt{d}{\sig}{\priv}$,
consider 
	\[
	M_C\eqdef\bigl\{\,
	\pin{r}{\mathcal{M}}
	\bigm| 
	r\text{ a }\priv\text{-colour}\,\bigr\}\ \subseteq \Atom{d}{\sig\cup\priv},\qquad \text{ where } \mathcal{M} =\textstyle \bigcup C.
	\]
	Then $M_C\subseteq C\setminus D$, for every $D\subseteq C$ such that $\MForm{d}{n, m}{D}$ has no uniform $\sig$-interpolant, and $|M_C|= 2^{|\priv|}$.
\end{lem}

\begin{proof}
	First, each
	$\pin{r}{\mathcal{M}}$
	shares the
	$\sig^{\uparrow\priv}$-cell 
	$C$, since pinning $\priv$ leaves $\sig$-formulas
	untouched; thus, $M_C\subseteq C$. 

 Next, suppose $\MForm{d}{n,m}{D}$, for $D\subseteq C$, has no uniform $\sigma$-interpolant. We show $M_C\cap D=\emptyset$. 
 Suppose, towards a contradiction, that $\pin{r}{\mathcal{M}}\in D$
 for some 
 $\priv$-colour~$r$. 
Then $\MForm{}{}{\pin{r}{\mathcal{M}}} \models\MForm{d}{n,m}{D}$, and so $\fpin{r}{(\MForm{}{}{\pin{r}{\mathcal{M}})}} \models \fpin{r}{\MForm{d}{n,m}{D}}$.
By Lemma~\ref{lem:constants:atomsets}(1), $\fpin{r}{(\MForm{}{}{\pin{r}{\mathcal{M}})}} \equiv \MForm{}{}{\mathcal{M}} = \CForm{}{}{C}$, and so 
 $\CForm{}{}{C}\models\fpin{r}{\MForm{d}{n,m}{D}}$.
On the other hand, as  $D \subseteq C$, we have $\MForm{d}{n,m}{D}\models\CForm{}{}{C}$. It follows that $\fpin{r}{\MForm{d}{n,m}{D}}\models\CForm{}{}{C}$ because $\CForm{}{}{C}$ is $\priv$-free and so $\fpin{r}{(\CForm{}{}{C})} =\CForm{}{}{C}$.
Thus,  
\begin{equation*}
\fpin{r}{\MForm{d}{n,m}{D}}\equiv\CForm{}{}{C}.
\end{equation*} 
Let $\chi$ be an arbitrary $\priv$-free formula with
$\MForm{d}{n,m}{D}\models\chi$. Then $\fpin{r}{\MForm{d}{n,m}{D}}\models\fpin{r}{\chi}$.
As $\fpin{r}{\chi} =\chi$, we get $\CForm{}{}{C}\models\chi$.
So,  formula $\CForm{}{}{C}$
 is a uniform $\sig$-interpolant of $\MForm{d}{n,m}{D}$, a contradiction.
 So, $M_C\cap D=\emptyset$.

	Finally, 
	atoms $\pin{r}{\mathcal{M}}$ and $\pin{r'}{\mathcal{M}}$
	for distinct $\priv$-colours $r\ne r'$, differ as some $q\in\priv$ has opposite
	values. 
	This gives $\smash{2^{|\priv|}}$ distinct members of $M_C$. 
\end{proof}
Observe that the proof actually shows that $\CForm{}{}{C}$ is a uniform $\sig$-interpolant of $\MForm{}{}{D}$ whenever $D\subseteq C\setminus M_C$ because every model of $\CForm{}{}{C}$ can be turned into a model of $\MForm{}{}{D}$ by interpreting $\priv$ according to the pinning function $r$.

\begin{thm}\label{thm:fsi}
	For every cell $C\in \CellAt{d}{\sig}{\priv}$,
	\[
	\overline{\dens}_{C}\ \le\ \frac{1}{2^{2^{|\priv|}}} .
	\]
\end{thm}

\begin{proof}
   Consider, for a cell $C$, the set $M_C$ of Lemma~\ref{lem:typefan}. We have $|M_{C}|= 2^{|\priv|}$.
	Let $D\subseteq C$ be any subset whose characteristic formula $\MForm{d}{n, m}{D}$ lacks a uniform
	$\sig$-interpolant. 
	By Lemma~\ref{lem:typefan}, 
	 $D\cap M_{C}=\emptyset$. Thus, $D\subseteq C\setminus M_{C}$, and so there are at most $2^{|C|-|M_{C}|}= 2^{|C|-\smash{2^{|\priv|}}}$ many such~$D$.
	Dividing by $2^{\abs{C}}$, we obtain
	\[
	\overline{\dens}_{C}
	\ = \  \frac{|\bigl\{D\subseteq C\bigm| \MForm{d}{n,m}{D}\text{ has no uniform }\sig\text{-interpolant}\bigr\}|}{2^{|C|}}
	\ \leq \ \frac{2^{|C|- 2^{|\priv|}}}{2^{|C|}} 
	\  = \ 2^{-2^{|\priv|}} . \qedhere
	\]
\end{proof}

We can now show the main result of this section for increasing forgotten signatures $\rho_0, \rho_1,\dots$:
\begin{thm}\label{thm:mtoinfty}
	For any $L$ and any fixed $n$, we have
$\smash{\overline{\dens}}^{\smash{d}}_{n,m}\ \le\ |\Atom{d}{\sig_n}|\cdot 2^{-2^{m}}$, and so
	$\lim\limits_{m\to\infty} \dens^d_{n,m} = 1$.
\end{thm}

\begin{proof}
	By Proposition~\ref{prop:second-part}, there are exactly $|\Atom{d}{\sig_n}|$-many $\sig_n^{\uparrow\priv_m}$-cells of $\Atom{d}{\sig_n\cup\priv_m}$.
	By the cell reduction principle
	(Theorem~\ref{thm:slicereduction}) and the upper bound on the number of counterexamples in a cell
	(Theorem~\ref{thm:fsi}), we obtain
	\[
	\dens_{n,m}^{\smash{d}}=\prod_{C\in \CellAt{d}{\sig_n}{\priv_m}}\dens_{C}
	\ \ge\ \bigl(1-2^{-2^{m}}\bigr)^{|\Atom{d}{\sig_n}|}
	\ \ge\ 1-|\Atom{d}{\sig_n}|\cdot 2^{-2^{m}},
	\]
	using $(1-x)^{k}\ge 1-kx$ for $x\in[0,1]$. Since $|\Atom{d}{\sig_n}|$ does not
	depend on $m$, the right-hand side tends to $1$.
\end{proof}

\subsection{Asymptotics for $m$ fixed and $n\to\infty$}\label{sec:ntoinfty}
\label{subsec:lift1}

Assume that $L$ lacks uniform interpolation for some formula of depth at most $d$. Then there are $n^*\geq 0$, \mbox{$m^*\geq 1$} and~$D_{m^*} \subseteq \Atom{d}{\sig_{n^*}\cup\priv_{m^*}}$ such that $\MForm{d}{n^*,m^*}{D_{m^*}}$ has no uniform $\sigma_{n^{\ast}}$-interpolant. $D_{m^*}$ serves as a seed counterexample to uniform interpolation, from which further counterexamples in any $\Atom{d}{\sig_n\cup\priv_m}$ with $n\geq n^*$ and $m\geq m^*$ are constructed.     
We then show $\dens^{\smash{d}}_{n,m}\to 0$ as $n\to\infty$, for any fixed $m\geq m^{\ast}$. 

Consider three pairwise disjoint signatures $\sig^*$, $\gsig$ and $\priv$, with $\sig = \sig^*\cup \gsig$.
Let $\uparrow\!\!\gsig$ be a pinning for $\gsig$ defined by taking $\uparrow\!\!\gsig (p) = \top$, for all $p\in \gsig$. We call $\uparrow\!\!\gsig$ a \emph{lifting}.
Then denote by 
$D^{\uparrow \tau}$ and $\afpin{\gsig}{\varphi}$ the results of pinning~$\gsig$ in $D$ and $\varphi$, respectively, with $\uparrow\!\!\gsig$.
We show that the lifting operation preserves non-existence of uniform interpolation.
\begin{lem}\label{lem:liftfail}
	For any $D\subseteq\Atom{d}{\sig^*\cup\priv}$, if $\MForm{d}{\sig^*, \priv}{D}$ does not have a uniform $\sig^*$-interpolant, then $\MForm{d}{\sig, \priv}{D^{\uparrow \tau}}$ does not
		have a uniform $\sig$-interpolant.
\end{lem}
\begin{proof} 	
Suppose $\MForm{d}{\sig\cup\tau,\priv}{D^{\uparrow\tau}}$ has a uniform $\sig$-interpolant
$\theta$. 
We have $\MForm{d}{\sig\cup\tau,\priv}{D^{\uparrow\tau}}\models\theta$, whence
$\MForm{d}{\sig,\priv}{D}\models\afpin{\tau}{\theta}$ because, by Lemma~\ref{lem:constants:atomsets}(1),
$\afpin{\tau}{\smash{\MForm{d}{\sig\cup\tau,\priv}{D^{\uparrow\tau}}}}\equiv\MForm{d}{\sig,\priv}{D}$.
Let $\psi$ be any $\priv$-free consequence of $\MForm{d}{\sig,\priv}{D}$.
Since $\tau$  is disjoint from $\sig^*\cup\priv$,
variables of $\psi$ outside $\sig^*$ may be renamed
so that $\psi$ contains no variable of $\tau$.
Then $\MForm{d}{\sig\cup\tau,\priv}{D^{\uparrow\tau}}\models\MForm{d}{\sig,\priv}{D}\models\psi$, and uniform
$\sig$-interpolation yields $\theta\models\psi$, whence 
$\afpin{\tau}{\theta}\models\psi$ as $\afpin{\tau}{\psi} = \psi$.
The renaming can be reversed, as $\afpin{\tau}{\theta}\models\psi$
uses only~$\sig^*$. Thus, $\afpin{\tau}{\theta}$ is a uniform
$\sig^*$-interpolant of $\MForm{d}{\sig^*,\priv}{D}$, a contradiction.
\end{proof}

On the other hand, since the flipping operation is involutive, it preserves both the existence and non-existence of uniform
interpolants.
\begin{lem}\label{lem:flip:inter}
For any $v \subseteq \sig$ and $D\subseteq \Atom{d}{ \sig\cup\priv}$, formula $\MForm{}{}{D}$ has a uniform $\sig$-interpolant  if and only if 
$\MForm{}{}{D^v}$ does.
\end{lem}
\begin{proof}
By Lemma~\ref{lem:flipping:atomsets}(1), if $\theta$ is a uniform $\sig$-interpolant of $\MForm{}{}{D}$, then $\theta^v$ is a uniform $\sig$-interpolant of
$\MForm{}{}{D^v}$: apply the same involutive substitution to each entailment,
including consequences in arbitrary additional variables.
\end{proof}

We now apply the lifting and flipping operations to 
show the main convergence result in this section. 
\begin{thm}\label{thm:ntoinfty}
	Let $d\geq 0$. If $\MForm{d}{n^*,m^*}{D}$, for $D\subseteq \Atom{d}{\sig_{n^*}\cup\priv_{m^{\ast}}}$, has no uniform $\sig_{n^{\ast}}$-interpolant,
	then, for every fixed $m\ge m^{\ast}$, there is
	a $\sigma_{n^{\ast}}^{\uparrow\rho_{m}}$-cell $C_{m}$ such that
\begin{equation*}	
\dens^d_{n,m}\ \leq \ \bigl(1-2^{-|C_{m}|}\bigr)^{2^{n}}, \qquad \text{ for all } n\geq n^{\ast}. 
\end{equation*}
Consequently, $\lim\limits_{n\to\infty} \dens^{\smash{d}}_{n,m} = 0$.
\end{thm}
\begin{proof}
Assume $\MForm{d}{n^*,m^*}{D}$ with $D\subseteq \Atom{d}{\sig_{n^*}\cup\priv_{m^{\ast}}}$, $n^\ast\geq 0$ and $m^* \geq 1$, has no uniform $\sig_{n^\ast}$-interpolant. Then, by Lemma~\ref{lem:slicedecomp}, there is
a cell $C_{m^{\ast}}\in \CellAt{d}{\sig_{n^{\ast}}}{\priv_{m^{\ast}}}$ 
such that $\MForm{d}{n^*,m^*}{C_{m^{\ast}} \cap D}$ has no
uniform $\sig_{n^{\ast}}$-interpolant. 
Consider now the \emph{trail} $(C_m)_{m\geq m^*}$ of $C_{m^*}$, the sequence defined by taking 
\[
C_m\ \  = \ \ \bigl\{\mathcal{M}\in\Atom{d}{\sig_{n^\ast}\cup\priv_m} \bigm| \mathcal{M} \subseteq \textstyle\bigcup C_{m^{\ast}} \bigr\} \ \ \in \ \ \CellAt{d}{\sig_{n^{\ast}}}{\priv_{m}}, \qquad\text{ for } m \geq m^*.
\]
Equivalently, $\mathcal{M}\in\Atom{d}{\sig_{n^\ast}\cup\priv_m}$ belongs to $C_{m}$ if some (equivalently, all) $(\M,w)\in \mathcal{M}$ satisfy the cell characteristic formula $\Gamma(C_{m^{\ast}})$.
By Proposition~\ref{prop:second-part},  $\bigcup C_{m}= \bigcup C_{m^{\ast}}$, for each $m \geq m^*$, and so
$C_{m}$ trivially inherits from $C_{m^{\ast}}$ a set $D_{m}\subseteq C_{m}$ such that $\MForm{d}{n^*,m}{D_{m}}$ has no uniform $\sig_{n^{\ast}}$-interpolant. 
Fix $m \geq m^*$ and consider any $n \geq n^*$. For every $v \subseteq \sig_n$, let
$C_{m}^{n,v} \eqdef (C_{m}^{\uparrow \tau_n})^v$, where $\tau_n \eqdef \sig_n\setminus \sig_{n^*}$. By Lemmas~\ref{lem:constants:atomsets} and \ref{lem:flipping:atomsets}, $C_{m}^{n,v}$ is a $\sig_n^{\uparrow\rho_m}$-cell, which,  by Lemmas~\ref{lem:liftfail} and \ref{lem:flip:inter},  contains a counterexample to uniform $\sig_n$-interpolation. Thus,
\[
\overline{\dens}_{C_{m}^{n,v}} \ \geq \ \frac{1}{2^{|C_{m}^{n,v}|}}, \quad \text{ for all } n\geq n^{\ast} \text{ and } v\subseteq \sig_{n}.
\] 
By Lemmas~\ref{lem:constants:atomsets} and \ref{lem:flipping:atomsets}, the size of the cells is invariant under flipping and lifting: 
$|C_{m}^{n,v}| = |C_m|$,  for all $n\geq n^{\ast}$ and $v\subseteq\sig_{n}$.
By the cell reduction principle (Theorem~\ref{thm:slicereduction}), as $0 \leq \dens_{C}\leq 1$, we obtain
\[
\dens^d_{n,m}\ \ =\ \prod_{C\in \CellAt{d}{\sig_n}{\priv_m}}\hspace*{-1em}\dens_{C} \ \ \leq \ \ \prod_{v\subseteq \sig_{n}}\dens_{C_{m}^{n,v}} \ \ \leq \ \ \Bigl(1-\frac{1}{2^{\abs{C_{m}}}}\Bigr)^{2^{n}}.
\]
Since $\abs{C_{m}}$ does not depend on $n$, the claim follows.
\end{proof}

We have determined the asymptotic behaviour of $\dens^d_{n,m}$ if either $n$ or $m$ is fixed. It would also be of interest to investigate what happens if both grow. 

\begin{exa}\label{exm:phase}
In the 3-fork example, we obtain the following sharp
transition when both signature sizes grow. Put
\[
n_c(m)\ \ \eqdef\ \ \frac{2^{4m}+2^{3m}}{6}.
\]
As $n,m\to\infty$, we have
\begin{equation*}
	\dens_{n,m}
	\ \ =\ \ e^{-(1+o(1))\,2^{3(n-n_c(m))}}.
\end{equation*}
Consequently, for any sequence $n=n(m)\to\infty$ as $m\to\infty$,
\[
\lim_{m\to\infty}\dens_{n(m),m}
\ \ =\ \ \begin{cases}
	1, & \text{if } n(m)-n_c(m)\to-\infty,\\
	e^{-2^{3t}}, & \text{if }  n(m)-n_c(m)\to t,\\
	0, & \text{if }  n(m)-n_c(m)\to+\infty.
\end{cases}
\]
The result shows that the transition from almost all formulas having uniform interpolants to almost none is extremely sharp. It occurs when the number of retained variables is about $n_c(m)$, where $m$ is the number of forgotten variables. Near this threshold, adding just a couple of retained variables can dramatically reduce the density: for example, for $m = 3$, we have $n_c(m) = 768$, and $\dens_{n,m}$ goes down from $\approx 0.88$ for~$n = 767$ to~$\approx 0.0003$ for~$n = 769$.
\end{exa}

\section{Craig Interpolants}\label{sec:craig}

A formula $\theta$ is a \emph{Craig interpolant} for formulas
$\varphi$ and $\psi$ if
\[
\signat(\theta)\subseteq
\signat(\varphi)\cap\signat(\psi),
\qquad
\varphi\models\theta,
\qquad
\psi\models\neg\theta.
\]
A logic $L$ has the \emph{Craig interpolation property (CIP)}
if every pair $\varphi,\psi$ with
$\varphi\wedge\psi\models\bot$ admits a Craig interpolant.
For quasi-normal modal logics, this is equivalent, by the
deduction theorem for $\models$, to the usual formulation
of CIP in terms of implications, also called
\emph{local interpolation}; see~\cite{DBLP:journals/corr/abs-2512-01600}.

As in the uniform interpolation setting, we fix a modal depth
$d\ge 0$ and disjoint signatures $\sig$ and~$\priv$. Let~$\priv'$ be a fresh disjoint copy of $\priv$.
We consider pairs of formulas $\varphi$, $\psi$ of depth at
most $d$ with
\[
\signat(\varphi)=\sig\cup\priv
\qquad\text{ and }\qquad
\signat(\psi)=\sig\cup\priv'.
\]
Thus, $\sig$ is precisely their shared signature, while
$\priv$ and $\priv'$ contain the variables occurring only
on the left and right, respectively. As formulas are counted
up to $L$-equivalence, requiring every variable in the
prescribed signatures to occur does not affect the count.

The fixed bijection from $\priv$ to $\priv'$ induces a
bijection $\Atom{d}{\sig\cup\priv}
\rightarrow
\Atom{d}{\sig\cup\priv'}$ on atoms: the result of renaming $\priv$ to $\priv'$ in $\mathcal{M}$ is denoted by $\mathcal M'$, and $D' = \{\mathcal M'\mid\mathcal M\in D\}$, for $D\subseteq\Atom{d}{\sig\cup\priv}$.

\begin{defi}\label{def:clashpair}
	Let $D_1,D_2\subseteq
	\Atom{d}{\sig\cup\priv}$.
	\begin{enumerate}[label=\textup{(\roman*)}]
		\item
		The pair $(D_1,D_2)$ is \emph{inconsistent}
		if $\MForm{d}{\sig\cup\priv}{D_1}\wedge 
		\MForm{d}{\sig\cup\priv'}{D_2'}\models\bot$.
		The set
		of all inconsistent pairs in $\Atom{d}{\sig\cup\priv}$ is denoted by $\SetClash^{\smash{d}}_{\sig,\priv}$.
		\item
		A \emph{Craig $\sig$-interpolant} for
		$(D_1,D_2)$ is a formula
		$\theta\in\ML_\sig$ such that
		\[
		\MForm{d}{\sig\cup\priv}{D_1}\models\theta,
		\qquad
		\theta\models\neg\MForm{d}{\sig\cup\priv'}{D_2'}.
		\]
		The
		set of all pairs in $\Atom{d}{\sig\cup\priv}$ admitting a Craig $\sig$-interpolant is denoted by $\SetCraig^{\smash{d}}_{\sig,\priv}$.
	\end{enumerate}
\end{defi}

Clearly, $
\SetCraig^{\smash{d}}_{\sig,\priv}
\subseteq
\SetClash^{\smash{d}}_{\sig,\priv}$.
In particular, pairs with an empty component belong to
both sets.
We next introduce a relation on atoms that describes
inconsistency of pairs. Two atoms
$\mathcal M_1,\mathcal M_2\in
\Atom{d}{\sig\cup\priv}$
are \emph{incompatible}, written
$\mathcal M_1\frown\mathcal M_2$, if
\[
\mathcal M_1\cap\mathcal M_2'=\emptyset.
\]
Otherwise, they are \emph{compatible}.

\begin{prop}\label{prop:clashcrit}
	Let $D_1,D_2\subseteq
	\Atom{d}{\sig\cup\priv}$.
	Then $(D_1,D_2)$ is inconsistent if and only if
	\[
	\mathcal M_1\frown\mathcal M_2,
	\qquad
	\text{for all }\mathcal M_1\in D_1
	\text{ and }\mathcal M_2\in D_2.
	\]
\end{prop}

\begin{proof}
	The conjunction $\Phi(D_1)\wedge\Phi(D_{2}')$ is
	satisfiable if and only if
	$\mathcal M_1\cap\mathcal M_2'\ne\emptyset$
	for some $\mathcal M_1\in D_1$ and
	$\mathcal M_2\in D_2$.
\end{proof}
\begin{exa}\label{ex:3-fork:craig}
The 3-fork formulas $\varphi$ and $\psi$
from Example~\ref{exm:formulas} illustrate
incompatibility. Their conjunction is
inconsistent: $\varphi$ requires at least
two successors satisfying $p$, while
$\psi$ requires at least two satisfying~$\neg p$, and the frame has only three
successors. On the other hand, the two
formulas have the same $\{p\}$-consequences.
Consequently, although
$\varphi\models\neg\psi$, there is no
Craig interpolant over $\{p\}$.
Equivalently, consider atom $\mathcal M_\varphi$ from Example~\ref{ex:3-forks-cells} and let $\mathcal M_\psi$ denote the atom corresponding to $\psi$ after renaming $q'$ to~$q$. Then $\mathcal{M}_{\psi}=\mathcal{M}_{pq,pq,\neg p\{q, \neg q\}}$ and  
$\mathcal M_\varphi\frown\mathcal M_\psi$ even though both atoms belong to the same~$\sig^{\uparrow\priv}$-cell.
\end{exa}

For every cell
$C\in\CellAt{d}{\sig}{\priv}$,
let $\SetClash_C$ and $\SetCraig_C$ denote the
sets of inconsistent pairs and pairs admitting a
Craig $\sig$-interpolant, respectively, among
$D_1,D_2\subseteq C$. 
\begin{lem}\label{lem:craig-slicedecomp}
	For all
	$D_1,D_2\subseteq\Atom{d}{\sig\cup\priv}$,
	\begin{enumerate}
		\item $(D_1,D_2)$ is inconsistent if and
	only if
	$(D_1\cap C,D_2\cap C)\in\SetClash_C$, 
	for every
	$C\in\CellAt{d}{\sig}{\priv}$\textup{;}
	\item  $(D_1,D_2)$ admits a Craig
	$\sig$-interpolant if and only if
	$(D_1\cap C,D_2\cap C)\in\SetCraig_C$,
	for every
	$C\in\CellAt{d}{\sig}{\priv}$.
    \end{enumerate}
\end{lem}
\begin{proof}
		(1) By Proposition~\ref{prop:clashcrit},
		the pair $(D_1,D_2)$ is inconsistent if and only if
		every $\mathcal M_1\in D_1$ is incompatible with
		every $\mathcal M_2\in D_2$.
		If $\mathcal M_1$ and $\mathcal M_2$ are compatible,
		then  $\mathcal M_1\cap\mathcal M_2'\ne\emptyset$ implies that $\mathcal M_1$ and $\mathcal M_2$ belong to the same $\sig^{\uparrow\priv}$-cell.
		Thus, $(D_1,D_2)$ is inconsistent if and only if
		$(D_1\cap C,D_2\cap C)$ is inconsistent for every
		$\sig^{\uparrow\priv}$-cell~$C$.
		
		(2) Recall from Proposition~\ref{prop:second-part}
		and the definition of cell characteristic formulas
		preceding Lemma~\ref{lem:slicedecomp} that
		$\Gamma(C)$ is a $\sig$-formula and, for every cell $C$,
		$\Phi(D_1)\land\Gamma(C)\equiv \Phi(D_1\cap C)$ and $\Phi(D_2')\land\Gamma(C)
		\equiv \Phi((D_2\cap C)')$.
		The second equivalence follows from the first
		by renaming $\priv$ to~$\priv'$, since
		$\Gamma(C)$ contains only variables from~$\sig$.
		Moreover, the disjunction of all $\Gamma(C)$
		is valid.
		Now, if $\theta\in\ML_\sig$ is a Craig
		$\sig$-interpolant for $(D_1,D_2)$, then $\theta\land\Gamma(C)$ is a Craig
		$\sig$-interpolant for
		$(D_1\cap C,D_2\cap C)$.
		Conversely, if, for every cell $C$,
		there is a Craig $\sig$-interpolant
		$\theta_C$ for $(D_1\cap C,D_2\cap C)$, then $
		\bigvee_{C\in\CellAt{d}{\sig}{\priv}}
		\bigl(\Gamma(C)\land\theta_C\bigr)$
		is a Craig $\sig$-interpolant for $(D_1,D_2)$.
	\end{proof}
\section{Density of Craig Interpolants}
We now introduce the density of Craig interpolants and
establish a cell reduction principle analogous to
Theorem~\ref{thm:slicereduction}.
\begin{thm}[cell reduction principle for Craig interpolation]
	\label{thm:craig-factor}
	\[
	|\SetClash^{\smash{d}}_{\sig,\priv}|
	=
	\prod_{C\in
		\CellAt{d}{\sig}{\priv}}
	|\SetClash_C|,
	\qquad
	|\SetCraig^{\smash{d}}_{\sig,\priv}|
	=
	\prod_{C\in
		\CellAt{d}{\sig}{\priv}}
	|\SetCraig_C|.
	\]
\end{thm}

\begin{proof}
	Since the cells partition
	$\Atom{d}{\sig\cup\priv}$,
	the map
	$
	(D_1,D_2)
	\mapsto
	\bigl((D_1\cap C,D_2\cap C)\bigr)_{
		C\in\CellAt{d}{\sig}{\priv}}
	$
	is a bijection from
	$
	2^{\smash{\Atom{d}{\sig\cup\priv}}}
	\times
	2^{\smash{\Atom{d}{\sig\cup\priv}}}
	$
	onto
	$
	\prod_{C\in
		\CellAt{d}{\sig}{\priv}}
	(2^C\times 2^C)$.
	By Lemma~\ref{lem:craig-slicedecomp},
	this bijection restricts to bijections
	\[
	\SetClash^{\smash{d}}_{\sig,\priv}
	\longrightarrow
	\prod_{C\in
		\CellAt{d}{\sig}{\priv}}
	\SetClash_C
	\quad \text{ and } \quad
	\SetCraig^{\smash{d}}_{\sig,\priv}
	\longrightarrow
	\prod_{C\in
		\CellAt{d}{\sig}{\priv}}
	\SetCraig_C.
	\]
	The result follows by taking cardinalities.
\end{proof}

We define the \emph{density of Craig interpolation}
as the proportion of inconsistent pairs that
admit a Craig interpolant:
	\[
	\DCraig^{\smash{d}}_{\sig,\priv}
	\eqdef
	\frac{|\SetCraig^{\smash{d}}_{\sig,\priv}|}
	{|\SetClash^{\smash{d}}_{\sig,\priv}|},
	\qquad
	\DCraig_C
	\eqdef
	\frac{|\SetCraig_C|}{|\SetClash_C|}.
	\]
The denominators are non-zero, since pairs
with an empty component are inconsistent.
By Theorem~\ref{thm:craig-factor},
\begin{equation*}
	\DCraig^{\smash{d}}_{\sig,\priv}
	=
	\prod_{C\in
		\CellAt{d}{\sig}{\priv}}
	\DCraig_C.
\end{equation*}
In particular,
$\DCraig^{\smash{d}}_{\sig,\priv}=1$ if and only if
$\DCraig_C=1$ for every cell $C\in
\CellAt{d}{\sig}{\priv}$.
Thus, at each fixed depth, every
inconsistent pair admits a Craig
interpolant if and only if this holds within every cell. We also write $\smash{\overline{\DCraig}}^{\smash{d}}_{\sig,\priv}\eqdef 1-\smash{\DCraig}^{\smash{d}}_{\sig,\priv}$ and $\overline{\DCraig}_{C}\eqdef 1-\DCraig_{C}$ for the
	densities of Craig interpolation \emph{failure}.

We next consider the asymptotic behaviour of $\DCraig^{\smash{d}}_{\sig_n,\priv_m}$ on increasing signatures (but fixed modal depth).
We abbreviate $\smash{\DCraig^{\smash{d}}_{\sig_n,\priv_m}}$ by $\DCraig^{\smash{d}}_{n,m}$.

\subsection{Asymptotics for $m$ fixed and $n\to\infty$}\label{subsec:lift7}

Suppose that, at the fixed depth $d$,
there is an inconsistent pair without a Craig
interpolant. Thus, for some $n^\ast\ge0$ and
$m^\ast\ge1$, there are
$D_1,D_2\subseteq
\Atom{d}{\sig_{n^\ast}\cup\priv_{m^\ast}}$
such that $(D_1,D_2)$ is inconsistent and
admits no Craig $\sig_{n^\ast}$-interpolant.
We show that, for every fixed $m\ge m^\ast$,
the density $\DCraig^{\smash{d}}_{n,m}$ tends to $0$
as $n\to\infty$.

We use the lifting and flipping operations
introduced in Section~\ref{subsec:lift1} (see also Section~\ref{sec:pinning:flipping}).
The following lemma shows that lifting
preserves both inconsistency and Craig
interpolation.

\begin{lem}[lifting for Craig  interpolation]\label{label:lift3}
	Let $\sig^*$, $\tau$ and $\priv$ be pairwise disjoint
 signatures. 
	For all $D_1,D_2\subseteq \Atom{d}{\sig^*\cup\priv}$,
	\begin{enumerate}[label=\textup{(\roman*)}]
		\item $(D_1,D_2)$ is inconsistent
		if and only if
		$(D_1^{\uparrow\tau},
		D_2^{\uparrow\tau})$ is inconsistent\textup{;}
		
		\item $(D_1,D_2)$ admits a Craig
		$\sig^*$-interpolant if and only if
		$(D_1^{\uparrow\tau},
		D_2^{\uparrow\tau})$ admits a Craig
		$(\sig^*\cup\tau)$-interpolant.
	\end{enumerate}
\end{lem}

\begin{proof}
	Since $\priv'$ is disjoint from $\tau$, lifting
	commutes with the renaming of $\priv$ to $\priv'$.
	Thus, by Lemma~\ref{lem:constants:atomsets}(1),
	$\Phi(D_1^{\uparrow\tau})\equiv\Phi(D_1)\land
		\bigwedge_{p\in\tau}\Box^{\le d}p$ and 
		$\Phi((D_2^{\uparrow\tau})')
		\equiv
		\Phi(D_2')\land
		\bigwedge_{p\in\tau}\Box^{\le d}p$.
	In particular,
	$\afpin{\tau}{\Phi(D_1^{\uparrow\tau})}\equiv \Phi(D_1)$ and $	\afpin{\tau}{\Phi((D_2^{\uparrow\tau})')}\equiv \Phi(D_2')$.
	Recall that, by Lemma~\ref{lem:constants}(1),
	pinning preserves entailment.
	
	\smallskip
	\noindent\textup{(i)}
	If $(D_1,D_2)$ is inconsistent, 
	then 
	$\Phi(D_1^{\uparrow\tau})\models\Phi(D_1)$ and
	$\Phi((D_2^{\uparrow\tau})')\models\Phi(D_2')$
	imply that $(D_1^{\uparrow\tau},D_2^{\uparrow\tau})$ is inconsistent.	
	Conversely, suppose that
	$(D_1^{\uparrow\tau},D_2^{\uparrow\tau})$
	is inconsistent. Applying the pinning substitution
	$\cdot^{\downarrow\tau}$ and using the
	equivalences above gives
	$\Phi(D_1)\land\Phi(D_2')\models\bot$.	
	Hence, $(D_1,D_2)$ is inconsistent.
	
	\smallskip
	\noindent\textup{(ii)}
	Suppose first that $\theta\in\ML_{\sig^*}$
	is a Craig $\sig^*$-interpolant for $(D_1,D_2)$.
	Then $\Phi(D_1)\models\theta$ and $\theta\models\neg\Phi(D_2')$.
	The equivalences established above imply
	$\Phi(D_1^{\uparrow\tau})\models\theta$ and $\theta\models
	\neg\Phi((D_2^{\uparrow\tau})')$.
	Since $\theta\in\ML_{\sig^*}
	\subseteq\ML_{\sig^*\cup\tau}$,
	it is also a Craig $(\sig^*\cup\tau)$-interpolant
	for $(D_1^{\uparrow\tau},D_2^{\uparrow\tau})$.
	
	Conversely, suppose that
	$\theta\in\ML_{\sig^*\cup\tau}$ is a Craig
	$(\sig^*\cup\tau)$-interpolant for
	$(D_1^{\uparrow\tau},D_2^{\uparrow\tau})$.
	Applying $\cdot^{\downarrow\tau}$ to both
	entailments and using
	Lemma~\ref{lem:constants}(1) together with
	the equivalences above, we obtain
	$
	\Phi(D_1)\models\afpin{\tau}{\theta}$ and 
	$\afpin{\tau}{\theta}\models\neg\Phi(D_2')$.
	Since $\theta\in\ML_{\sig^*\cup\tau}$,
	we have $\afpin{\tau}{\theta}\in\ML_{\sig^*}$.
	Thus $\afpin{\tau}{\theta}$ is a Craig
	$\sig^*$-interpolant for $(D_1,D_2)$.
\end{proof}

By Lemma~\ref{lem:flipping:atomsets}, flipping also preserves inconsistency
and Craig interpolation:
\begin{lem}\label{lem:flipping:craig}
Let $\sig$ and $\priv$ be disjoint signatures. For all $v\subseteq\sig$  and all $D_1,D_2\subseteq \Atom{d}{\sig\cup\priv}$,
	\begin{enumerate}[label=\textup{(\roman*)}]
		\item $(D_1,D_2)$ is inconsistent if and only if $(D_1^v,D_2^v)$ is inconsistent\textup{;}\
		\item $(D_1,D_2)$  admits a Craig $\sig$-interpolant if and only if $(D_1^v,D_2^v)$ does.
\end{enumerate}		
\end{lem}

We can now prove the Craig analogue of
Theorem~\ref{thm:ntoinfty}. 
\begin{thm}\label{thm:ntoinfty3}
	Let $d \geq 0$. If there is an inconsistent pair
	$
	D_1,D_2\subseteq
	\Atom{d}{
		\sig_{n^\ast}\cup\priv_{m^\ast}}$
	without a Craig $\sig_{n^\ast}$-interpolant,
	then, for every fixed $m\ge m^\ast$,
	there is a
	$\sig_{n^\ast}^{\uparrow\priv_m}$-cell
	$C_m$ such that
\begin{equation*}	
	\DCraig^{\smash{d}}_{n,m}
	\ \ \le \ \
	\bigl(1-2^{-2|C_m|}\bigr)^{2^n}, \qquad \text{ for all } n\ge n^\ast.
\end{equation*}
	Consequently,
    $\lim\limits_{n\to\infty}\DCraig^{\smash{d}}_{n,m}=0$.
    \end{thm}

\begin{proof}
	Assume that $D_{1},D_{2}\subseteq \Atom{d}{
		\sig_{n^\ast}\cup\priv_{m^\ast}}$
	is an inconsistent pair without a Craig $\sig_{n^\ast}$-interpolant. By Lemma~\ref{lem:craig-slicedecomp},
	there is a
	$\smash{\sig_{n^\ast}^{\uparrow\priv_{m^\ast}}}$-cell
	$C_{m^\ast}$ such that
	$
	(D_1\cap C_{m^\ast},
	D_2\cap C_{m^\ast})
	$
	is inconsistent and admits no
	Craig $\sig_{n^\ast}$-interpolant.
	Consider the trail $(C_m)_{m\geq m^*}$ of $C_{m^*}$ defined in the proof of Theorem~\ref{thm:ntoinfty} and recall
	that $\bigcup C_{m}= \bigcup C_{m^{\ast}}$, for each $m \geq m^*$.
	As in the proof of Theorem~\ref{thm:ntoinfty}, for every $m\geq m^\ast$, there are sets $D_{1,m},D_{2,m}\subseteq C_m$ 
	such that $(D_{1,m},D_{2,m})$ is inconsistent and admits no Craig
	$\sig_{n^\ast}$-interpolant.
	
	Fix $m\ge m^\ast$ and, for every $n\ge n^\ast$,
	consider
	$
	\tau_n\eqdef
	\sig_n\setminus\sig_{n^\ast}$.
	For each $v\subseteq\sig_n$, let
	as before $C_m^{n,v}
	\eqdef
	(C_m^{\uparrow\tau_n})^v$.
	By Lemmas~\ref{label:lift3} and~\ref{lem:flipping:craig},
	each cell $C_m^{n,v}$
	contains an inconsistent pair without
	a Craig $\sig_n$-interpolant.
	Consequently,
	\[
	|\SetCraig_{C_m^{n,v}}|
	\ \ \le \ \ 
	|\SetClash_{C_m^{n,v}}|-1.
	\]
	By Lemmas~\ref{lem:constants:atomsets} and \ref{lem:flipping:atomsets}, 
$|C_{m}^{n,v}| = |C_m|$. And, since
	$
	1\le
	|\SetClash_{C_m^{n,v}}|
	\le 2^{2|C_m|}$, 
	we obtain
	\[
	\smash{\overline{\DCraig}}_{C_m^{n,v}}
	\ \ \ge \ \
	2^{-2|C_m|}.
	\]
	By the cell reduction principle
	for Craig interpolation (Theorem~\ref{thm:craig-factor}),
	\[
	\DCraig^{\smash{d}}_{n,m}
		\ \ =
		\prod_{C\in
			\CellAt{d}{\sig_n}{\priv_m}}\hspace*{-1em}
		\DCraig_C \ \ 
		\le \ \ 
		\prod_{v\subseteq\sig_n}
		\DCraig_{C_m^{n,v}}
		\ \ \le\ \ 
		\bigl(1-2^{-2|C_m|}\bigr)^{2^n}.
	\]
	Since $m$ is fixed, $|C_m|$ does
	not depend on $n$, and the final
	expression tends to $0$ as
	$n\to\infty$.
\end{proof}

\subsection{Asymptotics for $n$ fixed and $m\to\infty$}\label{sec:craig-minfty}

We now consider the case in which the retained signature
is fixed and the number of non-shared variables tends to
infinity. Here we obtain results for locally tabular logics.
As in Section~\ref{sec:uniform-mtoinfty}, we aim to bound
the density of interpolation failure within individual cells.
For locally tabular logics, we drop the depth restriction
and work with all formulas over a finite signature.
We show that, if the incompatibility graphs of cells admit
profiles whose sizes are bounded by a function of the
retained signature size alone, then $\DCraig_{n,m}\to1$
for fixed $n$. We then apply this result to tabular
quasi-normal modal logics.

Fix a locally tabular quasi-normal modal logic $L$
and disjoint signatures $\sig$ and $\priv$.
We write $\Atom{}{\sig\cup\priv}$ for the finite
set of $(\sigma\cup\rho)$-atoms defined without a
modal depth restriction.
The characteristic formulas $\Phi(\mathcal M)$ of atoms~$\mathcal M$ and the cell characteristic formulas
$\Gamma(C)=\Phi(\bigcup C)$ are likewise defined
without a modal-depth restriction.
The following observation will be useful throughout.

\begin{lem}\label{lem:localfinitecellcharacteristic}
	Let $C\in \CellAt{}{\sig}{\priv}$ and let
	$D,D_1,D_2\subseteq C$ be non-empty. Then
	\begin{enumerate}
		\item $\Phi(D)$ has a uniform $\sigma$-interpolant
		if and only if $\Gamma(C)$ is a uniform
		$\sigma$-interpolant of~$\Phi(D)$\textup{;}
		\item the pair $(D_1,D_2)$ does not admit a
		Craig $\sigma$-interpolant.
	\end{enumerate}
\end{lem}

\begin{proof}
(1)	By local tabularity, Proposition~\ref{prop:second-part}
	applies without a depth bound. Thus, $\Gamma(C)$ is the
	characteristic formula of a $\sig$-atom.
	Since $D$ is non-empty and all its atoms belong to the
	same cell, $\Phi(D)$ and $\Gamma(C)$ have exactly the
	same $\sigma$-consequences. In particular,
\begin{equation*}	
	\Phi(D)\models\theta\qquad \text{ iff }\qquad \Gamma(C)\models\theta,\qquad \text{ for every } \theta\in\ML_\sig.
\end{equation*}
	If $\theta$ is a uniform $\sigma$-interpolant of
	$\Phi(D)$, then $\theta\models\Gamma(C)$, since
	$\Gamma(C)$ is a $\priv$-free consequence of $\Phi(D)$.
	The above equivalence gives
	$\Gamma(C)\models\theta$, so
	$\theta\equiv\Gamma(C)$.
	The converse is immediate.
	
	(2) Observe that any Craig $\sigma$-interpolant $\theta$
	would satisfy $\Gamma(C)\models\theta$ by the same
	equivalence. Since $\Phi(D_2')\models\Gamma(C)$
	and $D_2\ne\emptyset$, this contradicts
	$\theta\models\neg\Phi(D_2')$.
\end{proof}

We now count the pairs admitting a Craig interpolant
within a single cell.

\begin{thm}\label{thm:lftab-count}
	Let $L$ be a locally tabular quasi-normal logic.
	Then
	$|\SetCraig_C|=2^{\abs{C}+1}-1$
	for every $\sig^{\uparrow\priv}$-cell~$C$.
	In particular,
	\[
	|\SetCraig_{\sig,\priv}|
	\ \ =
	\prod_{C\in\CellAt{}{\sig}{\priv}}
	\bigl(2^{\,\abs{C}+1}-1\bigr).
	\]
\end{thm}

\begin{proof}
	The product formula follows from
	Theorem~\ref{thm:craig-factor}.
	We compute $|\SetCraig_C|$ for a fixed cell~$C$.
	By Lemma~\ref{lem:localfinitecellcharacteristic},
	no pair $(D_1,D_2)$ with non-empty
	$D_1,D_2\subseteq C$ admits a Craig interpolant.
	On the other hand, if $D_1=\emptyset$ or
	$D_2=\emptyset$, then $\bot$ or $\top$ is a
	Craig interpolant, respectively.
	Therefore, the pairs in $C$ with Craig interpolants
	are exactly those with an empty component, giving
	$
	|\SetCraig_C|
	=
	2^{|C|}+2^{|C|}-1
	=
	2^{|C|+1}-1$.
\end{proof}

To count the inconsistent pairs within a single cell, we consider the graph on $C$ induced by the incompatibility relation between atoms in $C$.
\begin{defi}[incompatibility graph]
	\label{def:compat-graph}
	The \emph{incompatibility graph} of a cell $C$
	is $H_C=(C,\frown)$, where $\frown$ is the
	incompatibility relation between atoms in $C$.
\end{defi}

\begin{exa}\label{exm:biclique}
In Example~\ref{ex:3-fork:craig}, the 3-fork atoms
$\mathcal M_\varphi$ and $\mathcal M_\psi$
form an edge of the incompatibility graph
of $\smash{C^{\rho}_{p,\{p,\neg p\}}}$, for $|\rho|\geq 1$.
Thus the singleton pair
$(\{\mathcal M_\varphi\},
\{\mathcal M_\psi\})$
is a non-empty ordered biclique in $\smash{C^{\rho}_{p,\{p,\neg p\}}}$.
It represents an inconsistent pair
without a Craig interpolant.
\end{exa}

Every atom is compatible with itself, so $H_C$
has no loops. By Proposition~\ref{prop:clashcrit},
an inconsistent pair $(D_1,D_2)$ with
$D_1,D_2\subseteq C$ is precisely an
\emph{ordered biclique} in $H_C$.
By Theorem~\ref{thm:lftab-count}, such a pair
fails Craig interpolation exactly when both
sides are non-empty. We count them as follows:
\begin{multline}\label{eq:defofNoCraig}
\NoCraig_{C}\eqdef|\SetClash_{C}|-|\SetCraig_{C}|\\
=\abs{\set{(D_{1},D_{2})\mid \text{ $D_{1},D_{2}$ non-empty ordered biclique in $H_{C}$}}}.
\end{multline}
Then $|\SetClash_{C}|=(2^{\abs{C}+1}-1)+\NoCraig_{C}$, and
\begin{equation}\label{eq:dcraig-nocraig}
	\DCraig_{C}=\frac{2^{\abs{C}+1}-1}{2^{\abs{C}+1}-1+\NoCraig_{C}},
	\qquad
	\smash{\overline{\DCraig}}_{C}=\frac{\NoCraig_{C}}{|\SetClash_{C}|}\ \le\ \frac{\NoCraig_{C}}{2^{\abs{C}}} .
\end{equation}

To bound $\NoCraig_C$, we replace $H_C$
by a smaller graph that preserves its
incompatibility relation. A graph homomorphism
is called \emph{strong} if it both preserves
and reflects edges.

\begin{defi}\label{def:profile}
	Let $C$ be a cell and let
	$H_P=(P,\frown_P)$ be a graph.
	A \emph{profile map} for $C$ is a surjective
	strong graph homomorphism
	$\pi\colon H_C\to H_P$.
	We call $H_P$ a \emph{profile graph} of $C$,
	or simply a \emph{profile}.
	For $p\in P$ and $Q\subseteq P$, put
	\[
	w(p)\eqdef|\pi^{-1}(p)|,
	\qquad
	w(Q)\eqdef\sum_{p\in Q}w(p).
	\]
\end{defi}

In a  graph $H_G = (G, \frown_G)$, the \emph{neighbourhood} of $g\in G$ is $N_G(g) \eqdef 
	\{g'\in G\mid g\frown_G g'\}$, and
	the \emph{common neighbourhood} of a non-empty
	$G'\subseteq G$ is $
	N_G(G')\eqdef
	\bigcap_{g\in G'} N_G(g)$.

The following lemma provides an upper bound for $\NoCraig_{C}$ in terms of ordered bicliques in $H_{P}$ and the weight $w$.
	\begin{lem}\label{lem:ineqprofile} Let $C$ be a cell and $H_{P}=(P,\frown_{P})$ a profile graph of $C$. Then
	\[
	\NoCraig_{C}
	\ \le\ \sum_{\substack{\emptyset\ne Q\subseteq P\\ N_{P}(Q)\ne\emptyset}}
	2^{\,w(Q)}\cdot 2^{\,w(N_{P}(Q))}.
\]
\end{lem}
\begin{proof}
Let $\pi\colon H_{C}\to H_{P}$ be a profile map for $C$. 
	Observe that $(D_{1},D_{2})$ is an ordered biclique in $H_{C}$ with $D_{1},D_{2}$ non-empty iff there exists a non-empty $Q\subseteq P$ such that $D_{1}\subseteq \pi^{-1}(Q)$, $N_{P}(Q)\ne\emptyset$, and $D_{2}\subseteq \pi^{-1}(N_{P}(Q))$. The inequality follows from the definition of the weights~$w$.
\end{proof}

Among all profiles of a cell, there is a coarsest
one, obtained by identifying atoms with the
same incompatibility neighbourhood.
\begin{defi}\label{def:canprof}	
	Let, for any cell $C$ and $\mathcal{M}_{1},\mathcal{M}_{2}\in C$,
	\[
	\mathcal M_{1}\sim_{C} \mathcal M_{2}\ \iff\ N_{C}(\mathcal M_{1})=N_{C}(\mathcal M_{2}).
	\]
	Call the quotient graph $H_{C}/_{\sim_{C}}$ defined by $\sim_{C}$ with domain $C/_{\sim_{C}}$ the \emph{canonical profile graph} of $C$ and $\pi_{C}\colon H_C\to H_C/_{\sim_C}$ the \emph{canonical profile map} for $C$. Its
	number of classes is $\mathrm{inc}(C)\eqdef\abs{C/_{\sim_{C}}}$.
\end{defi}
The canonical profile map $\pi_{C}$ is a
profile map. Moreover, every profile map
$\pi\colon H_C\to H_P$ refines the canonical
partition, so $\mathrm{inc}(C)\le |P|$.
The logic $L$ has CIP if and only if
$H_C$ has no edges for every cell $C$.
Equivalently, $\mathrm{inc}(C)=1$ for
every cell.
\begin{exa}	
The 3-fork gives a simple example of
a canonical profile. For
$C=\smash{C^{\rho}_{p,\{p,\neg p\}}}$ with
$\rho=\{q\}$, the atoms fall into three
canonical classes: those with two distinct
successor colours, those with three distinct
successor colours of which two refine $p$,
and those with three distinct successor
		colours of which two refine~$\neg p$.
		The only edge of the canonical profile
		joins the last two classes. In particular,
		$\mathrm{inc}(C)=3$, although $C$ contains
		more than three atoms.
\end{exa}	
We now introduce a hierarchy of bounds on the size of canonical profiles.
\begin{defi}
  A locally tabular quasi-normal modal logic $L$ has the \emph{uniformly bounded intra-cell incompatibility
  		property \textup{(}uniform BICIP\textup{)}} if there is a constant $F_{L}$ such that $\mathrm{inc}(C)\leq F_{L}$ for all $\sigma^{\uparrow\rho}$-cells $C$.
   $L$ has \emph{BICIP} if there is a function $f_{L}$ depending on $n=|\sigma|$ only such that $\mathrm{inc}(C)\leq f_{L}(n)$ for every $\sigma^{\uparrow\rho}$-cell $C$, independently of $\rho$.
 \end{defi}
We next show that the pinned atoms of Lemma~\ref{lem:typefan}
are isolated in the incompatibility graph. For a
$\sig^{\uparrow\priv}$-cell~$C$, put
$\mathcal M\eqdef\bigcup C\in\Atom{}{\sig}$; see Proposition~\ref{prop:second-part}. As in Lemma~\ref{lem:typefan},
for each $\priv$-colour $r$, define
$\pin{r}{\mathcal M}\eqdef
[\pin{r}{\M},w]_{\sig\cup\priv}$ for $(\M,w)\in\mathcal M$, and let
$
M_C\eqdef\{\pin{r}{\mathcal M}\mid
r\text{ a }\priv\text{-colour}\}$.
By Lemma~\ref{lem:constants}(2), applied at every modal
depth, this definition does not depend on the chosen
representative $(\M,w)$. Pinning leaves all $\sig$-formulas
unchanged, so $M_C\subseteq C$. Distinct $\priv$-colours
differ at the distinguished world, and hence
$|M_C|=2^{|\priv|}$.

\begin{lem}\label{lem:sat-universal}
	Every atom in $M_C$ is compatible with every atom in $C$.
\end{lem}
\begin{proof}
	Fix a $\priv$-colour $r$ and $\mathcal N\in C$.
	Choose $(\M,w)\in\mathcal M$, $(\N,u)\in\mathcal N$.
	As \mbox{$(\M,w)\equiv_{\sig}(\N,u)$},
	Lemma~\ref{lem:constants}(2), applied at every depth, gives
	$
	[\pin{r}{\N},u]_{\sig\cup\priv}
	=[\pin{r}{\M},w]_{\sig\cup\priv}
	=\pin{r}{\mathcal M}$.
	Let~$\mathfrak P$ be based on the same frame as $\N$,
	interpret the variables in $\priv$ constantly according to~$r$, and put $V_{\mathfrak P}(q')=V_{\N}(q)$ for each
	corresponding $q\in\priv$ and $q'\in\priv'$. All other
	valuations agree with $\N$. Then
	$
	[\mathfrak P,u]_{\sig\cup\priv}
	=\pin{r}{\mathcal M}$ and $[\mathfrak P,u]_{\sig\cup\priv'}=\mathcal N'$.
	Thus $\pin{r}{\mathcal M}\cap\mathcal N'\ne\emptyset$,
	which proves compatibility.
\end{proof}
Consequently, no atom of $M_C$ occurs in a non-empty
ordered biclique of $H_C$.
\begin{thm}\label{thm:bounded-profile}
	Let $L$ be a locally tabular quasi-normal
	modal logic with BICIP, and let $f_L$ be
	a corresponding bound. Then, for every
	$\sig^{\uparrow\priv}$-cell $C$,
	\[
	\smash{\overline{\DCraig}}_C
	\ \ \le \ \
	2^{f_L(|\sig|)-2^{|\priv|}}.
	\]
	Consequently, for every fixed $n\ge0$,
	$\lim\limits_{m\to\infty}\DCraig_{n,m}=1$.
\end{thm}
\begin{proof}
		Fix a $\sig^{\uparrow\priv}$-cell $C$
		and let $\pi_{C}\colon H_C\to H_P\eqdef H_{C}/{\sim_{C}}$ be its canonical
		profile map. By assumption,
		$|P|\le f_L(|\sig|)$.
		By Lemma~\ref{lem:sat-universal}, the atoms in $M_C$
		belong to a single canonical profile class $p_0\in P$
		with $w(p_0)\ge |M_C|=2^{|\priv|}$.
		Let $\emptyset\ne Q\subseteq P$ with
		$N_P(Q)\ne\emptyset$. Then $p_0\notin Q$, since
		otherwise $N_P(Q)=\emptyset$. Also,
		$p_0\notin N_P(Q)$, as $p_0$ has no neighbours.
		Since $H_P$ has no loops, $Q\cap N_P(Q)=\emptyset$.
		Consequently,
		\[
		w(Q)+w(N_P(Q))
		\ \ = \ \ w(Q\cup N_P(Q))
		\ \ \le \ \ |C|-w(p_0)
		\ \ \le \ \ |C|-2^{|\priv|}.
		\]
	We compute an upper bound for $\NoCraig_{C}$ using Lemma~\ref{lem:ineqprofile}:
		\[
		\NoCraig_{C}
		\ \le\ \sum_{\substack{\emptyset\ne Q\subseteq P\\ N_{P}(Q)\ne\emptyset}}
		2^{\,w(Q)+w(N_{P}(Q))}
		\ \le\ 2^{\abs{P}}\cdot 2^{\,\abs{C}-2^{|\rho|}}
		\ \le\ 2^{\,\abs{C}-(2^{|\rho|}-f_{L}(|\sigma|))}.
		\]
		Dividing by
		$|\SetClash_{C}|\ge 2^{\abs{C}}$ via~\eqref{eq:dcraig-nocraig},
		\[
		\smash{\overline{\DCraig}}_{C} \ \ =\ \ \frac{\NoCraig_{C}}{|\SetClash_{C}|}\ \ \le\ \ \frac{\NoCraig_{C}}{2^{\abs{C}}}\ \ \le\ \ 2^{\,f_{L}(|\sigma|)-2^{|\rho|}},
		\]
		as required. By the  cell reduction principle for Craig interpolation (Theorem~\ref{thm:craig-factor}),
		$
		\DCraig_{n,m}
		=
		\prod_{C}
		\DCraig_C
		$.
		Since $1-\prod_C x_C\le\sum_C(1-x_C)$
		for $x_C\in[0,1]$, we obtain
		\[
		\smash{\overline{\DCraig}}_{n,m}
			\ \ \le \ \ 
			\sum_{C \in
			\CellAt{}{\sig_n}{\priv_m}}\hspace*{-1em}\smash{\overline{\DCraig}}_C	\ \ \le \ \ 
			|\Atom{}{\sig_n}|\,
			2^{f_L(n)-2^m}.
		\]
		For fixed $n$, the right-hand side tends
		to $0$ as $m\to\infty$.
	\end{proof}
The canonical profile can be hard to compute or work with. We now introduce another profile with nodes that are directly obtained from atoms. Recall that for a locally tabular quasi-normal modal logic we can identify atoms $\mathcal{M}\in \Atom{}{\sigma\cup\rho}$ with a pointed extensional~$(\sigma\cup \rho)$-model $\mathfrak{S}_{\mathcal{M}}$ (Lemma~\ref{lem:standardmodel}). 
\begin{defi}\label{def:shape}
	The \emph{$\sig$-shadow} of an atom $\mathcal M\in \Atom{}{\sigma\cup\rho}$ is
	defined as $\sh(\mathcal M):=(\restrict{\mathfrak{S}_{\mathcal M}}{\sig},w_{\mathcal{M}})$, the pointed $\sig$-model with the same underlying pointed frame as $\mathfrak{S}_{\mathcal M}$ and with the valuation restricted to $\sig$.
\end{defi}
Note that $\sh(\mathcal M)$ need not be extensional:
forgetting $\rho$ may make distinct worlds indistinguishable
in~$\ML_\sigma$. We retain the underlying frame of the
standard $(\sigma\cup\rho)$-model and do not take a further
quotient.

\begin{defi}\label{def:cover}
	An $L$-\emph{cover} of two pointed
	$\sig$-models $(\M_1,w_1)$ and $(\M_2,w_2)$
	is a pointed $\sig$-model $(\M,w)$
	based on a frame for $L$, together with
	$\sig$-p-morphisms
	\[
	f_i\colon (\M,w)\to (\M_i,w_i),
	\qquad i=1,2.
	\]
\end{defi}

\begin{thm}\label{thm:cover} Let $L$ be a locally tabular quasi-normal modal logic. Let $C$ be a $\sigma^{\uparrow\rho}$-cell. Then $\mathcal M_{1},\mathcal M_{2}\in C$ are compatible in $H_{C}$ if and only if $(\sh(\mathcal M_{1}),\sh(\mathcal M_{2}))$ has an $L$-cover.
The cover may be chosen finite.
\end{thm}

\begin{proof}
	\noindent$(\Rightarrow)$
	Suppose that $\mathcal M_1$ and $\mathcal M_2$
	are compatible. Then
	$\mathcal M_1\cap\mathcal M_2'\ne\emptyset$.
	Choose a pointed model in this intersection, and let
	$(\M,w)$ be the finite standard model of its
	$(\sig\cup\priv\cup\priv')$-atom.
	By Lemma~\ref{lem:standardmodel}, there are
	pointed p-morphisms
	$
	f_1\colon (\M,w)\to
	(\mathfrak S_{\mathcal M_1},w_{\mathcal M_1})
	$
	over $\sig\cup\priv$, and
	$
	f_2\colon (\M,w)\to
	(\mathfrak S_{\mathcal M_2'},w_{\mathcal M_2'})
	$
	over $\sig\cup\priv'$.
	Restricting the valuations to $\sig$ gives
	$\sig$-p-morphisms $f_i\colon (\restrict{\M}{\sig},w)
	\to \sh(\mathcal M_i)$,
for $i=1,2$,
	where the second target is identified with
	$\sh(\mathcal M_2)$ via the renaming of $\priv$
	to $\priv'$. Thus
	$(\restrict{\M}{\sig},w)$ is a finite $L$-cover
	of the two shadows.
	
	\medskip
	\noindent$(\Leftarrow)$
	Suppose that the shadows have an $L$-cover
	$(\M,w)$, where $\M=(W,R,V)$, witnessed by
	$\sig$-p-morphisms
	$
	f_i\colon (\M,w)\to
	\sh(\mathcal M_i)$, $i=1,2$.
	Let $V_i$ be the valuation of
	$\mathfrak S_{\mathcal M_i}$.
	Define $\M^+=(W,R,V^+)$ by
	\[
	V^+(p)=
	\begin{cases}
		f_1^{-1}(V_1(p)),
		& p\in\priv,\\
		f_2^{-1}(V_2(q)),
		& p=q'\in\priv',\ q\in\priv,\\
		V(p),
		& p\notin\priv\cup\priv'.
	\end{cases}
	\]
	Since $\priv$ and $\priv'$ are disjoint and
	neither meets $\sig$, this defines a valuation
	extending $V$ on $\sig$.
	By construction, $f_1$ preserves the valuations
	of all variables in $\sig\cup\priv$, and $f_2$
	preserves those in $\sig\cup\priv'$ after
	renaming $\priv$ to~$\priv'$ in its target.
	Their underlying maps already satisfy the
	back and forth conditions. Hence they induce
	pointed p-morphisms
$f_1\colon
		(\restrict{\M^+}{\sig\cup\priv},w)
		\to
		(\mathfrak S_{\mathcal M_1},w_{\mathcal M_1})$ and
$f_2\colon
		(\restrict{\M^+}{\sig\cup\priv'},w)
		\to
		(\mathfrak S_{\mathcal M_2'},w_{\mathcal M_2'})$.
	By Lemma~\ref{lem:standardmodel}, it follows that
	$
	(\M^+,w)\in
	\mathcal M_1\cap\mathcal M_2'$.
	Thus, $\mathcal M_1$ and $\mathcal M_2$
	are compatible.
	
	Finally, the cover in the first implication is
	finite, and the converse construction does not
	require finiteness. This proves the last assertion.
\end{proof}
\begin{exa}
Consider again the 3-fork atoms defined by
$\varphi$ and $\psi$ in
Example~\ref{exm:formulas}.
Their $\{p\}$-shadows have three successors,
with two satisfying $p$ in the first shadow
and two satisfying $\neg p$ in the second.
A common p-morphic predecessor based on a
3-fork would need at least two successors of
each colour, which is impossible.
Thus, the shadows have no cover, as follows
from Theorem~\ref{thm:cover}.
\end{exa}

The notion of an $L$-cover is closely related
to Robinson's joint consistency property~\cite[Theorem~2.2.23]{modeltheory}.
Atoms in the same $\sigma^{\uparrow\rho}$-cell
have the same complete $\sigma$-theory, and
their compatibility means that, after
renaming the forgotten variables on one side,
the two theories have a common realisation.
Theorem~\ref{thm:cover} characterises this
joint realisability in terms of a common
p-morphic predecessor of their $\sigma$-shadows.
Thus, covers provide a frame-theoretic version
of a local Robinson consistency condition.

Define the \emph{shadow incompatibility graph}
of $C$ by
\[
\sh(H_C)
\eqdef
\bigl(\sh(C),
\frown^\sh\bigr),
\]
where $\sh(C) =  \{\sh(\mathcal M)\mid \mathcal M\in C\}$ and two shadows are related by
$\frown^\sh$ if they have no $L$-cover.
\begin{prop}\label{prop:shadow1}
	Let $C$ be a $\sigma^{\uparrow\rho}$-cell. Then $\shpi\colon H_{C}\to \sh(H_{C})$ is a profile map for $C$.
\end{prop}
As the shadow incompatibility graph lies between the cell incompatibility graph $H_C$ and the canonical profile $H_C/_{\sim_C}$, we have a strong graph homomorphism $\kappa\colon \sh(H_{C}) \rightarrow H_C/_{\sim_C}$ such that the canonical profile map $\pi_{C}$ is $\kappa\circ\shpi$.

We apply the shadow profile map $\shpi$ to show that all tabular modal logics have the uniform BICIP.
If $L$ is tabular, then $L=\mathrm{Log}(\mathcal{K})$ for the (up to isomorphism) finite class $\mathcal{K}$ of finite
rooted Kripke frames validating $L$. Assume
\[
\mathcal{K}=\set{(\F_{1},r_{1}),\dots,(\F_{k},r_{k})},\qquad \F_{i}=(W_{i},R_{i}),\ r_{i}\in W_{i}.
\]
\begin{thm}\label{thm:shapebound}
	Tabular quasi-normal modal logics have the uniform BICIP\textup{:} for every $\sig^{\uparrow\rho}$-cell $C$,
	\[
	\mathrm{inc}(C)\ \le\ F_{L}\ \eqdef\ \sum_{i=1}^{k} N_L^{N_i}
	\qquad N_{i}=\abs{W_{i}},\ \ N_{L}=\max_{i}N_{i}.
	\]
\end{thm}
\begin{proof}
	Assume $C$ is a $\sig^{\uparrow\rho}$-cell. Because the canonical profile is the coarsest profile, it is sufficient to show that $|\sh(C)|\leq F_{L}$. Recall that the shadow $\sh(\mathcal{M})$ of $\mathcal{M}\in C$ is obtained by taking the restriction $(\restrict{\mathfrak{S}_{\mathcal M}}{\sig},w_{\mathcal{M}})$ to $\sigma$ of the standard model $(\mathfrak{S}_{\mathcal M},w_{\mathcal{M}})$. Consider the $\sigma$-atom $\mathcal{A}=\bigcup C$ and note that its $\sigma$-shadow coincides with its standard model. By Lemma~\ref{lem:standardmodel}, we have a $\sigma$-p-morphism from $\sh(\mathcal{M})$ onto $\sh(\mathcal{A})$. Observe that once
	the underlying mapping of a p-morphism is fixed, the valuation 
	of $\sh(\mathcal{M})$ is uniquely determined by the valuation of $\sh(\mathcal{A})$. Because any $\sh(\mathcal{M})$ is based on some $(\F_{i},r_{i})\in \mathcal{K}$ and $\sh(\mathcal{A})$ is based on a frame in~$\mathcal{K}$, the number of different valuations $\sh(\mathcal{M})$ could have is bounded by the number $N_{L}^{N_{i}}$ of mappings from the underlying frame $(\F_{i},r_{i})$ of $\sh(\mathcal{M})$ to the frame underlying $\sh(\mathcal{A})$. The inequality follows.
\end{proof}
Theorems~\ref{thm:bounded-profile} and~\ref{thm:shapebound} imply the following result for tabular logics.
\begin{thm}\label{thm:craig-mtoinfty}
	For every tabular quasi-normal modal logic $L$ and every fixed $n\geq 0$, $\lim\limits_{m\to\infty}\DCraig_{n,m}= 1$.
\end{thm}

\section{Density after Existential Quantification}

Interpolants are independent of the names of forgotten
or non-shared variables and are invariant under natural
transformations, such as uniformly replacing any such
variable by its negation. In this section, we abstract
away from variable names and some of these transformations
by identifying input formulas whose existential
quantifications over all forgotten or non-shared variables
are logically equivalent over a given class of frames.

Throughout this section, we fix a class $\mathcal K$ of
pointed frames and assume that $L=\textup{Log}(\mathcal K)$.
Let $\varphi\in\ML_\tau$ and $\rho\subseteq\tau$.
The formula $\exists\vec{\rho}.\varphi$ is interpreted
in a model $\M=(\F,V)$ by setting
$\M,w\models\exists\vec{\rho}.\varphi$ if and only if
there exists a valuation $V'$ coinciding with $V$
on all variables outside $\rho$ such that
$\M',w\models\varphi$, where $\M'=(\F,V')$.
We write
$\exists\vec{\rho}.\varphi\equiv
\exists\vec{\rho}.\psi
$
if the two formulas are true at exactly the same
pointed models based on members of $\mathcal K$,
and denote their equivalence class by
$[\exists\vec{\rho}.\varphi]$.
Unlike ordinary modal equivalence, equivalence after
existential quantification may depend on the chosen
frame class since different classes of frames determining
the same logic can induce different equivalence
relations on existentially quantified formulas (see Example~\ref{exm:counter} below).
Thus, the counting measure considered in this section
depends not only on $L$ but also on $\mathcal K$.

Fix disjoint signatures $\sigma$ and $\rho$. 
The uniform $\sigma$-interpolants of a formula $\varphi\in\ML_{\sigma\cup\rho}$ depend only on the equivalence class $[\exists\vec{\rho}.\varphi]$. More precisely, if $\chi$ is a uniform $\sigma$-interpolant of $\varphi$ and $\exists\vec{\rho}.\psi\equiv\exists\vec{\rho}.\varphi$, then $\chi$ is also a uniform $\sigma$-interpolant of $\psi$.
Similarly, the consistency and Craig interpolants of pairs $\varphi\in\ML_{\sigma\cup\rho}$ and $\psi\in\ML_{\sigma\cup\rho'}$ depend only on the equivalence classes $[\exists\vec{\rho}.\varphi]$ and $[\exists\smash{\vec{\rho}}^{\,\prime\!}\!.\psi]$, respectively.
This allows us to define an alternative notion of interpolant density, obtained by counting existentially quantified formulas of the form $\exists\vec{\rho}.\varphi$ up to logical equivalence.
Fix a modal depth $d$. Let 
\[
\mathcal{Q}^{\smash{d}}_{\sigma,\rho} = \{[\exists \vec{\rho}.\varphi]\mid \varphi \in \ML^{d}_{\sigma\cup \rho}\},\qquad
\mathcal U^{\Pi,d}_{\sigma,\rho}\eqdef\bigl\{[\exists \vec{\rho}.\varphi]\in \mathcal{Q}^{\smash{d}}_{\sigma,\rho}\mid \varphi\text{ has a uniform }\sig\text{-interpolant}\bigr\}.
\]	
and define the \emph{density of uniform interpolation after existential quantification} as
\[
\dens_{\sigma,\rho}^{\Pi,d}\eqdef\frac{|{\mathcal U}_{\sigma,\rho}^{\Pi,d}|}{|\mathcal{Q}^{\smash{d}}_{\sigma,\rho}|} .
\]

Existentially quantified formulas are closed under neither  negation nor conjunction.
Normal forms are therefore less transparent than in $\ML$, and counting becomes correspondingly more involved.
In particular, we can no longer identify equivalence classes of existentially quantified formulas with all subsets of a set of atoms, as in the unquantified case.
Nevertheless, we use a representation based on the atoms introduced earlier.

Every $\varphi\in \ML^{d}_{\sigma\cup\rho}$ is equivalent to a unique $\MForm{d}{\sigma\cup\rho}{D}$ with $D\subseteq \Atom{d}{\sigma\cup \rho}$.
By distribution of existential quantification over disjunction, $\exists \vec{\rho}.\varphi$ is equivalent to a formula of the form
\[
\bigvee_{\mathcal{M}\in D}\exists\vec{\rho}.\MForm{d}{\sigma\cup\rho}{\mathcal{M}} \ \ \equiv \ \
\exists\vec{\rho}.\MForm{d}{\sigma\cup\rho}{D}
\]
with $D\subseteq \Atom{d}{\sigma\cup \rho}$.
We thus work with the equivalence classes $[\exists \vec{\rho}.\MForm{d}{\sigma\cup\rho}{D}]$ for $D\subseteq \Atom{d}{\sigma\cup \rho}$.
These classes form a join-semilattice with bottom element and join defined by
\begin{align*}
	\bot& \ \ := \ \ [\exists\vec{\rho}.\MForm{d}{\sigma\cup\rho}{\emptyset}],
	\\
	[\exists\vec{\rho}.\MForm{d}{\sigma\cup\rho}{D}]\vee
	[\exists\vec{\rho}.\MForm{d}{\sigma\cup\rho}{E}]
	&\ \ :=\ \ [\exists\vec{\rho}.\MForm{d}{\sigma\cup\rho}{D\cup E}].
\end{align*}

\begin{exa}[pinned atoms and existential quantification]
	Recall the set $M_C$ of pinned atoms from
	Lemma~\ref{lem:typefan}. For every $\rho$-colour $r$,
	Lemma~\ref{lem:constants:atomsets} gives
	\[
	\MForm{d}{\sigma\cup\rho}{\pin{r}{\mathcal M}}
	\ \ \equiv \ \ 
	\Gamma(C)\land
	\bigwedge_{q\in\rho}
	\Box^{\le d}(q\leftrightarrow r(q)),
	\qquad \textstyle\mathcal M=\bigcup C.
	\]
	Since $\Gamma(C)$ is $\rho$-free and the variables
	in $\rho$ can be interpreted as true or false according
	to $r$ on every frame, we obtain
	$
	\exists\vec{\rho}.
	\MForm{d}{\sigma\cup\rho}{\pin{r}{\mathcal M}}
	\equiv \Gamma(C)$.
	In fact, by distribution of existential quantification
	over disjunction, $
	\exists\vec{\rho}.
	\MForm{d}{\sigma\cup\rho}{D}
	\equiv \Gamma(C)$, for every non-empty $D\subseteq M_C$.
	Thus, under ordinary counting, the $2^{|\rho|}$
	distinct atoms in $M_C$ give rise to
	$\smash{2^{2^{|\rho|}}}-1$ pairwise non-equivalent formulas
	$\Phi(D)$ with $\emptyset\ne D\subseteq M_C$.
	After existential quantification, however, all these
	formulas belong to a single equivalence class.
	This illustrates why 
	Theorem~\ref{thm:fsi} does not extend to counting
	after existential quantification. Its proof exploits
	the independent choices among the atoms in $M_C$,
	which collapse to a single equivalence class after
	quantification. 
\end{exa}

Unlike in the unquantified case, we can now have
$\exists\vec{\rho}.\MForm{d}{\sigma\cup\rho}{\mathcal{M}_{1}}\models \exists\vec{\rho}.\MForm{d}{\sigma\cup\rho}{\mathcal{M}_{2}}$,
even for distinct $\mathcal{M}_{1},\mathcal{M}_{2}$, and such an entailment might only hold in one direction.
Observe, however, that, for any two different cells $C_{1},C_{2}\in \CellAt{d}{\sig}{\priv}$ and non-empty $D_{i}\subseteq C_{i}$, $i=1,2$, we still have
\[
\exists\vec{\rho}.\MForm{d}{\sigma\cup\rho}{D_{1}}\models \neg \exists\vec{\rho}.\MForm{d}{\sigma\cup\rho}{D_{2}}.
\]
This allows us to extend the cell factorisation to the counting of equivalence classes after existential quantification.
For any cell $C\in\CellAt{d}{\sigma}{\rho}$, let
\[
\mathcal{Q}_{C}= \{[\exists \vec{\rho}.\MForm{d}{\sigma\cup\rho}{D}]\mid D \subseteq C\}
,\quad
\mathcal U_{C}^{\Pi}\eqdef\bigl\{[\exists \vec{\rho}.\varphi]\in \mathcal{Q}_{C}\mid \varphi\text{ has a uniform }\sig\text{-interpolant}\bigr\}.
\]
The \emph{density of uniform interpolation in $C$ after existential quantification} is defined by
\[
\dens_{C}^{\Pi}\eqdef\frac{|{\mathcal U}_{C}^{\Pi}|}{|\mathcal{Q}_{C}|}.
\]
\begin{thm}[projective cell reduction principle]\label{thm:slicereduction5}
	$\displaystyle\dens_{\sigma,\rho}^{\Pi,d}\ =\ \prod_{C\in \CellAt{d}{\sig}{\priv}}\dens_{C}^{\Pi}$.
\end{thm}
\begin{proof}
	The cells partition
	$\Atom{d}{\sigma\cup\rho}$ and so consider the map
	\[
	[\exists\vec{\rho}.
	\MForm{d}{\sigma\cup\rho}{D}]
	\longmapsto
	\bigl(
	[\exists\vec{\rho}.
	\MForm{d}{\sigma\cup\rho}{D\cap C}]
	\bigr)_{C\in\CellAt{d}{\sig}{\priv}}.
	\]
	This map is well-defined, since
	\[
	\exists\vec{\rho}.\MForm{d}{\sigma\cup\rho}{D\cap C}
	\ \ \equiv \ \
	\Gamma(C)\land\exists\vec{\rho}.\MForm{d}{\sigma\cup\rho}{D}.
	\]
	It is also bijective: existentially quantified formulas supported on distinct cells are mutually inconsistent, and
	every element of $\mathcal Q_{\sigma,\rho}^{d}$ is
	the join of its cell components.
	By
	Lemma~\ref{lem:slicedecomp}, under this bijection, $[\exists \vec{\rho}.\MForm{d}{\sigma\cup\rho}{D}]\in\mathcal U_{\sigma,\rho}^{\Pi,d}$ iff
	$[\exists \vec{\rho}.\MForm{d}{\sigma\cup\rho}{C\cap D}]\in\mathcal U^{\Pi}_{C}$ for every cell $C\in \CellAt{d}{\sig}{\priv}$. Hence
	\[
	|\mathcal U_{\sigma,\rho}^{\Pi,d}|=\prod_{C\in\CellAt{d}{\sigma}{\rho}}|\mathcal U_{C}^{\Pi}|,
	\qquad
	\abs{\mathcal{Q}^{\smash{d}}_{\sigma,\rho}}=\prod_{C\in\CellAt{d}{\sigma}{\rho}} \abs{\mathcal{Q}_{C}},
	\]
	and the equation follows.
\end{proof}

The definition of Craig interpolation density after existential quantification is analogous to the uniform case, so we provide only a sketch. Fix again disjoint signatures $\sigma$ and $\rho$.
Let $\SetClash_{\sigma,\rho}^{\Pi,d}$ be the set of ordered pairs of the form
\[
\bigl([\exists \vec{\rho}.\MForm{d}{\sigma\cup\rho}{D_{1}}],
[\exists \vec{\rho}.\MForm{d}{\sigma\cup\rho}{D_{2}}]\bigr)
\]
with $D_{1},D_{2}\subseteq \Atom{d}{\sigma\cup\rho}$, such that
$\MForm{d}{\sigma\cup\rho}{D_{1}}$ and
$\MForm{d}{\sigma\cup\rho'}{D_{2}'}$
are inconsistent. As before, here $D_2'$ denotes the image of $D_2$ under the fixed renaming of $\rho$ to $\rho'$.
Let $\SetCraig_{\sigma,\rho}^{\smash{\Pi,d}}$ be the set of these pairs admitting a Craig $\sigma$-interpolant.
The corresponding cellwise sets $\SetClash_C^{\smash{\Pi}}$ and $\SetCraig_C^{\smash{\Pi}}$ are defined analogously for each cell
$C\in \CellAt{d}{\sig}{\priv}$.
As in the case of uniform interpolants, these definitions do not depend on the chosen representatives $D_1,D_2$.
One can show:

\begin{thm}[projective cell reduction principle for Craig interpolation]\label{thm:craig-factor5}
	\[
	|\SetClash_{\sigma,\rho}^{\Pi,d}|
	=
	\prod_{C\in \CellAt{d}{\sigma}{\rho}}
	|\SetClash^{\Pi}_{C}|,
	\qquad
	|\SetCraig_{\sigma,\rho}^{\Pi,d}|
	=
	\prod_{C\in\CellAt{d}{\sigma}{\rho}}
	|\SetCraig^{\Pi}_{C}|.
	\]
\end{thm}

The \emph{density of Craig interpolation after existential quantification} is defined by
\[
\DCraig_{\sigma,\rho}^{\Pi,d}
\eqdef
\frac{|\SetCraig_{\sigma,\rho}^{\Pi,d}|}
{|\SetClash^{\Pi,d}_{\sigma,\rho}|},
\qquad
\DCraig_C^\Pi
\eqdef
\frac{|\SetCraig_C^\Pi|}{|\SetClash_C^\Pi|}.
\]
By Theorem~\ref{thm:craig-factor5}, we obtain
\[
\DCraig_{\sigma,\rho}^{\Pi,d}
=
\prod_{C\in\CellAt{d}{\sigma}{\rho}}
\DCraig_C^\Pi.
\]

\subsection{Asymptotics for $m$ fixed and $n\rightarrow \infty$}

We first fix $\rho$ and let $\sigma$ grow. The analysis is essentially the same as in the ordinary, unquantified case, so we keep the exposition brief. Indeed, the modified counting method concerns the forgotten signature $\rho$, which remains fixed in this regime.

As before, let $\sigma_n=\{p_1,\ldots,p_n\}$ and $\rho_m=\{q_1,\ldots,q_m\}$ be disjoint signatures, and use numerical parameters $n,m$ in place of the signatures. For example, $\dens_{n,m}^{\Pi,d}$ denotes the density of formulas in $\ML^{\smash{d}}_{\sigma_n\cup\rho_m}$, counted up to equivalence after existential quantification over $\rho_m$, that admit uniform $\sigma_n$-interpolants.

The lifting and flipping constructions from the proofs of Theorems~\ref{thm:ntoinfty} and~\ref{thm:ntoinfty3} yield the same asymptotic behaviour in the quantified case. As before, we write $(C_m)_{m\ge m^\ast}$ for the trail of a cell $C_{m^\ast}$ containing a counterexample to interpolation.

\begin{thm}\label{thm:ntoinfty5}
	Let $L=\textup{Log}(\mathcal{K})$ and fix $d\geq 0$.
	
	(1) If $\Phi(D)$, for $D\subseteq C_{m^\ast}\in\CellAt{d}{\sig_{n^*}}{\priv_{m^\ast}}$, has no uniform $\sigma_{n^\ast}$-interpolant, 
	then, for every fixed $m\ge m^\ast$ and all $n\ge n^\ast$,
	\[
	\dens^{\Pi,d}_{n,m}
	\ \ \leq \ \ 
	\bigl(1-2^{-\abs{C_m}}\bigr)^{2^n}
	\xrightarrow[\ n\to\infty\ ]{}0.
	\]
	
	(2) If there is an inconsistent pair, for $D_1,D_2\subseteq C_{m^\ast}\in\CellAt{d}{\sig_{n^*}}{\priv_{m^\ast}}$, without a Craig $\sigma_{n^\ast}$-interpolant, 
	then, for every fixed $m\ge m^\ast$ and all $n\ge n^\ast$,
	\[
	\DCraig_{n,m}^{\Pi,d}
	\ \ \leq \ \
	\bigl(1-2^{-2\abs{C_m}}\bigr)^{2^n}
	\xrightarrow[\ n\to\infty\ ]{}0.
	\]
\end{thm}

\begin{proof}
	The proof follows the same argument as in Theorems~\ref{thm:ntoinfty} and~\ref{thm:ntoinfty3}. It suffices to observe that, for any $\sigma^{\uparrow\rho}$-cell $C$, the following hold:
	\begin{enumerate}[label=\textup{(\roman*)}]
		\item if some $\MForm{d}{\sigma\cup\rho}{D}$ with $D\subseteq C$ has no uniform $\sig$-interpolant, then
		$
		\smash{\overline{\dens}}_C^\Pi
		\geq \frac{1}{|\mathcal Q_C|}
		\geq \smash{2^{-|C|}}$;
		\item if $\SetClash_C\setminus\SetCraig_C\neq\emptyset$, then
		$
		\smash{\overline{\DCraig}}_C^\Pi\
		\geq \ \frac{1}{\abs{\mathcal Q_C}^2}
		\ \geq \ 2^{-2\abs C}$.
	\end{enumerate}
	Indeed, $\abs{\mathcal Q_C}\leq 2^{\abs C}$, and each hypothesis guarantees at least one failure among the corresponding equivalence classes.
	Applying the same lifting and flipping arguments as in Theorems~\ref{thm:ntoinfty} and~\ref{thm:ntoinfty3}, together with the projective cell reduction principles, yields the claimed bounds.
\end{proof}

\subsection{Asymptotics for $n$ fixed and $m\rightarrow \infty$}\label{subsection:asymqunat}

We now fix $\sigma$ and let $\rho$ grow. In this regime, the asymptotic behaviour of the density depends on the modal logic and can differ significantly from that of the ordinary density. As in the ordinary counting case, we focus on locally tabular logics and therefore drop the modal depth restriction. For every finite signature $\tau$, the set $\ML_\tau$ is then finite up to logical equivalence. The definitions and results from the bounded-depth setting carry over accordingly. For instance, for disjoint signatures $\sigma$ and $\rho$, the set
$
\mathcal{Q}_{\sigma,\rho}
=
\{[\exists\vec{\rho}.\varphi]\mid \varphi\in\ML_{\sigma\cup\rho}\}$
consists of the equivalence classes of existentially quantified formulas.

Fix disjoint signatures $\sigma$ and $\rho$, and let $C\in\CellAt{}{\sigma}{\rho}$ be a $\sigma^{\uparrow\rho}$-cell. Let $H_C=(C,\frown)$ be its incompatibility graph and let
$\pi_{C}\colon H_C\to H_{P}\eqdef H_C/{\sim_C}$ be the canonical profile map. Thus $|P|=\mathrm{inc}(C)$.
For $S\subseteq P$, define
\[
\QCS{S}
:=
\set{
	[\exists\vec\priv.\MForm{}{\sigma\cup\rho}{D}]
	\ \middle|\
	D\subseteq\pi_C^{-1}(S)
}.
\]
Note that $\bot=[\exists\vec\rho.\MForm{}{\sigma\cup\rho}{\emptyset}]$ belongs to $\QCS{S}$ for every $S\subseteq P$, including $S=\emptyset$.

As in the non-quantified case (Theorem~\ref{thm:lftab-count}), we start with counting the set $\SetCraig^{\Pi}_{C}$ of pairs $(x,y)\in \mathcal{Q}_{C}\times \mathcal{Q}_{C}$ for which a Craig interpolant exists. As these are exactly the pairs of the form $(x,\bot)$ or $(\bot,y)$ with $x,y\in \mathcal{Q}_{C}$ we obtain: 
\begin{thm}\label{thm:lftab-count1}
	Let $L=\textup{Log}(\mathcal{K})$ be a locally tabular quasi-normal logic. Then, for every $\sig^{\uparrow\rho}$-cell $C$,
	\begin{equation*}
		|\SetCraig_C^\Pi|  \ \ = \ \ 2\abs{\mathcal{Q}_{C}}-1.
	\end{equation*}
\end{thm}
The next observation allows us to bound the number of disjunctions built from formulas in $\QCS{S}$ and $\QCS{T}$ if $(S,T)$ is an ordered biclique in $H_{P}$.
\begin{lem}
	\label{lem:injective-join}
	If $(S,T)$ is an ordered biclique of $H_P$, then
	\[
		J_{S,T}\colon \QCS{S}\times\QCS{T}\longrightarrow\QC,
		\qquad
		J_{S,T}(x,y)=x\vee y,
	\]
	is injective.  Consequently, $|\QCS{S}| \cdot |\QCS{T}|\le |\mathcal{Q}_{C}|$.	
\end{lem}
\begin{proof}
		Let $x,x'\in\QCS{S}$ and
		$y,y'\in\QCS{T}$, and suppose that
		$x\vee y=x'\vee y'$.
		Since $(S,T)$ is an ordered biclique,
		every element of $\QCS{S}$ is inconsistent
		with every element of $\QCS{T}$.
		It follows that $x=x'$ and $y=y'$ and so $J_{S,T}$ is injective.
\end{proof}
We now assume that $L$ has BICIP. So assume that $\mathrm{inc}(C) = \abs{P}\leq f_{L}(n)$ for all $\sigma^{\uparrow\rho}$-cells, with $\abs{\sigma}=n$, independent of $m$. We use that then the number of non-empty ordered pairs of disjoint subsets
of $P$ is bounded by 
\begin{equation}\label{pp:eq:kappa}
	\kappa_{L}(n):=3^{f_{L}(n)}-2^{f_{L}(n)+1}+1.
\end{equation}
\begin{thm}\label{pp:thm:craig-floor}
	If $L=\textup{Log}(\mathcal{K})$ has BICIP, then 
	$\liminf\limits_{m\to\infty}\DCraig_{n,m}^\Pi>0$, for every $n\geq 0$.
\end{thm}
\begin{proof}
	Let $\mathcal{I}_{C}$ be the set of inconsistent pairs in
	$(\QC\setminus\{\bot\})\times(\QC\setminus\{\bot\})$. Then
	\[
	\NoCraig_{C}^{\Pi} \ \ =\ \ |\SetClash_{C}^{\Pi}|-|\SetCraig_{C}^{\Pi}|\ \ =\ \ \abs{\mathcal{I}_{C}}.
	\]
    We bound $\NoCraig_C^\Pi$ from above. Choose for $(x,y)\in\mathcal I_C$ any sets
	$D,E\subseteq C$ with $x=\exists\vec{\rho}.\MForm{}{\sigma\cup\rho}{D}$ and $y=\exists\vec{\rho}.\MForm{}{\sigma\cup\rho}{E}$. Then the ordered pair of images $(\pi_C[D],\pi_C[E])$ is an ordered biclique of non-empty sets in $H_P$ and $(x,y)\in \QCS{\pi_C[D]}\times \QCS{\pi_C[E]}$. We have shown 
	\[
		\mathcal I_C \ \
		\subseteq
		\bigcup_{\substack{\emptyset\ne S,\ T\subseteq\PC\\S\frown_PT}}
		\QCS{S}\times\QCS{T}.
	\]
	There are at most $\kappa_{L}(n)$ such ordered bicliques in $H_{P}$ and for each pair $(S,T)$ in this union, Lemma~\ref{lem:injective-join} gives
	$|\QCS{S}|\cdot |\QCS{T}|\le |\mathcal{Q}_{C}|$. Hence we obtain the upper bound
	\[
	\NoCraig_C^\Pi
	\ \ \le \ \ 
	\sum_{\substack{\emptyset\ne S,\ T\subseteq\PC\\S\frown_PT}}
	|\QCS{S}|\cdot |\QCS{T}|
	\ \ \le\ \ \kappa_{L}(n)|\mathcal{Q}_{C}|.
	\]
	Using Theorem~\ref{thm:lftab-count1}, we obtain
	\[
		\DCraig_C^\Pi
		\ \ = \ \ \frac{2|\mathcal Q_C|-1}{2|\mathcal Q_C|-1+\NoCraig_C^\Pi} \ \ \geq \ \ \frac{2|\mathcal Q_C|-1}{2|\mathcal Q_C|-1+\kappa_{L}(n) |\mathcal Q_C|}
		\ \ \geq\ \ \frac{1}{1+\kappa_{L}(n)}.
	\]
	Using projective cell reduction principle for Craig interpolation (Theorem~\ref{thm:craig-factor5}) and the fact that there are $|\Atom{}{\sig_n}|$ factors, independently of $m$, it follows that
	\[
		\DCraig_{n,m}^\Pi
		\ \ \ge \ \
		\left(\frac1{1+\kappa_{L}(n)}\right)^{|\Atom{}{\sig_n}|}.
	\]
    Hence we have obtained a lower bound for $\DCraig_{n,m}^\Pi$ that does not depend on $m$.
\end{proof}
Fix a cell $C_0\in \CellAt{}{\sig}{\emptyset}$ and consider its trail $(C_{m})_{m\geq 0}$. For $m\le\ell$, let $\iota_{m,\ell}\colon C_m\to C_\ell$
be the pinning operation on atoms that interprets in every model of every $\mathcal{M}\in C_{m}$ every variable in $\priv_\ell\setminus\priv_m$ as false.
	Then
	\begin{equation}\label{eq:constant-compatibility}
		\mathcal M\frown\mathcal N
		\quad\Longleftrightarrow\quad
		\iota_{m,\ell}(\mathcal M)
		\frown\iota_{m,\ell}(\mathcal N).
	\end{equation}
Indeed, a joint model on the left defines a joint model on the right by interpreting the variables in $\priv_\ell\setminus\priv_m$ as false on both sides. A joint model on the right
is also a joint model on the left. Let $P(m)$ be the canonical profile of $C_{m}$ and $\pi_{m}$ the canonical profile map for~$C_m$, for all $m\geq 0$.

\begin{lem}\label{lem:domination}
	Assume that $L=\textup{Log}(\mathcal{K})$ has BICIP witnessed by $f_{L}(n)$. Then there is $m_0 >0$ such that, for every $m\ge m_0$ and $D\subseteq C_m$,
	\begin{equation}\label{eq:domination-criterion}
		\Phi(D)\text{ has a uniform $\sig$-interpolant}
		\quad\Longleftrightarrow\quad
		D=\emptyset
		\ \text{ or }
		N_{P(m)}(\pi_{m}[D])=\emptyset.
	\end{equation}
\end{lem}

\begin{proof}
 For $m\le\ell$, choose $\mathcal A_p\in C_m$ with
 $\pi_m(\mathcal A_p)=p$, for each $p\in P(m)$.
 By \eqref{eq:constant-compatibility}, the map
 $p\mapsto\pi_\ell(\iota_{m,\ell}(\mathcal A_p))$
 is injective. Hence $|P(m)|$ is nondecreasing and,
 by the hypothesis, bounded by $f_L(|\sig|)$.
 Fix $m_0>0$ from which the $|P(m)|$ are constant.
 If $m_0\le m\le\ell$, then for every $\mathcal B\in C_\ell$
 there is $p\in P(m)$ with
 $\pi_\ell(\mathcal B)=\pi_\ell(\iota_{m,\ell}(\mathcal A_p))$.
 Thus, for all $\mathcal M\in C_m$,
 \begin{equation}\label{eq:trace-at-level-m}
  \iota_{m,\ell}(\mathcal M)\frown\mathcal B
  \quad\Longleftrightarrow\quad
  \mathcal M\frown\mathcal A_p.
 \end{equation}

 The case $D=\emptyset$ is immediate. Suppose $D\ne\emptyset$.
 Then $\Phi(D)$ has no uniform
 $\sig$-interpolant if and only if the cell characteristic formula $\Gamma(C_{m})$ is not a uniform $\sig$-interpolant of $\MForm{}{}{D}$ (by Lemma~\ref{lem:localfinitecellcharacteristic}) if and only if there is a
 $\priv_m$-free formula $\chi$ such that
 $\Phi(D)\models\chi$ and $\Gamma(C_{m})\not\models\chi$.
 Suppose such a $\chi$ exists. Renaming its variables
 outside $\sig$ if necessary, choose $\ell\ge m$
 so that $\chi$ uses only variables in
 $\sig\cup(\priv_\ell\setminus\priv_m)$.
 There is
 $(\M,w)\in\PMod_L$ with
 $\M,w\models\Gamma(C_{m})\wedge\neg\chi$.
 Put $\mathcal B=[\M,w]_{\sig\cup\priv_\ell}$;
 then $\mathcal B\in C_\ell$ and
 $\Phi(\mathcal B)\models\neg\chi$.
 For every $\mathcal M\in D$, the implication
 $\Phi(\mathcal M)\models\chi$ holds, also after
 renaming the variables of $\chi$ outside $\sig$.
 Therefore $\iota_{m,\ell}(\mathcal M)\frown\mathcal B$.
 By \eqref{eq:trace-at-level-m}, some $p\in P(m)$
 satisfies $\mathcal M\frown\mathcal A_p$ for every
 $\mathcal M\in D$. Hence
 $N_{P(m)}(\pi_m[D])\ne\emptyset$.

 Conversely, suppose $p\in N_{P(m)}(\pi_m[D])$.
 Let $\priv_m'$ be a fresh copy of $\priv_m$ and
 $\mathcal A_p'$ the corresponding renamed atom,
 using the prime notation from Section~\ref{sec:craig}.
 Since $\mathcal M\frown\mathcal A_p$ for every
 $\mathcal M\in D$, Proposition~\ref{prop:clashcrit} gives
 $
  \Phi(D)\models\neg\Phi(\mathcal A_p')$.
 On the other hand, $\Phi(\mathcal A_p')$ is
 consistent and entails $\Gamma(C_{m})$,
 because $\mathcal A_p\in C_m$. Thus
 $\Gamma(C_{m})\not\models\neg\Phi(\mathcal A_p')$.
 Taking $\chi=\neg\Phi(\mathcal A_p')$ proves that
 $\Phi(D)$ has no uniform $\sig$-interpolant.
\end{proof}

\begin{lem}\label{lem:finite-repair}
	Let $H_{P}=(P,\frown_{P})$ be a canonical profile, let $\emptyset\ne S\subseteq\PC$ and suppose
	$N_{P}(S)\not=\emptyset$. Then there is a non-empty $A\subseteq\PC$ such that $(S,A)$ is an ordered biclique of $H_{P}$ and $N_{P}(S\cup A)=\emptyset$.
\end{lem}
\begin{proof}
	Put $U_0=S$ and assume $U_{i}$ has been defined. If $N_{P}(U_i)\not=\emptyset$, choose
	$a_{i+1}\in N_P(U_i)$ and put $U_{i+1}=U_i\cup\{a_{i+1}\}$.  Irreflexivity implies $a_{i+1}\notin U_i$, for otherwise
	$a_{i+1}\frown_Pa_{i+1}$.  Thus, the cardinality of $U_{i}$ increases strictly at every	stage. Since $\PC$ is finite, the process terminates after at most
	$|\PC|$ steps at a set $U_k$ with $N_{\PC}(U_{k})=\emptyset$. Since $N_{\PC}(S)\not=\emptyset$, $k\ge1$.
	
	Put $A:=U_k\setminus S$.  Then $A$ is non-empty and
	$N_{P}(S\cup A)=\emptyset$ holds.  Every $a_i$ was chosen incompatible with all
	members of $S\subseteq U_{i-1}$, so $(S,A)$ is an ordered biclique of
	$H_P$. 
\end{proof}

\begin{thm}\label{thm:uniform-floor}
	If $L=\textup{Log}(\mathcal{K})$ has BICIP, then $\liminf\limits_{m\to\infty}\dens_{n,m}^\Pi>0$ for every $n\geq 0$.
\end{thm}
\begin{proof}
Fix $n\geq 0$, signature $\sigma$ with $|\sigma|=n$ and a cell $C_0\in \CellAt{}{\sig}{\emptyset}$. Fix $m_{0}$ supplied by
Lemma~\ref{lem:domination}. Consider any element $C=C_{m}$ of the trail $(C_m)_{m \geq 0}$  of $C_0$ for some $m\geq m_{0}$. 
	Let
	\[
	\mathcal G_C:=\set{x\in\QC\mid
		x\text{ has a uniform $\sig$-interpolant}},
	\qquad
	\mathcal B_C:=\QC\setminus\mathcal G_C.
	\]
	Let $\pi_{C}$ be the canonical profile map from $C$ to its canonical profile $P$. For every
	$x\in\mathcal B_C$, choose a non-empty representative $D_x\subseteq C$
	and put $S_x:=\pi_C[D_x]$.  By Lemma~\ref{lem:domination}, $N_{P}(S_x)\not=\emptyset$. Apply Lemma~\ref{lem:finite-repair} and choose non-empty $A_x$
	such that
	$(S_x,A_x)$ is an ordered biclique of $H_P$
	and $N_{P}(S_x\cup A_x)=\emptyset$.
	For every $a\in\PC$, choose $\mathcal M_a\in C$ with
	$\pi_C(\mathcal M_a)=a$, and define
	$
	e_a:=[\exists\vec\priv.\MForm{}{\sigma\cup\rho_{m}}{\mathcal M_a}]\in\QCS{\{a\}}$.
	For non-empty $A\subseteq\PC$, put
		$z_A:=\bigvee_{a\in A}e_a\in\QCS{A}$.
	For each ordered pair $(S,A)$ occurring among the chosen pairs
	$(S_x,A_x)$, define
	\[
	\mathcal B_{S,A}
	\eqdef\set{x\in\mathcal B_C\mid(S_x,A_x)=(S,A)}
	\]
	and
	\[
		F_{S,A}\colon \mathcal B_{S,A}\to \QC,
		\qquad
		F_{S,A}(x):=x\vee z_A.
	\]
	If $D_x$ represents $x$, then $F_{S,A}(x)$ is represented by
	$D_x\cup\set{\mathcal M_a\mid a\in A}$ and $\pi_{C}(D_x\cup\set{\mathcal M_a\mid a\in A})=S\cup A$. $N_{P}(S\cup A)=\emptyset$, so
	Lemma~\ref{lem:domination} gives
	$F_{S,A}(x)\in\mathcal G_C$.
	Moreover, $(S,A)$ is an ordered biclique of $H_P$.
	Lemma~\ref{lem:injective-join}, with the second
	coordinate fixed at $z_A$, shows that $F_{S,A}$ is injective.  Therefore $\abs{\mathcal B_{S,A}}\le\abs{\mathcal G_C}$.
	Every occurring $(S,A)$ is a non-empty ordered biclique and has disjoint
	sides.  There are at most $\kappa_{L}(n)$ such pairs.  Since the 
	$\mathcal B_{S,A}$ partition $\mathcal B_C$, we obtain $
	\abs{\mathcal B_C}\le\kappa_{L}(n)\abs{\mathcal G_C}$.
	It follows that
	\[
	\dens_C^\Pi
	\ \ = \ \
	\frac{\abs{\mathcal G_C}}
	{\abs{\mathcal G_C}+\abs{\mathcal B_C}}
	\ \ \ge\  \ \frac1{1+\kappa_{L}(n)}.
	\]
	
	As there are only $|\Atom{}{\sigma}|$ many cells $\CellAt{}{\sigma}{\emptyset}$, and we can choose as $m_{0}$ the maximum over all $m_{0}$ supplied by Lemma~\ref{lem:domination} for each cell, the projective cell reduction principle (Theorem~\ref{thm:slicereduction5}) implies
	\[
		\dens_{n,m}^\Pi
		\ \ \ge \ \
		\left(\frac1{1+\kappa_{L}(n)}\right)^{\abs{\Atom{}{\sigma}}}\ \ >\ \ 0.\qedhere
	\]
\end{proof}

We close this section by observing that, for every
tabular modal logic $L$ and fixed signature $\sigma$,
the set $\bigcup_{\rho}\mathcal Q_{\sigma,\rho}$
is finite, regardless of the chosen class of pointed frames
$\mathcal K$ with $L=\textup{Log}(\mathcal K)$.
\begin{lem}\label{lem:finite} Let $L=\textup{Log}(\mathcal{K})$ be a tabular quasi-normal modal logic. Then, for every $n\geq 0$ and every $\sigma_{n}^{\uparrow\rho_{m}}$-cell $C$, $\abs{\mathcal{Q}_{C}}\leq 2^{F_{L}}$ and so $\abs{\mathcal{Q}_{\sigma_{n},\rho_{m}}} \leq  2^{F_{L}|\Atom{}{\sigma_n}|}$ for all $m\geq 0$.
\end{lem}
	\begin{proof}
	 Fix a $\sigma_{n}^{\uparrow\rho_{m}}$-cell $C$. Assume first that we define logical equivalence of existentially quantified formulas using the class $\mathcal{K}_{L}$ of all rooted frames validating $L$. Hence $L$ is given in the same way as in Theorem~\ref{thm:shapebound}. Then every $x\in \mathcal{Q}_{C}$ is determined by the set of $\sigma_{n}$-shadows $\sh(C)$ in the sense that any two different $x,y\in \mathcal{Q}_{C}$ have different models in $\sh(C)$. It follows from the proof of Theorem~\ref{thm:shapebound} that $\abs{\mathcal{Q}_{C}}\leq 2^{F_{L}}$ with $F_{L}$ a constant depending only on $L$. The bound on $\mathcal{Q}_{\sigma_{n},\priv_{m}}$ follows directly from the cell reduction principle.
	 
	Now consider an arbitrary class $\mathcal K$ with
	 $L=\textup{Log}(\mathcal K)$. Since
	 $\mathcal K\subseteq\mathcal K_L$, logical equivalence
	 over $\mathcal K_L$ implies logical equivalence
	 over $\mathcal K$. Consequently, the number of
	 equivalence classes in $\mathcal Q_C$, defined
	 using equivalence over $\mathcal K$, is also
	 bounded by $2^{F_L}$.
\end{proof}

The stabilisation claim for tabular logics in Table~\ref{table:beh} follows from Lemma~\ref{lem:finite} and $\mathcal{Q}_{\sig,\rho_{m}} \subseteq \mathcal{Q}_{\sig,\rho_{m+1}}$, for all $m\geq 0$.
	\begin{exa}\label{ex:final}
	We analyse $\dens_{\sigma,\rho}^{\Pi}$ for the 3-fork. Assume $\rho\ne\emptyset$ (otherwise, it is trivial). As we have seen in Example~\ref{ex:33}, only cells of the form $C=\smash{C^{\priv}_{s_0, \{ s_1, s_2\}}}$ contain counterexamples to uniform $\sigma$-interpolation. After existential quantification, we obtain the same five non-logically equivalent formulas in~$\mathcal{Q}_{C}$:
	\begin{align*}
	\varphi_{11} & \ \ = \ \ \exists \vec{\rho}. \Phi(\mathcal{M}_{s_0r_0, \{s_1r_1, s_2r_2\}}), \\ 
	\varphi_{12} & \ \ = \ \ \exists \vec{\rho}.\Phi(\mathcal{M}_{s_0r_0, s_1r', s_2\{r_1, r_2\}}), \\ 
	\varphi_{21} & \ \ = \ \ \exists \vec{\rho}.\Phi(\mathcal{M}_{s_0r_0, s_1\{r_1, r_2\}, s_2r'}) 
	\end{align*}
	as well as $\bot$ and $\varphi_{12}\vee \varphi_{21}$.
	Then $\dens_{C}^{\smash{\Pi}}= 0.6$  since exactly $\varphi_{12}$ and $\varphi_{21}$ lack uniform $\sigma$-interpolants. Thus, we obtain
	\[
	\dens_{n,m}^{\Pi} \ \ = \ \ \left(\frac35\right)^{\frac12(2^n-1)2^{2n}}.
	\]
\end{exa}

\section{Case Study: Bounded-Depth Extensions of Grz}\label{sec:grz}

We consider the bounded-depth extensions $\Grzd$ of $\Grz$ introduced in the preliminaries, where $\Delta\geq 1$. They are all locally tabular and have CIP and uniform interpolation if and only  if $\Delta\leq2$~\cite[Chapter~8]{MGabbay2005-MGAIAD}.
We prove BICIP for every
$\Delta\geq1$. By Theorems~\ref{thm:mtoinfty}
and~\ref{thm:bounded-profile}, it follows that
$\lim\limits_{m\to\infty}\dens_{n,m}=\lim\limits_{m\to\infty}\DCraig_{n,m}=1$ for every $n\geq 0$ and $\Delta\geq 1$.
We also show that $\Grzd$ has uniform BICIP exactly when
$\Delta\leq3$, with $\mathrm{inc}(C)\leq 3$
for $\Delta=3$ and, by CIP, $\mathrm{inc}(C)=1$ for $\Delta\in\{1,2\}$. 

After existential quantification,
Theorems~\ref{thm:uniform-floor}
and~\ref{pp:thm:craig-floor} give positive lower bounds
for $\dens_{n,m}^{\Pi}$ and $\DCraig_{n,m}^{\Pi}$, respectively,
again for every fixed $n\geq 0$ and $\Delta\geq 1$, for any class~$\mathcal{K}$ with $\Grzd=\textup{Log}(\mathcal{K})$.
We further prove in this section that $\lim\limits_{m\to\infty}\dens_{n,m}^{\Pi}=1$ for any class $\mathcal{K}$ with $\Grzd=\textup{Log}(\mathcal{K})$ containing all finite tree-shaped frames of depth bounded by~$\Delta$. Whether $\lim\limits_{m\to\infty}\DCraig_{n,m}^{\Pi}=1$ on such $\mathcal{K}$ 
remains open.

Our main tool in this section is the shadow profile of cells introduced in Section~\ref{sec:craig-minfty} and the characterisation of compatibility of atoms via covers. 

For a finite poset $\F=(W,R)$, put
$R^+=R\setminus \{(w,w) \mid w\in W\}$.
Its \emph{depth} $\text{dpt}(\F)$ is the maximal number
of worlds in an $R^+$-chain; thus a singleton has depth $1$.
Write ${\uparrow}u=\{x\in W\mid uRx\}$.
A world $w$ is a \emph{root} if $W={\uparrow}w$.
As before, distinguished worlds are assumed to be roots.
A finite rooted pointed $\tau$-model on a poset is called
a \emph{$\tau$-poset}.

To establish BICIP, we seek
a bound $\mathrm{inc}(C)=|C/_{\sim_{C}}|\leq f_L(|\sigma|)$ for any $\sigma^{\uparrow\rho}$-cell $C$,
independent of $|\rho|$.
Recall that, for $\Grzd$, the shadows in $\sh(H_C)$ are $\sigma$-posets,
and 
$p\frown^{\sh}p'$ if and only if
$\{p,p'\}$ has no $\Grzd$-cover.
The canonical profile map $\pi_{C}$ for $C$ factors through the strong graph homomorphism $\kappa\colon\sh(H_C)\to H_{C}/_{\sim_{C}}$ so that 
\[
\pi_{C}=\kappa\circ\shpi.
\]
We let $P = C/_{\sim_{C}}$ and say that a shadow $p$ \emph{represents} $q\in P$ iff $\kappa(p)=q$.

A $\sigma$-poset $(\M,w)$ is \emph{tree-shaped} if its
order is the reflexive transitive closure of a finite
rooted tree with root $w$.
Every non-root world then has a unique immediate
predecessor, its \emph{parent}; its immediate successors
are its \emph{children}.
A tree-shaped $\sigma$-poset is \emph{normalised} if no
two distinct children $u_1,u_2$ of a common parent have
isomorphic rooted $\sigma$-submodels, that is, there is
no $\sigma$-isomorphism
$I\colon {\uparrow}u_1\to{\uparrow}u_2$ with $I(u_1)=u_2$.

To count normalised $\sigma$-posets up to isomorphism,
put $N=2^n$, and define
\[
B_1(n)=N,
\qquad
B_{k+1}(n)=N\cdot2^{B_k(n)}.
\]

\begin{thm}\label{thm:normalform}
	For $\Delta\geq1$,
	\begin{enumerate}
		\item the number of normalised $\sigma$-posets of depth
		bounded by $\Delta$ is, up to isomorphism, bounded by
		$B_\Delta(|\sigma|)$\textup{;}
		\item for any signature $\rho$ disjoint from $\sigma$
		and every $\sigma^{\uparrow\rho}$-cell $C$ in $\Grzd$\textup{:}
		for every $p\in\sh(C)$, there is a normalised
		$\sigma$-poset $q$ of depth $\leq\Delta$ such that
		$q\equiv_\sigma p$ and, for every $r\in\sh(C)$,
		the pair $\{p,r\}$ has a $\Grzd$-cover iff the pair
		$\{q,r\}$ has a $\Grzd$-cover.
	\end{enumerate}
\end{thm}
It follows that $\mathrm{inc}(C)$ is bounded by
$B_\Delta(|\sigma|)$ for any $\sigma^{\uparrow\rho}$-cell $C$.
To show this, take for every $p\in\sh(C)$ a normalised
$\normalised{p}$ satisfying Point~2.
If $\kappa(p_1)\ne\kappa(p_2)$, then there is
$r\in\sh(C)$ such that $\{r,p_1\}$ has a cover but
$\{r,p_2\}$ does not, or vice versa.
Hence $\{\normalised{p_1},r\}$ has a cover but $\{\normalised{p_2},r\}$
does not, or vice versa, and therefore
$\normalised{p_1}\not\cong \normalised{p_2}$.
Thus
\[
\mathrm{inc}(C)
\ \ =\ \ |\kappa[\sh(C)]|
\ \ \leq \ \ 
\bigl|\{[\normalised{p}]_{\cong}:p\in\sh(C)\}\bigr|
\ \ \leq \ \ B_\Delta(|\sigma|).
\]
We obtain:
\begin{thm}\label{thm:weaklyboundedGrz}
	For every $\Delta\geq1$, $\Grzd$ has BICIP\textup{:}
	$\mathrm{inc}(C)\leq B_\Delta(|\sigma|)$,  for any $\sigma^{\uparrow\rho}$-cell $C$.
\end{thm}
The proof of Theorem~\ref{thm:normalform} is in two steps. First, we associate with every $\sigma$-poset its tree-shaped unravelling~\cite[Chapter~3]{DBLP:books/daglib/0030819}. Let $(\M,w)$ with $\M=(W,R,V)$ be a $\sig$-poset. A \emph{branch} is a chain $\gamma=(x_{0}R^{+}x_{1}R^{+}\dots R^{+}x_{k})$ starting at the root $w=x_{0}$. Let $\text{end}_{\mathfrak{M}}(\gamma)$ map $\gamma$ to its \emph{endpoint} $x_{k}$. The \emph{unravelling} $U(\M,w)=((W^{U},R^{U},V^{U}),(w))$ is the $\sig$-model whose worlds $W^{U}$ are the branches of $(\M,w)$ and
	\[
	\gamma\,R^{U}\,\gamma' \text{ if $\gamma$ is a prefix of $\gamma'$},
	\qquad
	\gamma \in V^{U}(p) \text{ if $\text{end}_{\mathfrak{M}}(\gamma)\in V(p)$ },
	\]
    and with distinguished world the length-$0$ branch $(w)$. Then $R^{U}$ is a partial order, and since every
	branch begins with $w$, every world is $R^{U}$-reachable from $(w)$, so $U(\M,w)$ is a rooted pointed $\sig$-poset. 
\begin{lem}[unravelling]\label{lem:unravel}
	Let $(\M,w)$ be a $\sigma$-poset.
	Then $\text{dpt}(U(\M,w))=\text{dpt}(\M)$,
	$\text{end}_{\M}$ is a surjective $\sigma$-p-morphism,
	and, for every $\sigma$-poset $(\N,v)$, the pair
	$\{(\M,w),(\N,v)\}$ has a $\Grzd$-cover iff the pair
	$\{U(\M,w),(\N,v)\}$ has a $\Grzd$-cover.
\end{lem}
\begin{proof}
	That $\text{end}_{\M}$ is a surjective p-morphism and $\text{dpt}(U(\M,w))=\text{dpt}(\M)$ is immediate from
	the definition. If $(\Cov,r)$ covers $U(\M,w),(\N,v)$ then composing a witness p-morphism with $\text{end}_{\M}$ shows that it covers
	$(\M,w),(\N,v)$. Conversely, let $(\Cov,r)$ cover $(\M,w),(\N,v)$ via $f,g$. Consider
	$U(\Cov,r)$ with $\text{end}_{\Cov}\colon U(\Cov,r)\to(\Cov,r)$; then
	$\text{dpt}(U(\Cov,r))=\text{dpt}(\Cov)\le \Delta$ and $g\circ \text{end}_{\Cov}\colon U(\Cov,r)\to(\N,v)$. Define $q\colon U(\Cov,r)\to U(\M,w)$ by sending a branch
	$\gamma$ to the branch obtained from its $f$-image by
	deleting consecutive repetitions.
	Since $f$ is monotone and $\M$ is a poset, an image
	cannot recur after a different image.
	Thus, $q(\gamma)$ is a branch and
	\[
	\text{end}_{\M}(q(\gamma))
	=f(\text{end}_{\Cov}(\gamma)).
	\]
	It preserves $\sigma$-colours since $f$ does, and is monotone since deleting repetitions preserves the
	prefix order. For~(p3), let $\gamma$ end at $t_{0}$ and let $\delta$ extend $q(\gamma)$ by
	worlds $x_{1}R^{+}\dots R^{+}x_{j}$; applying (p3) for $f$ successively yields
	$t_{0}\,R^{+}\,t_{1}\,R^{+}\cdots R^{+}\,t_{j}$ in $\Cov$ with $f(t_{i})=x_{i}$, and $\gamma$ extended by
	$t_{1},\dots,t_{j}$ maps to $\delta$. So $q$ is a p-morphism and $U(\Cov,r)$ covers
	$U(\M,w),(\N,v)$.
\end{proof}
	Let $(\M,w)$ be tree-shaped. Suppose two distinct children $u_{1},u_{2}$ of a common parent~$z$ have isomorphic subtrees,
	witnessed by an isomorphism $I\colon{\uparrow}u_{1}\to{\uparrow}u_{2}$ of $\sig$-models
	with $I(u_{1})=u_{2}$. The \emph{sibling merge} of $u_{2}$ into $u_{1}$ is the
	$\sig$-model $(\M^{-},w)$ with worlds $W\setminus{\uparrow}u_{2}$ and $R$ and $V$ restricted. As ${\uparrow}u_{2}$ is upward closed and omits the root, $(\M^{-},w)$ is again a tree-shaped $\sig$-poset
	with $\text{dpt}(\M^{-})\le\text{dpt}(\M)$.

\begin{figure}[t]
	\centering
	\begin{tikzpicture}[x=1cm, y=1.25cm]
		\foreach \l/\y in {1/0, 2/1, 3/2, 4/3} {
			\node[lvlab] at (-1.2,\y) {level~\l};
			\foreach \p in {0, 4.3, 8.6} {
				\draw[lvl] (\p-0.35,\y) -- (\p+2.75,\y);
			}
		}
		
		\node[wnode] (Ar) at (1.2,0) {$c_0$};
			\node[wnode,  fill=greencolour] (Aa1) at (0,1) {$c_1$};
			\node[wnode, ,  fill=darkgreencolour] (Ab1) at (0,2) {$c'_1$};
			\draw[rel] (Ar) -- (Aa1);
			\draw[rel] (Aa1) -- (Ab1);
			\node[wnode,  fill=orangecolour] (Aa2) at (1.2,1) {$c_2$};
			\node[wnode,  fill=darkorangecolour] (Ab2) at (1.2,2) {$c'_2$};
			\draw[rel] (Ar) -- (Aa2);
			\draw[rel] (Aa2) -- (Ab2);
			\node[wnode, fill=bluecolour] (Aa3) at (2.4,1) {$c_3$};
			\node[wnode, fill=darkbluecolour] (Ab3) at (2.4,2) {$c'_3$};
			\draw[rel] (Ar) -- (Aa3);
			\draw[rel] (Aa3) -- (Ab3);
		
		\node[wnode] (Br) at (5.5,0) {$c_0$};
			\node[wnode,  fill=greencolour] (Bam1) at (4.3,1) {$c_1$};
			\node[wnode,  fill=greencolour] (Bap1) at (4.3,2) {$c_1$};
			\node[wnode, fill=darkgreencolour] (Bb1)  at (4.3,3) {$c'_1$};
			\draw[rel] (Br) -- (Bam1);
			\draw[rel] (Bam1) -- (Bap1);
			\draw[rel] (Bap1) -- (Bb1);
			\node[wnode,  fill=bluecolour] (Bam3) at (6.7,1) {$c_3$};
			\node[wnode,  fill=bluecolour] (Bap3) at (6.7,2) {$c_3$};
			\node[wnode, fill=darkbluecolour] (Bb3)  at (6.7,3) {$c'_3$};
			\draw[rel] (Br) -- (Bam3);
			\draw[rel] (Bam3) -- (Bap3);
			\draw[rel] (Bap3) -- (Bb3);
		\node[wnode, fill=orangecolour] (Ba2) at (5.5,1) {$c_2$};
		\node[wnode,  fill=darkorangecolour] (Bb2) at (5.5,2) {$c'_2$};
		\draw[rel] (Br) -- (Ba2);
		\draw[rel] (Ba2) -- (Bb2);
		
		\node[wnode] (Cr) at (9.8,0) {$c_0$};
			\node[wnode, fill=orangecolour] (Ca2)  at (9.8,1) {$c_2$};
			\node[wnode,  fill=darkorangecolour] (Cbm2) at (9.8,2) {$c'_2$};
			\node[wnode,  fill=darkorangecolour] (Cbp2) at (9.8,3) {$c'_2$};
			\draw[rel] (Cr) -- (Ca2);
			\draw[rel] (Ca2) -- (Cbm2);
			\draw[rel] (Cbm2) -- (Cbp2);
			\node[wnode, fill=bluecolour] (Ca3)  at (11,1) {$c_3$};
			\node[wnode,  fill=darkbluecolour] (Cbm3) at (11,2) {$c'_3$};
			\node[wnode,  fill=darkbluecolour] (Cbp3) at (11,3) {$c'_3$};
			\draw[rel] (Cr) -- (Ca3);
			\draw[rel] (Ca3) -- (Cbm3);
			\draw[rel] (Cbm3) -- (Cbp3);			
		\node[wnode,  fill=greencolour] (Ca1) at (8.6,1) {$c_1$};
		\node[wnode,  fill=darkgreencolour] (Cb1) at (8.6,2) {$c'_1$};
		\draw[rel] (Cr) -- (Ca1);
		\draw[rel] (Ca1) -- (Cb1);
		
		\node at (1.2,-0.85) {$\mathfrak{F}_{3}$};
		\node at (5.5,-0.85) {$\mathfrak{M}_{\{1,3\}}$};
		\node at (9.8,-0.85) {$\mathfrak{N}_{\{2,3\}}$};
	\end{tikzpicture}
	\caption{Depth-$4$ refinements of $\mathfrak F_3$.
			The common index $3$ forces a five-world chain in any cover
			of $\{\mathfrak M_{\{1,3\}},\mathfrak N_{\{2,3\}}\}$.}
		\label{fig:AFG}
\end{figure}
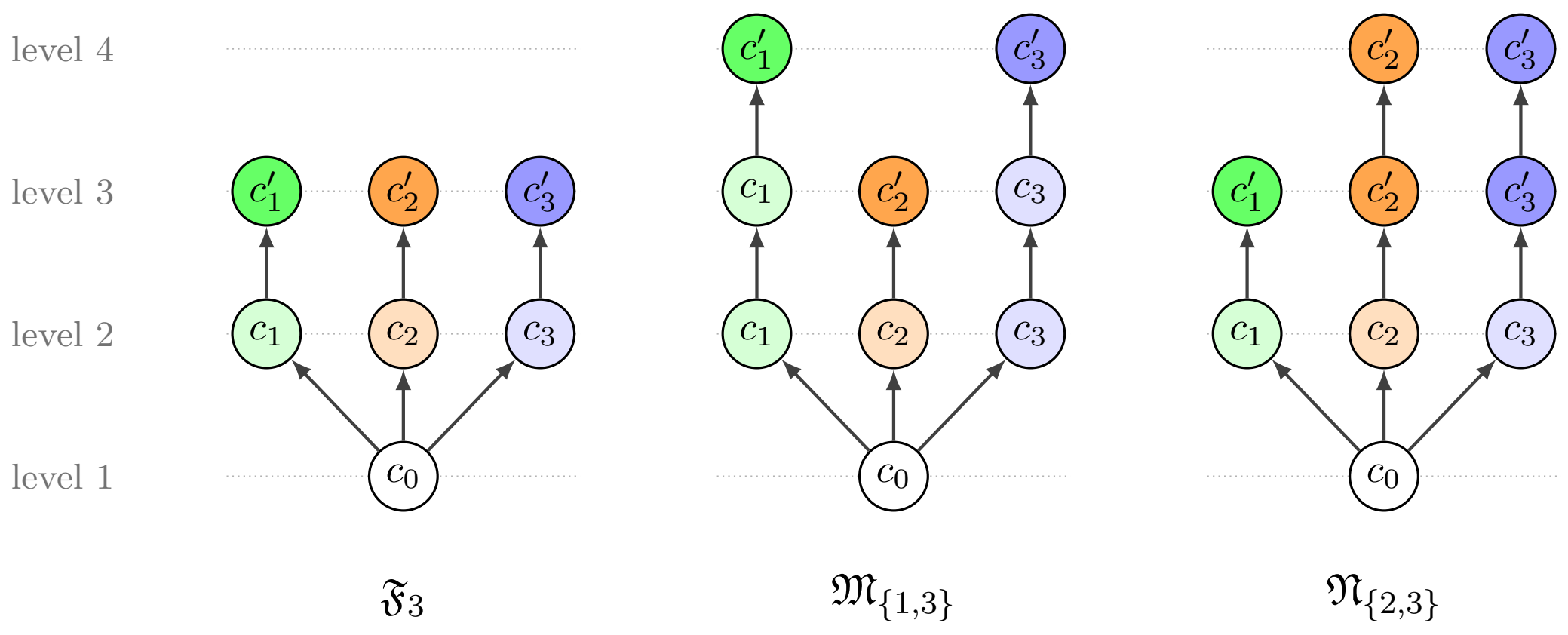

\begin{lem}[sibling merge]\label{lem:merge}
	Let $(\M^-,w)$ be the sibling merge of $u_2$ into $u_1$
	in a tree-shaped $\sigma$-poset $(\M,w)$.
	Then, for every $\sigma$-poset $(\N,v)$, the pair
	$\{(\M,w),(\N,v)\}$ has a $\Grzd$-cover iff the pair
	$\{(\M^-,w),(\N,v)\}$ has a $\Grzd$-cover.
\end{lem}
\begin{proof}
		Retain $z,u_1,u_2$ and the isomorphism
		$I\colon {\uparrow}u_1\to{\uparrow}u_2$.
		We write $\leq$ for the orders in the target models.
		
		$(\Rightarrow)$ Suppose $\{(\M,w),(\N,v)\}$ has a cover
		$(\Cov,u)$ via $f,g$.
		Define $e\colon (\Cov,u)\to(\M^-,w)$ by
		\[
		e(x)=
		\begin{cases}
			f(x),&f(x)\notin{\uparrow}u_2,\\
			I^{-1}(f(x)),&f(x)\in{\uparrow}u_2.
		\end{cases}
		\]
		Condition (p1) follows from $f$ and $I$.
		For (p2), an edge entering ${\uparrow}u_2$ from outside
		starts at or below $z$, and therefore also below $u_1$.
		For (p3), use $I^{-1}$ inside ${\uparrow}u_2$ and the
		original back witnesses elsewhere.
		Thus $e,g$ witness a cover of $\{(\M^-,w),(\N,v)\}$.
		
		$(\Leftarrow)$ Suppose $\{(\M^-,w),(\N,v)\}$ has a finite
		cover $(\Cov^-,u)$ via $f,g$, where
		$\Cov^-=((W^-,R^-),V^-)$.
		Put $X=f^{-1}({\uparrow}u_1)$ and let
		$X'=\{x'\mid x\in X\}$ be disjoint from $W^-$.
		Define $\Cov=((W,R),V)$ by
		\begin{align*}
			W={}&W^-\sqcup X',\\
			R={}&R^-
			\cup\{(x',y')\mid (x,y)\in R^-\cap(X\times X)\}\\
			&\hspace*{1.5em}\cup
			\{(x,y')\mid (x,y)\in R^-\cap((W^-\setminus X)\times X)\},\\
			V(p)={}&V^-(p)\cup\{x'\mid x\in V^-(p)\cap X\}
			\qquad(p\in\sigma).
		\end{align*}
		Since $X$ is upward closed, $R$ is reflexive, antisymmetric
		and transitive. There are no edges between $X$ and $X'$
		in either direction. Also $u\notin X$, and $u$ reaches
		every world of $X'$.
		
		Let $c\colon\Cov\to\Cov^-$ be given by $c(x)=x$ for $x\in W^-$
		and $c(x')=x$ for $x\in X$.
		Conditions (p1) and (p2) are immediate.
		For (p3), old worlds retain their old back witnesses;
		if $x'\in X'$ and $xR^-y$, then $y\in X$ and $x'Ry'$.
		Thus $c$ is a $\sigma$-p-morphism.
		Moreover, no chain meets both $X$ and $X'$, so $c$ is
		injective on chains. Hence
		\[
		\text{dpt}(\Cov)=\text{dpt}(\Cov^-)\leq\Delta.
		\]
		
		Define
		\[
		f'(x)=
		\begin{cases}
			f(x),&x\in W^-,\\
			I(f(y)),&x=y'\in X',
		\end{cases}
		\qquad
		g'=g\circ c.
		\]
		Then $g'$ is a $\sigma$-p-morphism.
		
		For $f'$, condition (p1) follows from $f$ and $I$.
		For (p2), only mixed edges need checking.
		If $xRy'$ with $x\in W^-\setminus X$ and $y\in X$, then
		\[
		f(x)\leq f(y),\qquad
		f(x)\notin{\uparrow}u_1,\qquad
		f(y)\in{\uparrow}u_1.
		\]
		Since $\M$ is tree-shaped, $f(x)\leq z\leq I(f(y))$.
		
		For (p3), let $f'(x)\leq a$.
		If $x=y'\in X'$, write $a=I(b)$ with $f(y)\leq b$.
		Choose $t$ with $yR^-t$ and $f(t)=b$.
		Then $t\in X$, $y'Rt'$ and $f'(t')=a$.
		If $x\in X$, then $a\in{\uparrow}u_1$, so an old back
		witness for $f$ suffices.
		If $x\in W^-\setminus X$ and $a\in\M^-$, again use an
		old back witness.
		Finally, if $x\in W^-\setminus X$ and
		$a\in{\uparrow}u_2$, then $f(x)\leq z$.
		Put $b=I^{-1}(a)$ and choose $y$ with $xR^-y$ and $f(y)=b$.
		Then $y\in X$, $xRy'$ and $f'(y')=a$.
		Thus $f'$ is a $\sigma$-p-morphism, and $f',g'$ witness
		the required $\Grzd$-cover.
\end{proof}	
\begin{proof}[Proof of Theorem~\ref{thm:normalform}]
	Point~1 follows by induction from the definition of the $B_k$:
	a normalised model is determined up to isomorphism by its
	root colour and the set of isomorphism types of its
	immediate successor submodels.
	
	For Point~2, unravel $p$ and repeatedly merge isomorphic
	sibling subtrees. Every merge removes worlds, so the
	construction terminates in a normalised model $\normalised{p}$.
	Its depth is at most $\Delta$, and
	Lemmas~\ref{lem:unravel} and~\ref{lem:merge} imply that,
	for every $r\in\sh(C)$,
	\[
	\{p,r\}\text{ has a $\Grzd$-cover}
	\quad\Longleftrightarrow\quad
	\{\normalised{p},r\}\text{ has a $\Grzd$-cover}.
	\]
	The endpoint and folding maps are $\sigma$-p-morphisms,
	so $\normalised{p}\equiv_\sigma p$.
\end{proof}
We have shown the following consequence of  Theorems~\ref{thm:weaklyboundedGrz} and \ref{thm:bounded-profile}.
\begin{cor}\label{cor:dcraig}
	For $n\geq 0$ and $\Delta\geq 1$, $\Grzd$ satisfies $\lim\limits_{m\to\infty}\DCraig_{n,m}=1$. 
\end{cor}
We next show that $\Grzd$ has uniform BICIP
iff $\Delta\leq3$. For $\Delta\leq2$, this follows from CIP.
For $\Delta=3$, we show that $\mathrm{inc}(C)\leq3$ for
every cell $C$, with equality for a suitable cell. The result indicates that, for $\Delta=3$, exactly two `formula types’ are responsible for the failure of CIP. We begin by reproving that $\Grz_{[2]}$ enjoys CIP.
\begin{lem}\label{lem:grz-height-two}
	If $(\M,w)\equiv_\sigma(\N,v)$ and
	$\text{dpt}(\M),\text{dpt}(\N)\leq2$, then
	$\{(\M,w),(\N,v)\}$ has a cover of depth at most $2$.
\end{lem}
\begin{proof}
	It follows from $(\M,w)\equiv_\sigma(\N,v)$ that the $\sigma$-colours of $w$ and $v$ coincide and that the sets of $\sigma$-colours of maximal worlds also coincide.
	Take a new root $t$ of $\sigma$-colour of $w$ and, above $t$,
	one maximal world $(x,y)$ for every pair of maximal
	worlds with the same $\sigma$-colour.
	Define $f(t)=w$, $g(t)=v$, $f(x,y)=x$ and $g(x,y)=y$.
	Every proper successor of either target root is maximal
	and occurs in such a pair.
	Thus $f,g$ satisfy (p1)--(p3).
	The construction also applies if either target is a singleton.
\end{proof}
We use the following construction.
Let $p=(\M,w)$ and $q=(\N,v)$ have depth at most $3$
and $w$ and $v$ the same $\sigma$-colour.
Suppose that every proper successor in either model can
be matched with a $\sigma$-equivalent world in the other
model, with both generated models of depth at most $2$.
Take a depth-$2$ cover of each matched pair, using
Lemma~\ref{lem:grz-height-two}, and put a new root below
their disjoint union.
The component maps extend to $\sigma$-p-morphisms onto
$p,q$: the matched pairs supply (p3) at the new root.
Thus $\{p,q\}$ has a cover of depth at most $3$.
\begin{thm}\label{thm:boundedGrz}
	For $\Delta\geq1$, $\Grzd$ has uniform BICIP
	iff $\Delta\leq3$.
	For $\Delta=3$, the bound $\mathrm{inc}(C)\leq3$ is sharp.
\end{thm}
\begin{proof}
	For $\Delta=1$, $\sh(C)$ has a single element, and so $\mathrm{inc}(C)=1$, for every cell $C$.
	At $\Delta=2$, Lemma~\ref{lem:grz-height-two} and
	Theorem~\ref{thm:cover} give the same conclusion.
	In particular, we reproved that these two logics have CIP.
	
	Let $\Delta=3$ and $C\in\CellAt{}{\sigma}{\rho}$.
	Take the $\sigma$-atom $\mathcal{A}=\bigcup C$
	and let $(\mathfrak S_{\mathcal A},r)$ be its standard model.
	By Lemma~\ref{lem:standardmodel}, every $\sigma$-shadow $(\M,w)$ of $\mathcal{M}\in C$
	has a $\sigma$-p-morphism
	$f\colon (\M,w)\to(\mathfrak S_{\mathcal A},r)$.
	
	If $\text{dpt}(\mathfrak S_{\mathcal A})=1$, then all worlds
	in all shadows have the same $\sigma$-colour.
	Hence $\mathrm{inc}(C)=1$.
	The same conclusion holds if
	$\text{dpt}(\mathfrak S_{\mathcal A})=3$:
	no non-root world can be mapped by $f$ to $r$, since (p3) would then
	give a four-world chain. Thus equal images under the maps $f$
	to $\mathfrak S_{\mathcal A}$ provide the required matches.
	
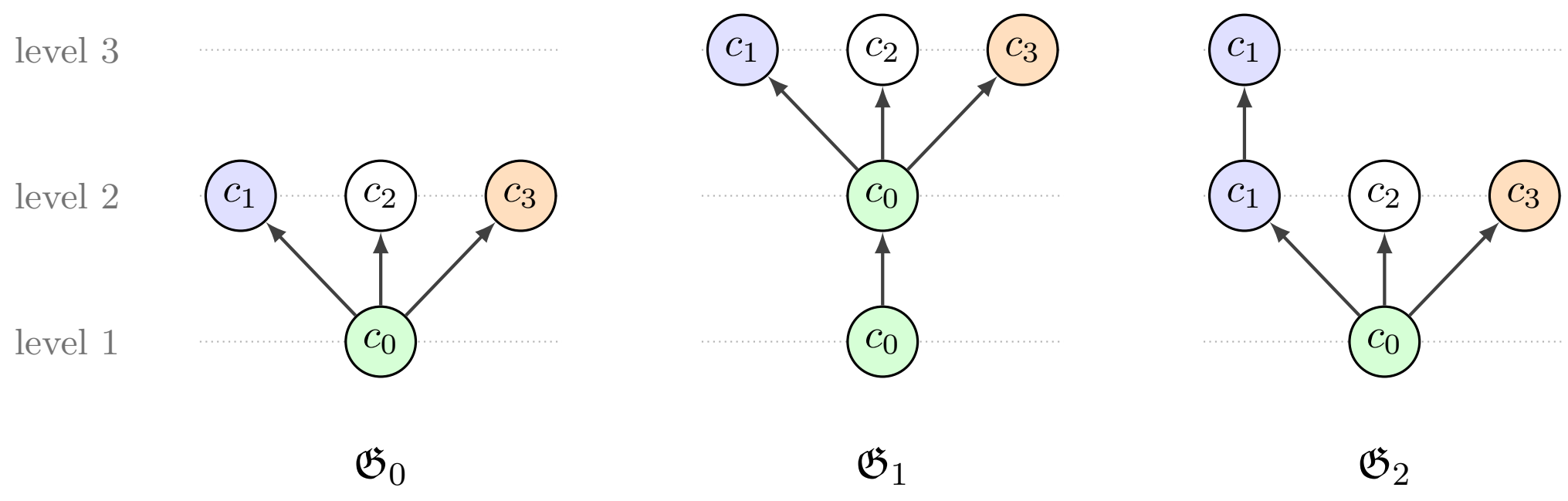
\begin{figure}[t]
	\centering%
	\begin{tikzpicture}[x=1cm, y=1.25cm]
		\foreach \l/\y in {1/0, 2/1, 3/2} {
			\node[lvlab] at (-0.9,\y) {level~\l};
			\foreach \p in {0, 4.3, 8.6} {
				\draw[lvl] (\p-0.35,\y) -- (\p+2.75,\y);
			}
		}
		
		\node[wnode,fill=greencolour] (Ar) at (1.2,0) {$c_0$};
			\node[wnode, fill=bluecolour] (Aa1) at (0,1) {$c_1$};
			\draw[rel] (Ar) -- (Aa1);
			\node[wnode] (Aa2) at (1.2,1) {$c_2$};
			\draw[rel] (Ar) -- (Aa2);
			\node[wnode,  fill=orangecolour] (Aa3) at (2.4,1) {$c_3$};
			\draw[rel] (Ar) -- (Aa3);
		
		\node[wnode, fill=greencolour] (Br0) at (5.5,0) {$c_0$};
		\node[wnode, fill=greencolour] (Br1) at (5.5,1) {$c_0$};
		\draw[rel] (Br0) -- (Br1);
			\node[wnode, fill=bluecolour] (Ba1) at (4.3,2) {$c_1$};
			\draw[rel] (Br1) -- (Ba1);
			\node[wnode] (Ba2) at (5.5,2) {$c_2$};
			\draw[rel] (Br1) -- (Ba2);
			\node[wnode,  fill=orangecolour] (Ba3) at (6.7,2) {$c_3$};
			\draw[rel] (Br1) -- (Ba3);
		
		\node[wnode, fill=greencolour] (Cr) at (9.8,0) {$c_0$};
		\node[wnode, fill=bluecolour] (Ca1m) at (8.6,1) {$c_1$};
		\node[wnode, fill=bluecolour] (Ca1p) at (8.6,2) {$c_1$};
		\draw[rel] (Cr) -- (Ca1m); 
		\draw[rel] (Ca1m) -- (Ca1p);
			\node[wnode] (Ca2) at (9.8,1) {$c_2$};
			\draw[rel] (Cr) -- (Ca2);
			\node[wnode,  fill=orangecolour] (Ca3) at (11.0,1) {$c_3$};
			\draw[rel] (Cr) -- (Ca3);
		
		\node at (1.2,-0.85) {$\mathfrak{G}_{0}$};
		\node at (5.5,-0.85) {$\mathfrak{G}_{1}$};
		\node at (9.8,-0.85) {$\mathfrak{G}_{2}$};
	\end{tikzpicture}
	\caption{A standard $\sigma$-model $\mathfrak G_0$ of depth $2$
			and root- and lower-stutter refinements.
			The pair $\{\mathfrak G_1,\mathfrak G_2\}$ has no
			$\Grz_{[3]}$-cover.}
		\label{fig:Grz3}
\end{figure}

	Suppose now $\text{dpt}(\mathfrak S_{\mathcal A})=2$.
	Its worlds are $r,v_1,\ldots,v_k$, where the $v_i$ are
	maximal and $c_1,\ldots,c_k$, the $\sigma$-colours of $v_1,\dots,v_k$, are pairwise distinct.
	For $k=3$, this is $\mathfrak G_0$ in Fig.~\ref{fig:Grz3}.
	A $\sigma$-shadow has a \emph{root stutter} if $f(x)=r$ for some
	non-root $x$, and a \emph{lower stutter} if $f(x)=v_i$
	for some $x$ with a proper successor.
	Figure~\ref{fig:Grz3} illustrates these in
	$\mathfrak G_1$ and $\mathfrak G_2$, respectively.
	
	The only incompatible pairs are those in which one shadow
	has a root stutter and the other has a lower stutter but
	no root stutter.
	Indeed, if both shadows have root stutters, or neither
	does, every proper world has a $\sigma$-equivalent proper
	match in the other shadow.
	If one shadow has neither stutter, its depth is at most~$2$,
	and its root can also be used as a match.
	The preceding construction therefore covers all these pairs.
	
	For the remaining case, suppose a cover exists.
	Lift a non-root world mapping to $r$ in the first shadow.
	Its image in the second shadow must be the root:
	the two cover maps induce the same image in the extensional
	model $\mathfrak S_{\mathcal A}$.
	Lifting a lower stutter from this world then gives a
	four-world chain, a contradiction.
	
	Hence, there are at most three distinct neighbourhoods,
	corresponding to shadows with neither stutter, with a root
	stutter, and with a lower stutter but no root stutter.
	Thus $\mathrm{inc}(C)\leq3$.
	The bound is attained by the three models in
	Fig.~\ref{fig:Grz3}, choosing pairwise distinct $\sigma$-colours for $r$, $v_1$, $v_2$ and $v_3$.
	Assign distinct $\rho$-colours within each model, using
	one common signature $\rho$.
	The expansions are extensional standard models of atoms
	in one cell, and their shadows have distinct neighbourhoods.
	
	For $\Delta=4$, let $\mathfrak F_k$ have branches
	$r<v_i<u_i$, $1\leq i\leq k$, with pairwise distinct
	$\sigma$-colours; take $2^{|\sigma|}\geq2k+1$.
	For $X,Y\subseteq\{1,\ldots,k\}$, let $\mathfrak M_X$
	introduce an $v_i$-stutter iff $i\in X$, and let
	$\mathfrak N_Y$ introduce a $u_i$-stutter iff $i\in Y$.
	Figure~\ref{fig:AFG} shows $k=3$, $X=\{1,3\}$ and $Y=\{2,3\}$.
	Then
	\[
	\{\mathfrak M_X,\mathfrak N_Y\}
	\text{ has a $\Grz_{[4]}$-cover}
	\quad\Longleftrightarrow\quad X\cap Y=\emptyset.
	\]
	If $X\cap Y=\emptyset$, combine the stutters in one model;
	each branch has at most one stutter, so its depth is at
	most $4$. Collapsing stutters gives the cover maps.
	If $i\in X\cap Y$, lifting the two $v_i$-worlds and then
	the two $u_i$-worlds forces a five-world chain.
	The distinct $\sigma$-colours prevent these lifts from
	moving to another branch.
	
	Choose one $\rho$ with $2^{|\rho|}\geq3k+1$ and assign
	distinct $\rho$-colours within each witness model.
	Their expansions are extensional standard models of atoms
	in one cell $C$, since collapsing stutters gives
	$\sigma$-p-morphisms onto~$\mathfrak F_k$.
	If $X\ne X'$ and, say, $i\in X\setminus X'$, then
	$\mathfrak N_{\{i\}}$ is incompatible with $\mathfrak M_X$
	but compatible with $\mathfrak M_{X'}$.
	Thus the~$\mathfrak M_X$ represent $2^k$ distinct profile
	nodes, and $\mathrm{inc}(C)\geq2^k$.
	
	For $\Delta>4$, prepend a chain of $\Delta-4$ worlds
	with distinct new $\sigma$-colours to all witness models.
	The disjoint-stutter covers then have depth at most $\Delta$,
	while overlapping stutters force depth at least $\Delta+1$.
	The same construction therefore gives unbounded profile
	sizes for every fixed $\Delta\geq4$.
\end{proof}

\subsection{Uniform Interpolation Density after Existential Quantification}

We show that for any class $\mathcal{K}$ of rooted frames determining $\Grzd$ and containing all finite tree-shaped frames of depth at most $\Delta$, we have $\lim_{m\to \infty}\dens_{n,m}^{\Pi}= 1$, for any fixed $n$.\footnote{This excludes, for instance, the class $\mathcal{K}$ consisting of a single perfect tree of infinite outdegree and depth~$\Delta$. This tree determines $\Grzd$ but defines a rather different equivalence relation.}
To this end, we first show a general sufficient condition for $\dens_{n,m}^{\Pi}\to 1$ for logics with BICIP. We use the notation from Section~\ref{subsection:asymqunat}.

\begin{lem}\label{thm:uniform-one}
	Let $L=\textup{Log}(\mathcal{K})$ be a locally tabular quasi-normal modal logic with BICIP. Assume that, with logical equivalence taken over $\mathcal{K}$, for every $n\geq0$ there is a function $L_n(m)>0$
	with $L_n(m)\to\infty$
	as $m\to\infty$, such that the following holds for all sufficiently large $m$:
	for every $\sigma_n^{\uparrow\rho_m}$-cell $C$ and every
	$q\in P$ with $P$ the canonical profile of $C$ and  $N_P(q)\ne\emptyset$, 
	\[
	|\QCS{\{q\}}|\ \ \geq \ \ L_n(m).
	\]
	Then $\lim\limits_{m\to\infty}\dens_{n,m}^{\Pi}=1$ for every $n\geq 0$.
\end{lem}
\begin{proof}
	Fix $n\geq 0$. For each $C_{\sigma_{n}}\in \CellAt{}{\sigma_{n}}{\emptyset}$, take the stabilisation
	index supplied by Lemma~\ref{lem:domination}, and let
	$m_0(n)$ be their maximum.
	Fix $m\geq m_0(n)$ sufficiently large for the hypothesis
	to hold. Also fix a $\sigma_{n}^{\uparrow\rho_{m}}$-cell $C$ with canonical profile $P$. 
	
	Let $S\subseteq P$ be non-empty with $N_{P}(S)\not=\emptyset$. Choose
	$q_S\in N_{P}(S)$. Then $(S,\{q_S\})$ is an ordered biclique of~$H_{P}$.
	Lemma~\ref{lem:injective-join} gives an injection
	$\QCS{S}\times \QCS{\{q_{S}\}}\to\QC$.
	Consequently,
	\begin{equation}\label{eq:small-bad-region}
		\abs{\QCS{S}}
		\ \ \le\ \ \frac{\abs{\mathcal{Q}_{C}}}{\abs{\QCS{\{q_{S}\}}}}
		\ \ \le\ \ \frac{\abs{\mathcal{Q}_{C}}}{L_{n}(m)}.
	\end{equation}
	Let $\mathcal{B}_{C}$ be the set of $x\in \mathcal{Q}_{C}$ without a uniform $\sigma_{n}$-interpolant. By Lemma~\ref{lem:domination},
	\[
	\mathcal{B}_{C}
	\subseteq
	\bigcup_{\substack{\emptyset \neq S \subseteq P\\
			N_{P}(S)\not=\emptyset}}
	\mathcal{Q}_{C}[S],
	\]
	and, with $2^{|P|}\leq 2^{f_{L}(n)}$, we obtain from \eqref{eq:small-bad-region} the following:
	\[
	\abs{\mathcal B_{C}}
	\leq 2^{f_{L}(n)}\frac{\abs{\mathcal{Q}_{C}}}{L_{n}(m)}.
	\]
	Dividing by $\abs{\mathcal{Q}_{C}}$ shows
		\[
			\smash{\overline{\dens}}_{C}^\Pi
			\ \ \le\ \ \frac{2^{f_{L}(n)}}{L_{n}(m)}.
		\]
	By the projective cell reduction principle (Theorem~\ref{thm:slicereduction5}),
	\[
 \smash{\overline{\dens}}_{n,m}^{\Pi}
	\ \ =\ \ 1-\hspace*{-1em}\prod_{C\in\CellAt{}{\sigma_{n}}{\rho_{m}}}\hspace*{-1em}\dens_C^{\Pi}
	\ \ \leq \hspace*{-0.5em}\sum_{C\in\CellAt{}{\sigma_{n}}{\rho_{m}}}\hspace*{-1em}\smash{\overline{\dens}}_C^{\Pi}
	\ \ \leq\ \ |\Atom{}{\sigma_n}|\frac{2^{f_L(n)}}{L_n(m)}
	\longrightarrow0.\qedhere
	\]
\end{proof}

We verify the conditions of Lemma~\ref{thm:uniform-one}
for any class $\mathcal{K}$ of rooted frames determining $\Grzd$ and containing all finite tree-shaped frames of depth at most $\Delta$.
We set $f_L(n)=B_\Delta(n)$, fix $\Delta\geq1$, $\mathcal{K}$, $n\geq0$, and let $K_n$ bound
the numbers of worlds in all normalised $\sigma_n$-posets
of depth at most $\Delta$.
Such a bound exists by Theorem~\ref{thm:normalform}.

Fix an arbitrary integer $v>0$.
It suffices to provide $m_v$ such that, for all $m\geq m_v$, every
$\sigma_n^{\uparrow\rho_m}$-cell $C$, and every
$q\in P$ with $P$ the canonical profile of $C$ and $N_P(q)\ne\emptyset$ it holds that $|\QCS{\{q\}}|>v$.
Put
\[
m_v=
\left\lceil\log_2((v+2)K_n)\right\rceil.
\]

Let $m\geq m_v$, let $C$ be such a cell, and choose
$p=(\M',w)\in\sh(H_C)$ with $\kappa(p)=q$.
By Theorem~\ref{thm:normalform}, there is a normalised
$\sigma_n$-poset $(\M,w)$ with
$(\M,w)\equiv_{\sigma_n}p$
such that, for every $r\in\sh(C)$,
\[
\{(\M,w),r\}\text{ has a cover}
\quad\Longleftrightarrow\quad
\{p,r\}\text{ has a cover}.
\]
At this stage, $(\M,w)$ need not itself belong to $\sh(C)$.

Since $N_P(q)\ne\emptyset$, $\M$ has a non-root world $u$.
Indeed, if $\M$ were a singleton of colour~$a$, every
world in every shadow in $C$ would have colour $a$.
Each shadow would then map constantly onto $\M$ and
would itself cover its pair with $\M$, contradicting
$N_P(q)\ne\emptyset$.

Put $U={\uparrow}u$ and let $\M[u;k]$ replace $U$ by
$k$ disjoint isomorphic sibling copies.
For $1\leq k\leq v+1$,
\[
|\M[u;k]|
\  =\ |W\setminus U|+k|U|
\ \leq \ (v+2)K_n
\ \leq\ 2^m.
\]
Assign distinct $\rho_m$-colours to its worlds.
The resulting model $(\M[u;k]^\bullet,w)$ is extensional,
hence the standard model of an atom $\mathcal A_k$.
The folding map gives
\[
(\M[u;k],w)\equiv_{\sigma_n}(\M,w)
\equiv_{\sigma_n}p,
\]
so $\mathcal A_k\in C$ and
$\sh(\mathcal A_k)=(\M[u;k],w)$.
Repeated sibling merging reduces $\M[u;k]$ to $\M$.
Thus, for every $r\in\sh(C)$,
\[
\{\sh(\mathcal A_k),r\}\text{ has a cover}
\quad\Longleftrightarrow\quad
\{p,r\}\text{ has a cover}.
\]
Consequently,
$\pi_{C}(\mathcal A_k)
=\kappa(\sh(\mathcal A_k))=q$.
Put
$
f_k=
[\exists\vec\rho_m.
\MForm{}{\sigma_n\cup\rho_m}{\mathcal A_k}]
\in\QCS{\{q\}}$. 
We claim that $f_k\ne f_\ell$ whenever $k<\ell$. The equivalence classes $f_{k},f_{\ell}$ are defined on the class $\mathcal{K}$ containing all tree-shaped frames of depth at most $\Delta$.
Hence, the $\sigma_n$-model $(\M[u;k],w)$ satisfies $f_k$ and is based on a frame in $\mathcal{K}$.
If it satisfied $f_\ell$, some $\rho_m$-valuation on
the same frame would realise $\mathcal A_\ell$.
By Lemma~\ref{lem:standardmodel}, there would be a
surjective p-morphism onto $\M[u;\ell]^\bullet$.
But
\[
|\M[u;k]| \ =\ |W\setminus U|+k|U|
\ <\ |W\setminus U|+\ell|U|\ =\ |\M[u;\ell]|,
\]
a contradiction. Hence $|\QCS{\{q\}}|>v$.

For sufficiently large $m$, take
$v=\lfloor2^m/K_n\rfloor-2$.
Then $m\geq m_v$, so the hypothesis of
Lemma~\ref{thm:uniform-one} holds with
\[
L_n(m)=\left\lfloor\frac{2^m}{K_n}\right\rfloor-1
\longrightarrow\infty.
\]
\begin{cor}\label{cor:grz}
	For any $\Delta\geq 1$ and any class $\mathcal{K}$ of rooted frames determining $\Grzd$ and containing all finite tree-shaped frames of depth at most $\Delta$\textup{:} if $n\geq0$ is fixed, then
	$\lim\limits_{m\to\infty}\dens_{n,m}^{\Pi}=1$.
\end{cor}
For $\DCraig_{n,m}^{\Pi}$,
Theorem~\ref{pp:thm:craig-floor} gives a positive lower
bound depending on $n$ and $\Delta$.
Whether $\DCraig_{n,m}^{\Pi}\to1$ for fixed $n,\Delta$
and $m\to\infty$ remains open.

\section{The Difference Logic $\Diff$}

We consider the logic $\Diff$ of modal formulas valid in
\emph{difference frames} $(W,\ne)$. Thus,
$w\models\Diamond p$ if $p$ holds at some world
different from $w$. We show that
$\lim_{m\to\infty}\DCraig_{n,m}=1$, whereas
on difference frames $\lim_{m\to\infty}\DCraig^{\Pi}_{n,m}
=\lim_{m\to\infty}\dens^{\Pi}_{n,m}=0$.

Not every frame validating $\Diff$ is a difference frame.
In fact, let $\mathcal{K}_{\not=}$ be the class of all difference frames and $\mathcal{K}_{\Diff}$ be the class of frames $(W,R)$ with $R\supseteq \{(w,v)\in W^{2}\mid w\ne v\}$. 
\begin{lem}[Segerberg~\cite{Segerberg1980-SEGANO-2}] $\Diff=\textup{Log}(\mathcal{K}_{\Diff})=\textup{Log}(\mathcal{K}_{\ne})$, and a connected frame $(W,R)$ validates $\Diff$ iff $(W,R)\in \mathcal{K}_{\Diff}$.
\end{lem}
Note that the inclusion $\textup{Log}(\mathcal{K}_{\Diff})\supseteq\textup{Log}(\mathcal{K}_{\ne})$
follows from the observation that a reflexive
world can be obtained as a p-morphic image of at least two distinct irreflexive worlds.

The logic $\Diff$ is locally tabular. Hence, by
Lemma~\ref{lem:standardmodel}, its $\tau$-atoms can
be identified with $\tau$-standard pointed models.
In such models, each $\tau$-colour is realised at
most once, since worlds of the same colour satisfy
the same formulas in $\ML_\tau$. Consequently, a
colour~$c$ satisfying $c\wedge\Diamond c$ must be
represented by a reflexive world, and standard
pointed models need not be based on difference frames.
For our purposes, it is nevertheless more convenient
to work with difference frames and record whether a
colour is realised by zero, one, or at least two worlds.
We therefore identify a $\tau$-atom with a pair
$(F,c_0)$, where $F$ maps each $\tau$-colour $c$
to an element of $\{0,1,\ge2\}$, and $c_0$ is
the $\tau$-colour of the distinguished world, with
$F(c_0)\ne0$.

Note that a characteristic formula of $(F,c_{0})$ is given by
\[
\Phi(F,c_0) \ \ = \ \
c_0 \ \
\land\bigwedge_{F(c)=0}\neg(c\vee\Diamond c)\ \
\land\bigwedge_{F(c)=1}\hspace*{-0.25em}
\bigl((c\vee\Diamond c)\land\neg\Diamond(c\land\Diamond c)\bigr) \ \
\land\bigwedge_{F(c)=\ge2}\hspace*{-0.5em}\Diamond(c\land\Diamond c).
\]
As before, let
$\sigma=\{p_1,\ldots,p_n\}$ and
$\rho=\{q_1,\ldots,q_m\}$ be disjoint signatures,
and put $N=2^n$, $M=2^m$, and $K=2^{n+m}$.
There are $K$ choices for the distinguished colour,
two choices for its multiplicity, and three choices
for each of the remaining $K-1$ colours. Hence
\begin{equation}
	|\Atom{}{\sig\cup\priv}|=2K3^{K-1},
	\qquad
	|\Atom{}{\sig}|= 2N3^{N-1}.
	\label{eq:number-slices}
\end{equation}

Fix a cell $C\in\CellAt{}{\sigma}{\rho}$.
Its atoms can be regarded as refinements of the
$\sigma$-atom $(\atomcolours_\sigma,c_\sigma)$ underlying $C$.
We describe these refinements in detail.
Partition the $\sigma$-colours into
$\atomcolours_\sigma^0$, $\atomcolours_\sigma^1$, and
$\atomcolours_\sigma^{\ge2}$, according to whether their multiplicity
in $\atomcolours_\sigma$ is zero, exactly one, or at least two. 
We call the colours in $\atomcolours_\sigma^{\ge2}$
\emph{heavy} and those in~$\atomcolours_\sigma^1$ \emph{light}.
Each atom in $C$ is determined by a family of
\emph{refinement maps} $\delta_\scol$, for $\scol\in \atomcolours_\sigma^{\ge2}\cup \atomcolours_\sigma^1$, which are functions of the form
\[
\{\rho\text{-colours}\}\ \ \to\ \ \{0,1,\ge2\},
\]
together with a choice of a $\rho$-colour $\rcol_\sigma$
for the distinguished world, satisfying
$\delta_{c_\sigma}(\rcol_\sigma)\ne0$.

For a refinement map $\delta\colon \{\rho\text{-colours}\}\to\{0,1,\ge2\}$, set
\[
\lambda(\delta)
\ \ \eqdef \ \ 
\abs{\{\rcol\mid\delta(\rcol)=1\}} \
+ \
2\abs{\{\rcol\mid\delta(\rcol)=\ge2\}}.
\]
The value $\lambda(\delta_\scol)$ is the least
possible number of worlds whose $(\sigma\cup\rho)$-colours refine $\scol$
according to $\delta_\scol$. To define a member of $C$, we require
$\lambda(\delta_\scol)\ge2$ for
$\scol\in \atomcolours_\sigma^{\ge2}$ and
$\lambda(\delta_\scol)=1$ for $\scol\in \atomcolours_\sigma^1$:  a heavy colour is realised at least twice, while  a light
colour has exactly one refinement, realised at a
unique world.

We call a refinement map $\delta$ \emph{rigid} if  $\delta(\rcol)\in\{0,1\}$ for all $\rho$-colours $\rcol$ and
\emph{flexible} otherwise. If $\delta_\scol$ is rigid, then $\lambda(\delta_\scol)$ is the exact number of worlds that satisfy $(\sigma\cup \rho)$-colours refining~$\scol$ according to $\delta_\scol$; if $\delta_\scol$ is flexible (and so $\scol\in \atomcolours_{\sigma}^{\ge2}$), then any $k\in [\lambda(\delta_\scol),\infty)$ can be the number of worlds that satisfy $(\sigma\cup \rho)$-colours refining~$\scol$. Thus, for a refinement map $\delta$, define the interval specifying a possible number of worlds satisfying the refinements:
\[
I(\delta)=
\begin{cases}
	\{\lambda(\delta)\},&\delta\text{ rigid},\\
	[\lambda(\delta),\infty),&\delta\text{ flexible}.
\end{cases}
\]

We now construct a profile of the incompatibility
graph of $C$. It records, for each heavy
$\sigma$-colour, the possible numbers of worlds
realising that colour, while disregarding the
individual $\rho$-refinements and the choice of
the distinguished world. Observe that, for a rigid~$\delta_\scol$, the value of $\lambda(\delta_\scol)$ is bounded by $2\leq \lambda(\delta_\scol)\leq M$, while for a flexible $\delta_\scol$, it is bounded by $2\leq \lambda(\delta_\scol) \leq 2M$. So, consider the set of possible values of $I(\delta_\scol)$:
\[
\intervals_M = \intervals_M^{\mathsf{R}} \cup \intervals_M^{\mathsf{F}}, \text{ where }
\intervals_M^{\mathsf{R}} =  \{\{\lambda\}\mid 2\le\lambda\le M\} \text{ and }
 \intervals_M^{\mathsf{F}} = \{[\lambda,\infty)\mid 2\le\lambda\le2M\}.
\]
Let $P=(\intervals_M)^{\atomcolours_\sigma^{\ge2}}$ be the set of mappings $p$ from $\atomcolours_\sigma^{\ge2}$ to $\intervals_{M}$.
For $\scol\in \atomcolours_\sigma^{\ge2}$, we call each $p(\scol)\in \intervals_M^{\mathsf{R}}$ a \emph{rigid component of $p$} 
and each $p(\scol)\in \intervals_M^{\mathsf{F}}$ a \emph{flexible component of $p$}.
Define incompatibility in~$P$ by setting
$p\frown^P q$ if
$p(\scol)\cap q(\scol)=\emptyset$ for some
$\scol\in \atomcolours_\sigma^{\ge2}$.
The graph $H_P=(P,\frown^P)$ is called the
\emph{box-profile graph} of~$C$.
We define the profile map $\pi\colon C \to P$ as follows.
Using $I$, we can map any $\mathcal{M}\in C$ with refinement mappings $(\delta_\scol)_{\scol\in \atomcolours_{\sigma}^{\ge2}}$  to an  element of $P$ by setting
\[
\pi(\mathcal M)
\ \ =\ \ \bigl\{ \scol \mapsto I(\delta_\scol) \bigm| \scol\in\atomcolours_{\sigma}^{\ge2} \bigr\}
\ \ \in \  (\intervals_{M})^{\atomcolours_{\sigma}^{\ge2}}.
\]
Observe that, for $\mathcal M,\mathcal N\in C$ with refinement mappings
$(\delta_\scol^{\mathcal M})_{\scol\in \atomcolours_\sigma^{\ge2}}$ and $(\delta_\scol^{\mathcal N})_{\scol\in \atomcolours_\sigma^{\ge2}}$, we have
\[
\mathcal M\frown\mathcal N
\quad\Longleftrightarrow\quad
I(\delta_\scol^{\mathcal M})\cap I(\delta_\scol^{\mathcal N})=\emptyset
\quad\text{for some }\scol\in \atomcolours_\sigma^{\ge2}.
\]
Now one can easily show the following.
\begin{lem}\label{lem:profile-bound}
	$\pi$ is a profile map for $C$ onto $(P,\frown^{P})$, 
and  $|P|=(3M-2)^h\le(3M)^h$, where $h=\abs{\atomcolours_{\sigma}^{\ge2}}$.
\end{lem}

As before, the weights of a profile $p\in P$ and set $Q\subseteq P$ for mapping $\pi$ are defined as $w(p)=\abs{\pi^{-1}(p)}$ and   $w(Q)=\sum_{p\in Q}w(p)$, respectively. To compute these weights,
we first compute weights $w_{1}$ and $w_{1}^{\bullet}$ for the component mappings $I$ and then lift these to $\pi$. Let
\[
w_1(J)\eqdef \abs{I^{-1}(J)}, \qquad \text{for } J\in\intervals_M.
\]
Abbreviate 
\begin{align*}
\mathcal R^{\mathsf R}_\lambda & \ =\  I^{-1}(\{\lambda\}) \ =\  \{\delta \text{ is a rigid refinement map} \mid \lambda(\delta)=\lambda\},
&& \text{for } \{\lambda\} \in \intervals_M^{\mathsf{R}},\\
\mathcal R^{\mathsf F}_\lambda & \  = \  I^{-1}([\lambda,\infty)) \ = \  \{\delta \text{ is a flexible refinement map} \mid \lambda(\delta)=\lambda\},
&& \text{for }  [\lambda, \infty) \in \intervals_M^{\mathsf{F}},
\end{align*}
so that 
\[
w_1(\{\lambda\})=|\mathcal R^{\mathsf R}_\lambda|,
\qquad
w_1([\lambda,\infty))=|\mathcal R^{\mathsf F}_\lambda|.
\]
Counting refinement maps gives 
\begin{equation}
	w_1(\{\lambda\})=\binom M\lambda,
	\qquad
	w_1(\intervals_M^{\mathsf{F}}) =  \sum_{\lambda=2}^{2M}w_1([\lambda,\infty))
	=3^M-2^M,
	\qquad
	w_1([2,\infty))=M.
	\label{eq:ray2-weight}
\end{equation}
To take into account the distinguished $\sigma$-colour $c_{\sigma}$, we also count \emph{marked refinements}, i.e., pairs $(\delta,\rcol)$ such that $\rcol$ is a $\rho$-colour with $\rcol\in\supp(\delta)$, where $\supp(\delta) = \{ r\mid \delta(r) \ne 0 \}$:
\[
w_1^\bullet(\{\lambda\})=\bigl|\{(\delta,\rcol):\delta\in\mathcal R^{\mathsf R}_\lambda,\ \rcol\in\supp(\delta)\}\bigr|,
\quad
w_1^\bullet([\lambda,\infty))=\bigr|\{(\delta,\rcol):\delta\in\mathcal R^{\mathsf F}_\lambda,\ \rcol\in\supp(\delta)\}\bigr|.
\]
Then 
\[
w_1^\bullet(\{\lambda\})=\sum_{\delta\in\mathcal R^{\mathsf R}_\lambda}\abs{\supp(\delta)}=\lambda\binom M\lambda,
\quad
w_1^\bullet([\lambda,\infty))=\sum_{\delta\in\mathcal R^{\mathsf F}_\lambda}\abs{\supp(\delta)}.
\]
As $\abs{\supp(\delta)}\ge1$ in general and
$\abs{\supp(\delta)}=1$ on $\mathcal R^{\mathsf F}_2$,
\begin{equation}
	w_1^\bullet(\{\lambda\})\ge w_1(\{\lambda\}),
	\qquad
	w_1^\bullet([\lambda,\infty))\ge w_1([\lambda,\infty)),
	\qquad
	w_1^\bullet([2,\infty))=M.
	\label{eq:marked-basic}
\end{equation}

We lift these weights to $\pi$. Fix a box-profile 
$q\in P$ and consider
the inverse image
\[
\pi^{-1}(q)
=
\{\mathcal M\in C\mid\pi(\mathcal M)=q\}.
\]
The refinement maps for distinct $\sigma$-colours can
be chosen independently, subject only to the choice
of the refinement of the distinguished world.
Consequently,
\begin{equation*}
	\pi^{-1}(q) \ \ \cong \ \ \displaystyle
		\prod_{\scol\in \atomcolours_{\sigma}^{1}}2^\rho
		\times
	\begin{cases}
		\prod_{\scol\in \atomcolours_{\sigma}^{\ge2}}I^{-1}(q(\scol)),
		& \hspace{-2mm}c_{\sigma}\in \atomcolours_{\sigma}^{1},\\[4mm]
		\displaystyle
		\bigl\{(\delta,\rcol)\mid
		\delta\in I^{-1}(q(c_{\sigma})),\
		\rcol\in\supp(\delta)\bigr\}
		\times \hspace{-3mm}
		\prod_{\scol\in \atomcolours_{\sigma}^{\ge2}
			\setminus\{c_{\sigma}\}}\hspace{-1.5em}I^{-1}(q(\scol)),
		& \hspace{-2mm}c_{\sigma}\in \atomcolours_{\sigma}^{\ge2}.
	\end{cases}
\end{equation*}
Indeed, if $c_{\sigma}$ is light, its unique
refinement determines the colour of the
distinguished world. If $c_{\sigma}$ is heavy,
an atom requires, in addition to its refinement
maps, a choice of a $\rho$-colour $\rcol$ with
$\delta_{c_{\sigma}}(\rcol)\ne0$ for the
distinguished world. These are exactly the
choices recorded by the marked refinements.

Taking cardinalities in the above formula and writing $\ell=|\atomcolours_{\sigma}^{1}|$, we obtain
\begin{equation}
	w(q) \ \ = \ \ \abs{\pi^{-1}(q)} \ \ = \ \ M^\ell\times
	\begin{cases}
		\displaystyle\prod_{\scol\in \atomcolours_{\sigma}^{\ge2}}w_1(q(\scol)), & c_{\sigma}\in \atomcolours_{\sigma}^{1},\\[3mm]
		w_1^\bullet(q(c_{\sigma}))\times\displaystyle\prod_{\scol\in \atomcolours_{\sigma}^{\ge2}\setminus\{c_{\sigma}\}}\hspace*{-0.5em}w_1(q(\scol)), & c_{\sigma}\in \atomcolours_{\sig}^{\ge2}.
	\end{cases}
	\label{eq:box-weight}
\end{equation}
This concludes the computation of individual weights. 

As before,  for $Q\subseteq P$, the common neighbourhood of $Q$ in $P$ is denoted by $N_{P}(Q)$, while its complement by $
\Nc_{P}(Q)\eqdef P\setminus N_{P}(Q)$. Set $g(Q)=w(\Nc_{P}(Q))-w(Q)$. Then $w(N_{P}(Q)) = \abs{C}-w(\Nc_{P}(Q))$ and so
\begin{equation}\label{eq:deficit-identity}
	w(Q)+w(N_{P}(Q)) \ =\ |C|-g(Q).
\end{equation}
We use the function $g(Q)$ to bound the
number of inconsistent pairs without a
Craig interpolant within cells~$C$, $\NoCraig_C=|\SetClash_C|-|\SetCraig_C|$; see~\eqref{eq:defofNoCraig}. 
\begin{lem}\label{lem:deficit-count}
	Let $\mathcal G\subseteq2^P\setminus\{\emptyset\}$ and assume that every non-empty ordered biclique in $P$ can be oriented so that its first side belongs to $\mathcal G$.  Then
	\begin{equation}
		\NoCraig_C
		\ \ \le \ \
		2^{\abs{C}+1}\sum_{Q\in\mathcal G}2^{-g(Q)}.
		\label{eq:deficit-sum}
	\end{equation}
	In particular, if $g(Q)\ge\Delta$ for every $Q\in\mathcal G$, then
	\begin{equation}
		1-\DCraig_C
		\ \ \le\ \ 
		2^{\abs P+1-\Delta}.
		\label{eq:deficit-uniform}
	\end{equation}
\end{lem}
\begin{proof}
	The first inequality follows from Lemma~\ref{lem:ineqprofile} and \eqref{eq:deficit-identity}
	with a factor $2$ to take into account the orientation of bicliques.
	Divide by $2^{\abs{C}}$ and use~\eqref{eq:dcraig-nocraig} for the
	second inequality.
\end{proof}

We now determine a sufficiently large $\Delta$ for an appropriate $\mathcal{G}$. Assume $\atomcolours_{\sigma}^{\ge2}\ne\emptyset$, that is, $h\geq 1$.
For a box-profile $p$, let 
\begin{equation*}	
\flexcompcol{p} = \{ \scol \mid p(\scol) \text{ is flexible} \}\qquad\text{ and }\qquad \rigcompcol{p} = \{ \scol \mid p(\scol)  \text{ is rigid} \}.
\end{equation*}
Call a box-profile $p$ \emph{somewhere rigid} if $|\rigcompcol{p}| > 0$, otherwise call it \emph{everywhere flexible}. Let $\somerigid$ be the set
of non-empty $Q\subseteq P$ such that each $p\in Q$ is somewhere rigid.

\begin{lem}\label{lem:point-side}
	Every non-empty ordered biclique in $P$ can be oriented so that its first side belongs to $\somerigid$. 
\end{lem}
\begin{proof}
	Clearly $q_{1}\not\frown^{P}q_{2}$ for any two everywhere flexible
	$q_{1},q_{2}\in P$. Hence an ordered biclique in $P$ cannot contain an everywhere flexible profile on both sides. 
	Orient it so that the first side contains none.
\end{proof}
We are in a position now to prove the main lower bound on $g(Q)$ and $\NoCraig_{C}$.
\begin{thm}\label{thm:max-ray-deficit}
	Assume $h=\abs{\atomcolours_{\sigma}^{\ge2}}$, $\ell= \abs{\atomcolours_{\sigma}^{1}}$,  and $M=2^m\ge2$. For every $Q\in \somerigid$,
	\begin{equation}
		g(Q)
		\ \ \ge\ \ 
		\frac{1}{4^{h-1}}M^{\ell+2h-1}.
		\label{eq:uniform-deficit}
	\end{equation}
\end{thm}

\begin{proof}
	Choose
	$q\in Q$ with
	\[
	k=\abs{\flexcompcol{q}}
	\]
	maximal among $Q$. Since $Q\in \somerigid$, we have $|\rigcompcol{q}| > 0$. 
	The \emph{shell} $\mathcal E(q)\subseteq P$ consists of the box-profiles $p\in P$ satisfying the following:
	\begin{enumerate}[label=(\roman*)]
		\item $\scol\in \flexcompcol{p}$, for every $\scol\in\flexcompcol{q}$;
		\item $q(\scol)\subseteq p(\scol)$, for every $\scol\in\rigcompcol{q}$;
		\item $\scol\in \flexcompcol{p}$, for at least one $\scol\in \rigcompcol{q}$.
	\end{enumerate}
	
\begin{cla}\label{claim:1}
$\mathcal E(q)\subseteq\Nc(Q)\setminus Q$. Consequently,
		$g(Q)\ge w(\mathcal E(q))$.
\end{cla}

\begin{proof}
First observe that, for every $p\in\mathcal E(q)$, $p(\scol)\cap q(\scol)\ne\emptyset$ for all $\scol\in \atomcolours_{\sigma}^{\ge2}$. Indeed, if $q(\scol)$ is flexible, then, by (i), $p(\scol)$ is flexible and $p(\scol)\cap q(\scol)\ne\emptyset$ follows. If $q(\scol)=\{\lambda\}$, then, by~(ii), either $p(\scol)$ is the same rigid component or $p(\scol)=[\mu,\infty)$ with $\mu\le\lambda$, in which case
		$\lambda\in [\mu,\infty)$.  Hence $p\in\Nc(\{q\})\subseteq\Nc(Q)$.
		
		By (i) and (iii),  every member $p$ of the shell has at least $(k+1)$-many flexible components,
		whereas maximality of $q$ says that no member of $Q$ has more than $k$. Thus $\mathcal E(q)\cap Q=\emptyset$.
\end{proof}
		
    \medskip
	We next show a lower bound for the weight of the shell. 
		
\begin{cla}\label{claim:2}
		Let $L= w_1(\intervals_M^{\mathsf{F}})$.   For any box-profile $q$ with  $|\flexcompcol{q}| = k$ and $|\rigcompcol{q}| > 0$, we have
		\begin{equation}
			w(\mathcal E(q))
			\ \ \ge \ \ 
			M^{\ell + 1} L^k \binom{M}{2}^{h - k -1}.
			\label{eq:shell-rho}
		\end{equation}
\end{cla}

\begin{proof}
We say that a rigid component $\{\lambda\}\in\intervals_M^{\mathsf{R}}$  is \emph{compatible} with a component $J\in \intervals_M$ if $\{\lambda\} \subseteq J$. Observe that $\{\lambda\}\subseteq[\mu,\infty)$ iff $\mu\le\lambda$. So, for each $\{\lambda\}\in\intervals_M^{\mathsf{R}}$, we introduce the ordinary weights of, respectively, rigid, flexible and all components compatible with $\{\lambda\}$:
		\[
		\rigweight_{\{\lambda\}}= \hspace*{-0.25em}\sum_{\substack{J\in\intervals_M^{\mathsf{R}}\\\{\lambda\}\subseteq J}}\hspace*{-0.3em}w_1(J) =  w_1(\{\lambda\}),
		\quad
		\flxweight_{\{\lambda\}}= \hspace*{-0.25em}\sum_{\substack{J\in\intervals_M^{\mathsf{F}}\\ \{\lambda\}\subseteq J}}\hspace*{-0.3em}w_1(J) = \sum_{\mu=2}^{\lambda}w_1([\mu,\infty)),
		\quad
		\totweight_{\{\lambda\}}= \rigweight_{\{\lambda\}}+\flxweight_{\{\lambda\}}.
		\]
		We can describe the shell as follows:
\begin{equation}
			\mathcal E(q)=\Bigl(\prod_{\scol\in \atomcolours_{\sigma}^{\ge2}}P_\scol\Bigr)\setminus\mathcal{D},
			\label{eq:shell-set}
		\end{equation}
where 
\begin{itemize}
\item $P_\scol = \intervals_M^{\mathsf{F}}$, for each $\scol \in \flexcompcol{q}$, see~(i);
\item $P_\scol = \bigl\{ J \in \intervals_M \bigm| q(\scol) \subseteq J \bigr\}$, for each $\scol \in \rigcompcol{q}$, see~(ii);
\item and $\mathcal{D} =\bigl\{p\in\textstyle\prod_{\scol\in \atomcolours_{\sigma}^{\ge2}}P_\scol\mid p(\scol)=q(\scol), \text{ for all } \scol \in\rigcompcol{q} \bigr\}$, see~(iii).
\end{itemize}		
		
		Assume first $c_{\sigma}\in\atomcolours_{\sigma}^{1}$, so,
		by~\eqref{eq:box-weight}, the weight $w(p)=M^\ell\prod_{\scol\in \atomcolours_{\sigma}^{\ge2}}w_1(p(\scol))$ is
		multiplicative over $\scol\in \atomcolours_{\sigma}^{\ge2}$. Hence, for any product family
		$\prod_{\scol}X_\scol$ with $X_\scol\subseteq\intervals_M$,
		\[
		w\Bigl(\prod_{\scol\in \atomcolours_{\sigma}^{\ge2}}X_\scol\Bigr)=M^\ell\prod_{\scol\in \atomcolours_{\sigma}^{\ge2}}w_{1}(X_\scol),
		\qquad
		w_{1}(X_\scol):=\sum_{x\in X_\scol}w_1(x).
		\]
		By the definitions of $P_\scol$, $L$ and $\totweight_{\{\lambda\}}$, we have
		$w_{1}(P_\scol)=L$, for each $\scol \in \flexcompcol{q}$, and $w_{1}(P_\scol)= \totweight_{q(\scol)}$, for each $\scol \in \rigcompcol{q}$.
		On $\mathcal{D}$, the weight of each $\scol \in\flexcompcol{q}$ is $L$ and of each $\scol \in\rigcompcol{q}$  is 
		$\rigweight_{q(\scol)}$:  thus, $w(\mathcal{D})=M^\ell L^k\prod_{\scol \in\rigcompcol{q}}\rigweight_{q(\scol)}$.
		Subtracting, using $\mathcal{D}\subseteq\prod_\scol P_\scol$ and~\eqref{eq:shell-set}, we obtain
		\begin{equation}
			w(\mathcal E(q))
			\ \ = \ \ M^\ell L^k\Bigl[\prod_{\scol \in\rigcompcol{q}}\totweight_{q(\scol)} \ \ -\prod_{\scol \in\rigcompcol{q}}\rigweight_{q(\scol)}\Bigr].
			\label{eq:shell-product}
		\end{equation}
		So far we assumed $c_\sigma\in \atomcolours_\sigma^1$, so the 
		weight $w_1$ applies. We now check that $c_\sigma\in \atomcolours_\sigma^{\ge2}$ can only increase~$w(\mathcal E(q))$, so that \eqref{eq:shell-product} stays valid as a
		lower bound. Recall from \eqref{eq:box-weight} that, if $c_\sigma\in \atomcolours_\sigma^{\ge2}$, then the
		component $p({c_\sigma})$ uses~$w_1^\bullet$ in place of $w_1$, all others are unchanged.
		
		\emph{Case $c_\sigma\in\flexcompcol{q}$.} Then its factor changes from
		$w_{1}(P_{c_\sigma})=L$ to
		$w_{1}^\bullet(P_{c_\sigma}):=w_1^\bullet(\intervals_M^{\mathsf{F}})
		\ge w_1(\intervals_M^{\mathsf{F}})=L$
		by~\eqref{eq:marked-basic}. This same factor multiplies both $w(\prod_\scol P_\scol)$ and $w(\mathcal{D})$, and so
		\eqref{eq:shell-product} holds as a lower bound.
		
		\emph{Case $c_\sigma\in\rigcompcol{q}$.} Consider the products over the remaining rigid components (empty products $=1$ when $\rigcompcol{q} = \{c_\sigma\}$):
		\[
		\product{\totweight} \ =\hspace*{-0.5em}\prod_{\scol \in\rigcompcol{q}\setminus \{c_\sigma\} }\hspace*{-1em} \totweight_{q(\scol)},\qquad
		\product{\rigweight}\ =\hspace*{-0.5em}\prod_{\scol \in\rigcompcol{q}\setminus \{c_\sigma\}}\hspace*{-1em}\rigweight_{q(\scol)};
		\]
		then $\product{\totweight}\ge \product{\rigweight}\ge0$ since $\totweight_{q(\scol)}\ge \rigweight_{q(\scol)}\ge0$, for each $\scol\in\rigcompcol{q}$. The $k$ coordinates in
		$\flexcompcol{q}$ contribute $L^k$. Partition $\mathcal E(q)$ by the value of $p(c_\sigma)$:
		\begin{itemize}
			\item if $p(c_\sigma) \in\intervals_M^{\mathsf{R}}$, then, by (iii), another rigid component must be flexible, so those
			$(h-k -1)$ components contribute $\product{\totweight}-\product{\rigweight}$, and $p(c_\sigma)$ contributes $\smash{\rigweight}^\bullet = w_1^\bullet(q(c_\sigma))$, giving weight
			$M^\ell L^k\,\smash{\rigweight}^\bullet (\product{\totweight}-\product{\rigweight}) \geq M^\ell L^k\, \rigweight_{q(c_\sigma)}(\product{\totweight}-\product{\rigweight})$;
			\item if $q(c_\sigma) \subseteq p(c_\sigma)\in\intervals_M^{\mathsf{F}}$, then, by (iii), the other $(h - k -1)$
			components range without constraints~($\product{\totweight}$), and $p(c_\sigma)$ contributes $\smash{\flxweight}^\bullet = \sum_{J\in\intervals_M^{\mathsf{F}}, q(c_\sigma)\subseteq J}w_1^\bullet(J)$, giving weight
			$M^\ell L^k\,\smash{\flxweight}^\bullet \product{\totweight} \geq M^\ell L^k\, \flxweight_{q(c_\sigma)}\product{\totweight}$.
		\end{itemize}
		Hence, we obtain the following lower bound for $w(\mathcal E(q))$, cf.~\eqref{eq:shell-product}:
\begin{multline*}
		w(\mathcal E(q)) 
\qquad \ge\qquad M^\ell L^k\,\Bigl[\rigweight_{q(c_\sigma)}(\product{\totweight}-\product{\rigweight})+\flxweight_{q(c_\sigma)}\product{\totweight}\Bigr] \qquad =\ \ \\ M^\ell L^k\,\Bigl[\underbrace{(\rigweight_{q(c_\sigma)}+\flxweight_{q(c_\sigma)})}_{= \totweight_{q(c_\sigma)}}\product{\totweight}-\rigweight_{q(c_\sigma)}\product{\rigweight}\Bigr]
		\qquad = \ \ M^\ell L^k\,\Bigl[\prod_{\scol \in\rigcompcol{q}}\hspace*{-0.5em}\totweight_{q(\scol)}-\prod_{\scol \in\rigcompcol{q}}\hspace*{-0.5em}\rigweight_{q(\scol)}\Bigr].
\end{multline*}

Observe now that, for any $\totweight_i$ and $\rigweight_i$   with $\totweight_i \geq \rigweight_i\geq 0$, we have the following:
		\[
		\prod_{i=1}^r \totweight_i-\prod_{i=1}^r \rigweight_i
		\ \ = \ \ \sum_{j=1}^r \Bigl[\Bigl(\prod_{i<j}\totweight_i\Bigr)\cdot (\totweight_j-\rigweight_j)\cdot \Bigl(\prod_{i>j}\rigweight_i\Bigr)\Bigr] \ \ \geq \ \ \bigl(\prod_{i<r}\totweight_i\bigr)\cdot (\totweight_r-\rigweight_r);
		\]
the inequality holds because the summands are non-negative, and so the last one does not exceed the total.
Also, for every $\{\lambda\}\in \intervals_M^{\textsf{R}}$, we have, cf.~\eqref{eq:ray2-weight}:
		\begin{equation}\label{eq:weight-compatible}
		\rigweight_{\{\lambda\}}=\binom M\lambda,\qquad
		 \flxweight_{\{2\}} \geq M,\qquad
		\flxweight_{\{\lambda\}} \ge M^2 \geq M, \text{ for } \lambda\geq 3,\qquad
		\totweight_{\{\lambda\}}\ge \binom{M}{2}.
		\end{equation}		
		Indeed,
		$\flxweight_{\{2\}}\ge w_1([2,\infty))=M$ and $\totweight_{\{2\}}\geq \rigweight_{\{2\}} = \binom{M}{2}$.  For $\lambda\ge3$, we have $\flxweight_{\{\lambda\}}\ge w_1([2,\infty)) + w_1([3,\infty)) = M + M(M-1) = M^2$,
		  whence
		$\totweight_{\{\lambda\}}\ge \binom{M}{2}$.
With these observations at hand, we derive the required lower bound for $w(\mathcal E(q))$:
\begin{multline*}
w(\mathcal E(q))  \ \ \geq \ \ M^\ell L^k\,\Bigl[\hspace*{-0.25em}\prod_{\scol \in\rigcompcol{q}}\hspace*{-0.75em}\totweight_{q(\scol)}-\hspace*{-0.5em}\prod_{\scol \in\rigcompcol{q}}\hspace*{-0.75em}\rigweight_{q(\scol)}\Bigr]  \ \ \geq\\  M^\ell L^k\,\Bigl[\bigl(\hspace*{-1.5em}\prod_{\scol \in\rigcompcol{q}\setminus \{\scol_0\}}\hspace*{-2em}\totweight_{q(\scol)}\bigr)\cdot \underbrace{(\totweight_{q(\scol_0)}-\rigweight_{q(\scol_0)})}_{=\flxweight_{q(\scol_0)}}\Bigr]  \ \ \geq \ \ M^\ell L^k \cdot \binom{M}{2}^{h - k -1} \!\!\! \cdot M,
\end{multline*}
where $\scol_0$ is any element of $\rigcompcol{q}$ and $|\rigcompcol{q}| = h - k$.
\end{proof}		
		
	Claims~\ref{claim:1} and~\ref{claim:2} give
	\[
	g(Q) \ \ \ge \ \ M^{\ell+ 1} L^k \binom{M}{2}^{h - k -1}.
	\]
	Since $M=2^m$ with $m\ge1$, we have $2^m\ge2m$ (true for $m=1,2$, then inductively
	from $2^{m+1}\ge4m\ge2(m+1)$), which is exactly $2^M\ge M^2$, and so
	$L=3^M-2^M\ge2^M\ge M^2$.
	Also, since $\frac{M(M-1)}2\ge \frac{M^2}{4}$, we have, using $k\geq 0$,
	\[
	g(Q)
	\ \ \ge \ \ M^{\ell + 1} M^{2k} 
	\Bigl(\frac{M^{2}}{4}\Bigr)^{h - k-1}
	\ \ =\ \ \frac{1}{4^{h - k-1}}M^{\ell+2h-1}
	\ \ \ge\ \ \frac{1}{4^{\,h-1}}M^{\ell+2h-1}.\qedhere
	\]
\end{proof}

\begin{thm}\label{thm:ordinary-limit}
	For every fixed $n$, $\lim\limits_{m\to\infty}\DCraig_{n,m}=1$.
\end{thm}
\begin{proof}
	Fix $\sigma=\{p_{1},\ldots,p_{n}\}$ and let $\rho_{m}=\{q_{1},\ldots,q_{m}\}$ for $m\geq 1$. Observe that for any $\sigma^{\uparrow\rho_{m}}$-cell $C$,
	$h=\abs{\atomcolours_{\sigma}^{\ge2}}$ and $\ell= \abs{\atomcolours_{\sigma}^{1}}$
	are determined by the restriction $\atomcolours_{\sigma}$ and do not depend on $m$. By the cell reduction principle for Craig interpolation (Theorem~\ref{thm:craig-factor}), it then suffices to show that, for all $\sigma^{\uparrow\rho_{m}}$-cells
	$C$ with fixed $h$ and $\ell$, we have $1-\DCraig_{C}\leq f_{h,\ell}(n,m)$ for a function $f_{h,\ell}(n,m)$ converging to $0$ for $n$ fixed and $m\to\infty$.
	
	We make a case distinction according to the value of $(h,\ell)$.
	
	\medskip\noindent
	\emph{Case $h\ge1$ and $(h,\ell)\ne(1,0)$}.
	By Lemma~\ref{lem:point-side}, every ordered biclique in $P$ can be oriented so that the first side is in $\somerigid$. Now, Lemma~\ref{lem:deficit-count},
	Theorem~\ref{thm:max-ray-deficit}, and $\abs{P}\leq (3M)^{h}$ (Lemma~\ref{lem:profile-bound}) give
	\begin{equation}
		1-\DCraig_C
		\ \ \le \ \
		2^{\,1+(3M)^h-\frac{1}{4^{h-1}}M^{\ell+2h-1}}.
		\label{eq:nonexceptional-slice}
	\end{equation}
	For fixed $h,\ell$ in this range,
	\[
	\frac{\frac{1}{4^{h-1}}M^{\ell+2h-1}}{(3M)^h}
	\ \ = \ \
	\frac{1}{4^{h-1}3^h}M^{\ell+h-1}
	\longrightarrow\infty,
	\]
	because $\ell+h-1\ge1$.  Thus~\eqref{eq:nonexceptional-slice} tends to $0$, as required.

\medskip\noindent
\emph{Case $(h,\ell)=(1,0)$}.
	Then there is exactly one $\scol\in \atomcolours_{\sigma}^{\ge2}$ and $\atomcolours_{\sigma}^{1}=\emptyset$. A rigid
	box-profile takes the form $\{\lambda\}\in\intervals_M^{\mathsf{R}}$. 
	Hence any $Q\in \somerigid$ is a non-empty subset of $\intervals_M^{\mathsf{R}}$.
	Let $\lambda^*$ be its largest value.  Then
	$\Nc_{P}(Q)
	=Q\cup\{[\mu,\infty) \mid 2\le\mu\le\lambda^*\}$.
By the definitions of $g$ and $\flxweight$,
\[
g(Q)\ \ =\ \ \sum_{\mu=2}^{\lambda^*}w_1^\bullet([\mu,\infty)) \ \
\ge\ \ \flxweight_{\{\lambda^*\}}.
\]
For $Q = \{\{2\}\}$, there are exactly
\begin{multline*}
B \ \ =\ \ 2\cdot (2^{w_1^\bullet(\{2\})}-1)\cdot (2^{|C|-w_1^\bullet(\{2\})-w_1^\bullet([2, \infty))}-1) \\ 
 \ \ = \ \ 2\cdot (2^{M^2-M} - 1)\cdot (2^{|C| - M^2} - 1) \ \ = \ \ 2^{|C| + 1-M}(1 - 2^{-(M^2-M)})(1 - 2^{-(|C|-M^2)})
\end{multline*}
non-empty ordered bicliques that have a side supported by $\{2\}$:
the only profiles compatible with $\{2\}$ are $\{2\}$ and
$[2,\infty)$. So, $\NoCraig_C\geq B$. 
Every other biclique can be oriented with a rigid side whose
maximum is $\lambda^*\ge3$. There are exactly $2^{\lambda^*-2}$ non-empty sets $Q$ of rigid box-profiles with largest value $\lambda^*$, and the number of pairs supported on $Q$ and $N_P(Q)$ is bounded by $2^{|C|-g(Q)}$.
By~\eqref{eq:weight-compatible} with $\lambda \geq 3$ and  Lemma~\ref{lem:deficit-count},
\begin{multline*}
B \ \ \leq \ \ \NoCraig_C 
\ \ \leq \ \ B \ + \  2^{|C|+1}\sum_{\lambda=3}^{M}2^{\lambda-2-\flxweight_{\{\lambda\}}}\\[-6pt]
\ \ \le \ \ B \ + \ 2^{|C|+1}(M-2)2^{M-2-M^2}\ \ \leq \ \ 2^{|C|+1 - M}\Bigl(1 + \frac{M-2}{2^{(M-1)^2 +1}}\Bigr).
\end{multline*}
Dividing both sides by $2^{|C| +1-M}$ and considering $\eta\eqdef  2^{M- (|C| + 1)}\cdot  \NoCraig_C$, we obtain 
\begin{equation*}
\Bigl(1 - \frac{1}{2^{M^2-M}}\Bigr)\Bigl(1 - \frac{1}{2^{|C|-M^2}}\Bigr) \ \ \leq \  \  \eta  \ \ \leq \ \ 1 + \frac{M-2}{2^{(M-1)^2+1}}.
\end{equation*}
%
%
%
%
%
By~\eqref{eq:dcraig-nocraig}, we have $2^M\smash{\overline{\DCraig}}_C = \eta/ (1 - 2^{-(|C| + 1)} + 2^{-M}\eta)$, and so 
$\lim\limits_{m\to\infty}2^M\smash{\overline{\DCraig}}_C=1$;
in particular, $\DCraig_C\to1$, as required.
%
%



\medskip
\noindent
\emph{Case $h=0$}. Then $\atomcolours_{\sigma}^{\ge2}=\emptyset$ and any two atoms in $C$ are compatible. It follows that $\DCraig_C=1$.
\end{proof}

\subsection{Density in $\Diff$ after Quantification}

We now consider the densities $\dens^\Pi_{n,m}$
and $\DCraig^\Pi_{n,m}$. We show that, for fixed
$n$, both converge to $0$ as $m\to\infty$ when
equivalence after existential quantification is
defined over difference frames.
Equivalence of existentially quantified modal
formulas over $\mathcal K_{\ne}$ does not imply
equivalence over $\mathcal K_{\Diff}$, as the
following example illustrates.

\begin{exa}\label{exm:counter}
	Consider the formulas
	$\varphi=\blacklozenge^{\geq 2}q
		\wedge\Diamond\blacklozenge^{=1}\neg q$ and $\psi
		=\blacklozenge^{= 1}q
		\wedge\Diamond\blacklozenge^{\geq 2}\neg q$, where $\blacklozenge^{\geq 2}\chi = \chi \land \Diamond\chi$ and $\blacklozenge^{=1} \chi = \chi \land \neg \Diamond \chi$. 
	Over $\mathcal K_{\ne}$, the formulas
	$\exists q.\varphi$ and $\exists q.\psi$
	are equivalent, since both hold precisely on
	pointed difference frames with at least three
	worlds.
	By contrast, let $W=\{0,1\}$ and
	$R=\{(0,0),(0,1),(1,0)\}$. 
	Then $(W,R)$ validates $\Diff$, but
	$(W,R),0\models\exists q.\varphi$ and
	$(W,R),0\not\models\exists q.\psi$.
	Thus the two formulas are not equivalent over
	$\mathcal K_{\Diff}$.
\end{exa}

Hence, the choice between $\mathcal K_{\ne}$
and $\mathcal K_{\Diff}$ affects the counting
of equivalence classes after existential
quantification. Since difference frames provide
the intended semantics for $\Diff$, we restrict
attention to this class throughout the section.

To show convergnce to $0$ as $m\to\infty$, by the projective cell reduction principles, it suffices to define a sequence $C_{m}$ of $\sigma^{\uparrow \rho_{m}}$-cells whose densities tend to $0$.

Fix $\sigma$ and a $\sigma$-colour $c_0$ and let
$(F_0,c_0)$ be the $\sigma$-atom with
$F_0(c_0)=\ge2$ and $F_0(c)=0$
for all $c\ne c_0$.
For $m\ge1$, let $C_m$ be the
$\sigma^{\uparrow\rho_m}$-cell with
$\sigma$-reduct $(F_0,c_0)$.
In the notation above, this cell has
$(h,\ell)=(1,0)$.
For the asymptotic argument we may assume $m\geq1$, so $M=2^{m}\geq2$.

For $D\subseteq C_m$, the formula
$\exists\vec{\rho}_m.
\MForm{}{\sig\cup\priv_m}{D}$
holds in a pointed $\sigma$-model
if all worlds have colour $c_0$ and
its cardinality satisfies the constraint
$\exists\vec{\rho}_m.\Phi(\mathcal{M})$ associated with some atom~$\mathcal{M}\in D$. As each constraint $\exists\vec{\rho}_{m}.\Phi(\mathcal{M})$ belongs to $\intervals_M$, in $\mathcal{K}_{\ne}$ the equivalence classes in
$\mathcal Q_{C_m}$ correspond to unions of sets in $\intervals_M$. 
Let $\mathcal S_M$ be the set of these unions.
\begin{lem}\label{lem:projective-classes} In $\mathcal{K}_{\ne}$, $|\mathcal{S}_{M}|=|\mathcal{Q}_{C_{m}}|= (M+1)2^{M-1}$.
\end{lem}
\begin{proof}
	A union $U\in \mathcal{S}_{M}$ containing no interval of the form $[\lambda,\infty)$ is an arbitrary subset of
	$\{2,\ldots,M\}$, giving $2^{M-1}$ possibilities. A union $U$ containing an interval $[\lambda,\infty)$ has a least  threshold $\mu\in\{2,\ldots,2M\}$ such that $[\mu,\infty)\subseteq U$. Then $\mu-1\not\in U$, while points below $\mu-1$ may
	be chosen freely. The thresholds $2\le \mu\le M+1$ therefore give
	\[
	1+1+\sum_{\mu=4}^{M+1}2^{\mu-3}\ \ =\ \ 2^{M-1}
	\]
	unions. Each of the remaining $M-1$ thresholds allows an arbitrary subset of $\{2,\ldots,M\}$ and hence gives $2^{M-1}$ unions. Thus there
	are $M2^{M-1}$ unions containing an interval of the form $[\lambda,\infty)$, and the claim follows.
\end{proof}

\begin{prop}\label{prop:critical-projective}
	For the cell $C_{m}$, 
	\[
	\DCraig_{C_{m}}^\Pi
	\ \ \le \ \ 3(M+1)\left(\frac23\right)^M,
	\qquad
	\dens_{C_{m}}^\Pi
	\ \ =\ \ \frac{2}{(M+1)2^{M-1}}.
	\]
\end{prop}

\begin{proof}
	By Theorem~\ref{thm:lftab-count1}, $|\SetCraig_{C_{m}}^{\Pi}|= 2|\mathcal{S}_{M}|-1$. All pairs of 
	disjoint non-empty subsets of the $(M-1)$-element set
	$\{2,\ldots,M\}$ correspond to inconsistent pairs of formulas. There are
	$
	3^{M-1}-2\cdot2^{M-1}+1$ 
	such pairs. Consequently
	\[
	|\SetClash_{C_{m}}^\Pi|
	\ \ \ge \ \ (2|\mathcal{S}_{M}|-1)+3^{M-1}-2\cdot2^{M-1}+1
	\ \ =\ \ 3^{M-1}+M2^M.
	\]
	Therefore, using Lemma~\ref{lem:projective-classes},
	\[
	\DCraig_{C_{m}}^\Pi
	\ \ \le \ \ \frac{2|\mathcal{S}_{M}|-1}{3^{M-1}+M2^M}
	\ \ \le \ \ \frac{(M+1)2^M}{3^{M-1}}
	\ \ =\ \ 3(M+1)\left(\frac23\right)^M.
	\]
	For uniform interpolation, the bottom and top elements of $\mathcal Q_{C_{m}}$
	have uniform $\sigma$-interpolants. We show that no other element of $\mathcal Q_{C_{m}}$ does. Assume 
	$\emptyset\ne U\subsetneq\{2,3,\ldots\}$ is represented by a union in
	$\mathcal S_M$ and choose $k\notin U$. 
Choose $D\subseteq C_m$ representing $U$. Then $\exists\vec\priv_m.\MForm{}{\sig\cup\priv_m}{D}$ holds precisely in the pointed models based on difference frames satisfying the cell characteristic formula $\Gamma(C_{m})$ and whose number of worlds belongs to $U$. Take a formula $\psi_{k}$ using symbols from $\rho_{k}'$ with $(\sigma \cup \rho_{m})\cap \rho_{k}'=\emptyset$ such that all pointed models based on difference frames satisfying $\psi_{k}$ have exactly $k$ worlds. By definition, $\Phi(D)\models \neg\psi_{k}$ but $\Gamma(C_{m})\not\models\neg\psi_{k}$. It follows that $\Gamma(C_{m})$ is not a uniform $\sigma$-interpolant of $\MForm{}{\sig\cup\priv_m}{D}$, and so by Lemma~\ref{lem:localfinitecellcharacteristic} there is none.
\end{proof}

\begin{thm}\label{thm:projective-limit}
	For every fixed $n$, $\lim\limits_{m\to\infty}\DCraig^\Pi_{n,m}
	=
	\lim\limits_{m\to\infty}\dens^\Pi_{n,m}=0$.
\end{thm}

\begin{proof}
	By the two projective cell reduction principles,
	Theorems~\ref{thm:craig-factor5} and~\ref{thm:slicereduction5}, both
	densities are products of cell factors in $[0,1]$. 
	Hence each product is bounded above by $\DCraig_{C_{m}}^\Pi$
	and $\smash{\dens_{C_{m}}^\Pi}$, respectively.
	Proposition~\ref{prop:critical-projective} gives the two
	bounds, and both tend to zero.
\end{proof}

\section{Conclusion}
We have obtained initial results on the density of formulas with Craig and uniform interpolants for quasi-normal modal logics. Many questions remain open. We begin with those arising directly from our results.
\begin{enumerate}
	\item Is there any (quasi-)normal modal logic with $\lim_{m\to\infty}\DCraig_{n,m}\not=1$ for some fixed $n$? If no general results can be obtained, how do standard modal logics lacking CIP, such as \textsf{K4.3}, \textsf{S4.3}, weak \textsf{K4}, and product modal logics, behave?
	
	\item Do we have $\lim_{m\to\infty}\DCraig^{\Pi}_{n,m}=1$ for the class of all frames validating $\Grzd$ when $\Delta=3$ or, more generally, when $\Delta\geq 3$?
	How do other standard modal logics, such as those listed under Point~1, behave with respect to $\lim_{m\to\infty}\DCraig^{\Pi}_{n,m}$ and $\lim_{m\to\infty}\dens^{\Pi}_{n,m}$?
	
	\item Inspired by Maksimova's classification of CIP for normal extensions of $\Sfour$~\cite{Maksimova1978,Maksimova1980,Maksimova1981,DBLP:journals/corr/abs-2604-22020}, can we also classify normal extensions of $\Sfour$ according to the behaviour of $\lim_{m\to\infty}\DCraig^{\Pi}_{n,m}$ and/or (uniform) BICIP?
	
	\item In this article, we have always kept one parameter fixed. A natural next question is to determine \emph{phase transition windows} when the retained and forgotten signatures grow simultaneously. For the logic of the~3-fork, this has been done in Example~\ref{exm:phase} (see also Example~\ref{ex:final}), and our quantitative estimates also already provide partial information. For instance, for ordinary uniform interpolation, the density tends to $1$ whenever $|\Atom{}{\sigma_n}|2^{-2^m}\to 0$, while it tends to $0$ whenever $2^n2^{-|C_{m}|}\to\infty$. Thus, the present results already delimit regions on either side of a possible phase transition, but they are generally not sharp enough to identify the window. Determining these regions more precisely and whether sharp thresholds exist for natural (classes of) modal logics remains an open problem.
	
	\item Throughout, the modal depth bound $d$ has been kept fixed. Allowing $d$ to grow together with the signatures introduces a further asymptotic parameter. Note that the number and size of atoms and cells can depend strongly on $d$.
\end{enumerate}

The results presented in this paper can easily be extended to quasi-normal modal logics with multiple modal operators of arbitrary arity.
More care is needed for classical modal logics without the normality axiom. For these logics, all proofs still go through, except for the proof of Lemma~\ref{lem:constants:atomsets} about pinning. Thus, the asymptotic behaviour of uniform and Craig interpolants when the number of retained variables grows remains open. On the other hand, it is easy to see that if the universal modality (a modal operator $\Box_{u}$ such that $w\models \Box_{u}p$ iff $p$ is true in all worlds) is definable in a classical modal logic, then the lifting lemmas can be proved as before. This applies, for instance, to temporal modal logics such as linear temporal logic (LTL).

A critical property on which some of our proofs rely is that any propositional variable can be interpreted in the same way as $\top$ or $\bot$ and can be uniformly replaced by its negation without affecting satisfiability. This is not the case, for instance, for modal logics with nominals, and their asymptotic behaviour could therefore be rather different.

It would be interesting to investigate alternative counting measures. For instance, how do the densities change if formulas are counted modulo symmetries of the retained variables, such as flips and permutations, if the number of disjuncts in their normal forms is restricted, or if one measures interpolation failure at the level of individual cells rather than globally?
  
Moving towards non-modal logics, the techniques developed in this paper can be directly applied to some natural fragments of first-order logic without function symbols and lacking CIP, such as the $2$-variable fragment~\cite{DBLP:conf/lics/JungW21}. Non-classical logics, such as super-intuitionistic and many-valued logics, might require rather different approaches~\cite{DBLP:journals/corr/abs-2512-01600}.

\bibliographystyle{alphaurl}
\bibliography{kwz,local,main,related_work_density}
\end{document}